\documentclass[amsmath,aps,prl,reprint,longbibliography,superscriptaddress]{revtex4-2}
\usepackage{graphics}
\usepackage{graphicx}
\usepackage{amsmath}
\usepackage{amssymb}
\usepackage{enumerate}
\usepackage{xfrac}
\usepackage[caption=false]{subfig}
\usepackage[normalem]{ulem}
\usepackage{color}
\usepackage{txfonts}
\usepackage[mathscr]{euscript}
  \usepackage[pdftex,
        pdflang={en-US},
        colorlinks=true,
	    pdfauthor={Pontus Laurell},
	    pdftitle=[Extraction of the interaction parameters for RuCl3 from neutron data using machine learning]{hyperref}
\usepackage{verbatim}
\usepackage{gensymb}
\usepackage{ulem}
\pdfmapfile{=md-chr7v.map}

\begin{document}

%%%%%%%%%%%%%%%%%%%%%%%%%%%%%% Mandatory DOE disclaimer
%\setcounter{page}{0}
%
%Notice: This manuscript has been authored by UT-Battelle, LLC, under contract DE-AC05-00OR22725 with the US Department of Energy (DOE). The US government retains and the publisher, by accepting the article for publication, acknowledges that the US government retains a nonexclusive, paid-up, irrevocable, worldwide license to publish or reproduce the published form of this manuscript, or allow others to do so, for US government purposes. DOE will provide public access to these results of federally sponsored research in accordance with the DOE Public Access Plan (http://energy.gov/downloads/doe-public-access-plan).
%
%\newpage
%\quad
%\pagebreak

%\title{High-dimensional modeling of the proximate Kitaev spin liquid material RuCl$_3$}
\title{Extraction of interaction parameters for \texorpdfstring{$\alpha$-RuCl$_3$}{RuCl3} from neutron data using machine learning}
%\title{\anj{A Machine-Learning workflow to tackle the interaction parameters for \texorpdfstring{$\alpha$-RuCl$_3$}{RuCl3} from inelastic neutron scattering}}
\author{Anjana M. Samarakoon}
% \email[]{asamarakoon@anl.gov}
\email{asamarakoon@anl.gov}
\affiliation{Shull-Wollan Center, Oak Ridge National Laboratory, Oak Ridge, Tennessee 37831, USA}
\affiliation{Neutron Scattering Division, Oak Ridge National Laboratory, Oak Ridge, Tennessee 37831, USA}
\affiliation{Materials Science Division, Argonne National Laboratory, Lemont, Illinois 60439, USA}

% \footnote{asamarakoon@anl.gov}
% \thanks{Current address: Materials Science Division, Argonne National Laboratory, Argonne, IL, USA.}
\author{Pontus Laurell}
\affiliation{Center for Nanophase Materials Sciences, Oak Ridge National Laboratory, Oak Ridge, Tennessee 37831, USA}
\affiliation{Computational Sciences and Engineering Division, Oak Ridge National Laboratory, Oak Ridge, Tennessee 37831, USA}
\affiliation{Department of Physics and Astronomy, University of Tennessee, Knoxville, Tennessee 37996, USA.}
\author{Christian Balz}
\affiliation{Neutron Scattering Division, Oak Ridge National Laboratory, Oak Ridge, Tennessee 37831, USA}
\affiliation{ISIS Neutron and Muon Source, Rutherford Appleton Laboratory, Didcot, OX11 0QX, United Kingdom}
\author{Arnab Banerjee}
\affiliation{Neutron Scattering Division, Oak Ridge National Laboratory, Oak Ridge, Tennessee 37831, USA}
\affiliation{Department of Physics and Astronomy, Purdue University, West Lafayette, Indiana 47906, USA}
\author{Paula Lampen-Kelley}
\affiliation{Department of Materials Science and Engineering, University of Tennessee, Knoxville, Tennessee 37996, USA}
\affiliation{Materials Science and Technology Division, Oak Ridge National Laboratory, Oak Ridge, Tennessee 37831, USA}
\author{David Mandrus}
\affiliation{Department of Materials Science and Engineering, University of Tennessee, Knoxville, Tennessee 37996, USA}
\affiliation{Materials Science and Technology Division, Oak Ridge National Laboratory, Oak Ridge, Tennessee 37831, USA}
\author{Stephen E. Nagler}
\affiliation{Neutron Scattering Division, Oak Ridge National Laboratory, Oak Ridge, Tennessee 37831, USA}
\affiliation{Quantum Science Center, Oak Ridge National Laboratory, Oak Ridge, Tennessee 37831, USA}
\author{Satoshi Okamoto}
\affiliation{Materials Science and Technology Division, Oak Ridge National Laboratory, Oak Ridge, Tennessee 37831, USA}
\affiliation{Quantum Science Center, Oak Ridge National Laboratory, Oak Ridge, Tennessee 37831, USA}
\author{D. Alan Tennant}
\affiliation{Shull-Wollan Center, Oak Ridge National Laboratory, Oak Ridge, Tennessee 37831, USA}
\affiliation{Department of Physics and Astronomy, University of Tennessee, Knoxville, Tennessee 37996, USA.}
\affiliation{Materials Science and Technology Division, Oak Ridge National Laboratory, Oak Ridge, Tennessee 37831, USA}
\affiliation{Quantum Science Center, Oak Ridge National Laboratory, Oak Ridge, Tennessee 37831, USA}
\date{\today}

%%%%%%%%%%%%%%%%%%%%%%%%%%%%%%%%%%%%%%%%%%%%%%%
% Abstract
%%%%%%%%%%%%%%%%%%%%%%%%%%%%%%%%%%%%%%%%%%%%%%%
% This is on the long side for a PRL abstract.
\begin{abstract}
	Single crystal inelastic neutron scattering data contain rich information about the structure and dynamics of a material. Yet the challenge of matching sophisticated theoretical models with large data volumes is compounded by computational complexity and the ill-posed nature of the inverse scattering problem. Here we utilize a novel machine-learning-assisted framework featuring multiple neural network architectures to address this via high-dimensional modeling and numerical methods. A comprehensive data set of diffraction and inelastic neutron scattering measured on the Kitaev material $\alpha-$RuCl$_3$ is processed to extract its Hamiltonian. Semiclassical Landau-Lifshitz dynamics and Monte-Carlo simulations were employed %at a temperature much higher than the expected quantum gap 
	to explore the parameter space of an extended Kitaev-Heisenberg Hamiltonian. A machine-learning-assisted iterative algorithm was developed to map the uncertainty manifold to match experimental data; a non-linear autoencoder used to undertake information compression; and Radial Basis networks utilized as fast surrogates for diffraction and dynamics simulations to predict potential spin Hamiltonians with uncertainty. Exact diagonalization calculations were employed to assess the impact of quantum fluctuations on the selected parameters around the best prediction.
\end{abstract}
\maketitle

\begin{figure}[!htb]
\centering\includegraphics[width=\columnwidth]{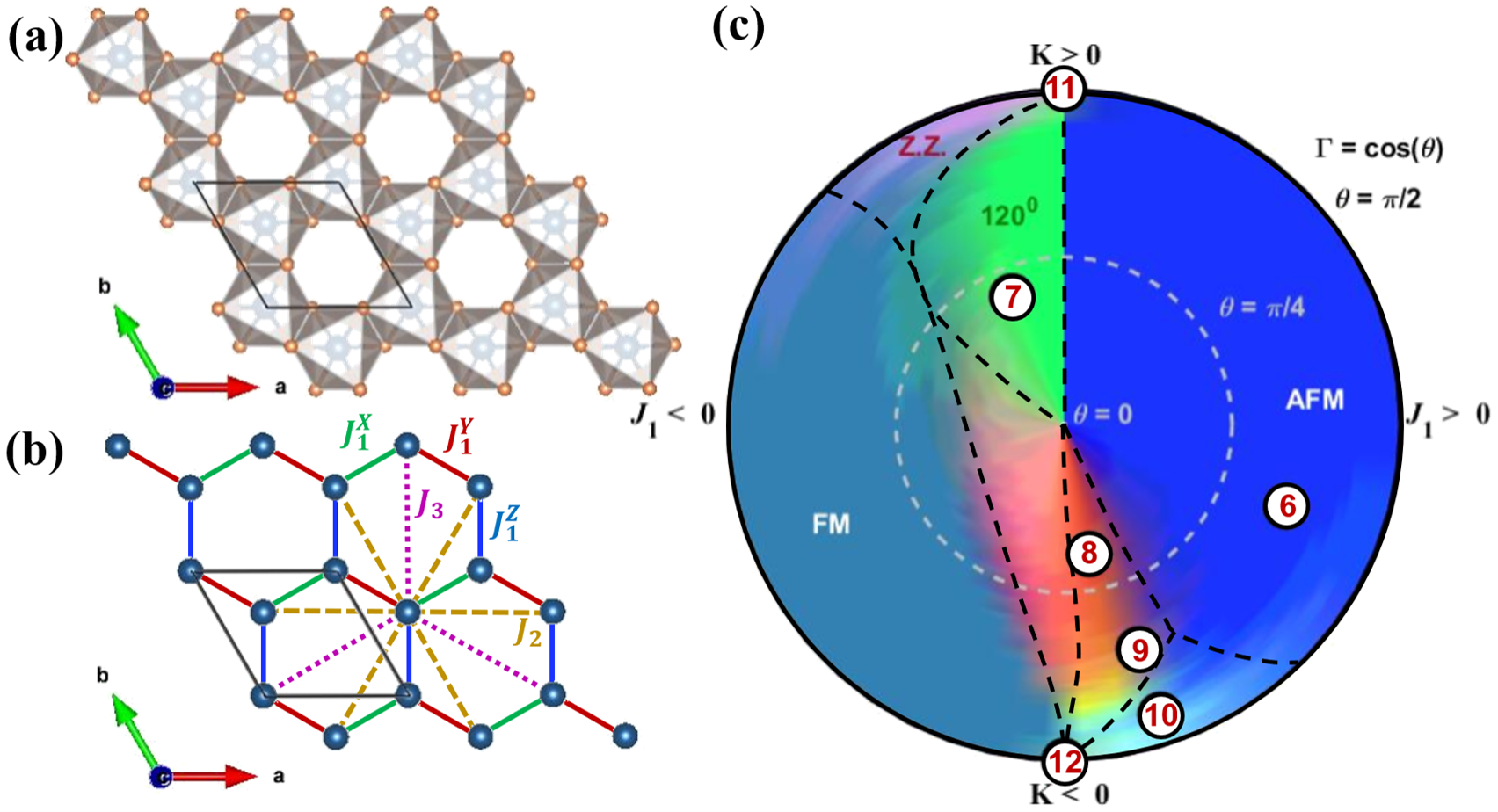}
% \centering\includegraphics[width=\columnwidth]{RuCl3Figure_01.png}
%\centering\includegraphics[width=0.99\textwidth]{sqw_rb2mnf4.jpg}
	\caption{
	(a) $\alpha$-RuCl$_3$ crystal structure, consisting of Ru sites at centers of  edge-sharing Cl octahedra. 
	(b) The magnetism is due to Ru$^{3+}$ ions, which form a honeycomb lattice. Nearest, second-nearest, and third-nearest neighbor bonds are indicated by solid, dashed, and dotted lines respectively. Anisotropic nearest-neighbor interactions are considered as given in Eq.~\eqref{eq:hamiltonian}.
	%\anj{(b) The Ru-ions the only magnetic species in the system are on the honeycomb lattice. A spin-Hamiltonian up to 3rd nearest neighbor (NN) interaction is considered as marked in panel (b). (Solid) The 1st NN exchanges are considered to be anisotropic as given in Eq. \ref{eq:hamiltonian}.  }
%     Three different anisotropic  exchange tensors ($J_1^p$)  along the principle axes ( $p = X$, $Y$ and $Z$ ) perpendicular to each other and the Ru-O-Ru-O plaquettes, are assigned for the 1st NN bonds. Both (dashed) the 2nd NN and (dotted) the 3rd NN  exchnages are treated as isotropic.
%     (b) Nearest (solid), second-nearest (dotted), and third-nearest (dashed) neighbor bonds on the honeycomb lattice.  \anj{The  }
%     The Kitaev coupling on ``X'' bonds (green) produces an $S_i^xS_j^x$ spin-spin interaction, and similarly on other bonds.
	(c) Machine-learned phase diagram for the $J_1-K-\Gamma$ model at $T=1$~K, 
	with theoretical phase boundaries from Ref.~\cite{PhysRevLett.112.077204} overlaid in black dashed lines. 
	The different colors, which represent different structure factors $S(\textbf{Q})$, are predicted by a trained neural network as explained in the main text.
	%\anj{The color code in which different colors represent different structure factors $S(Q)$,  is predicted by a trained Neural Network (NN) as explained later in the main text.} 
	Numbered points correspond to $S(\textbf{Q})$ shown in Fig.~\ref{fig:phasediagram}(b). The identified phases include ferromagnetic [FM], N\'eel [AFM], $120^\degree$, and zigzag [Z.Z.] orders. 
	}
\label{fig:lattice}
\end{figure}

\emph{Introduction.}---%
Highly frustrated quantum systems are important routes to realizing exotic ground states and excitations. They are proposed to host states ranging from long-range entangled quantum spin liquids (QSLs) with nonlocal excitations to quantum spin ices with emergent photons \cite{Balents2010, Lacroix2011, Savary2017}. Recently, the two-dimensional honeycomb spin-1/2 material $\alpha$-RuCl$_3$ [Fig. \ref{fig:lattice}(a)] has garnered particular attention following being reported \cite{PhysRevB.90.041112, PhysRevLett.114.147201, Banerjee2016, Banerjee2017, Do2017, PhysRevLett.118.107203, Banerjee2018, Balz2019} as a leading candidate \cite{Winter2017, Takagi2019, doi:10.7566/JPSJ.89.012002} for realization of the Kitaev model---an exactly solvable QSL Hamiltonian \cite{Kitaev2006, Hermanns2018}. The Kitaev model is a spin network with competing bond-dependent interactions and hosts a topological QSL ground state that supports two types of fractionalized excitations: visons, which are excitations of the emergent flux, and deconfined Majorana fermions. These quasiparticles are predicted to show non-Abelian statistics, %mutual non-Abelian statistics, with respect to one another,
suggesting potential applications in e.g. topological quantum computing \cite{RevModPhys.80.1083}. Recently, theoretical propositions have been made for interferometers utilizing their braiding statistics as a precursor to undertaking quantum operations \cite{PhysRevX.10.031014, PhysRevLett.126.177204}. Meanwhile, however, the experimental situation regarding the quasiparticles in $\alpha$-RuCl$_3$  remains inconclusive, primarily due to difficulties in determining the precise nature of the spin couplings in the material and to what extent these destabilize the QSL state in zero and applied magnetic fields.

\begin{figure*}[!htb]
\centering\includegraphics[width=0.95\textwidth]{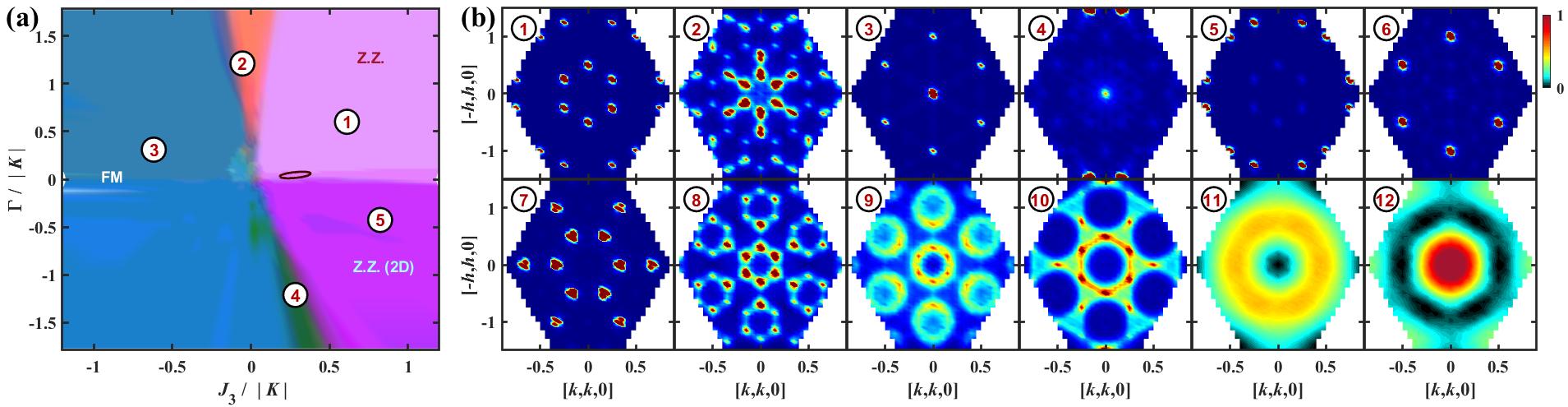}
%\centering\includegraphics[width=0.99\textwidth]{sqw_rb2mnf4.jpg}
	\caption{
	(a) Machine-learned phase map varying $\Gamma/|K|$ and $J_3/|K|$  through the optimal solution for $\alpha$-RuCl$_3$ at fixed $J_1/|K|=-0.1$ and $J_2/|K|=0$, with $K<0$. Labeled phases include ferromagnetic [FM], zigzag [Z.Z.] and planar zigzag [Z.Z. (2D)] orders. The ellipsoidal approximation to the optimal solution is marked in panel (a) in dark-red. Due to uneven sampling of the IMA process from which the data was taken, the prediction accuracy varies in parameter space resulting in a blotchy appearance. 
	The different colors, which represent different structure factors $S(\textbf{Q})$, are predicted by a trained neural network as explained in the main text.
    %\anj{The color code in which different colors represent different structure factors $S(Q)$,  is predicted by a trained Neural Network (NN) as explained later in the main text.}
	%is worse in some regions and the splotchs here and there in the phase map are prediction artifacts.
	Panel (b) shows the surrogate predicted neutron structure factor, $S^{\rm Sur}(\textbf{Q}),$ at numbered locations indicated in (a) and Fig.~\ref{fig:lattice}(c). Corresponding spin structures for the long-range ordered phases labeled 1, 3-7 are given in the SM \cite{SuppMat}. Note that Z.Z. and Z.Z. (2D) have similar long-range order, but along different spin-orientation directions. 
	This occurs because, as discussed in Ref.~\cite{PhysRevB.94.064435}, when $K<0$ the spin orientation in zigzag ground states depends on the sign of $\Gamma$.
    Due to the neutron spin polarization factor, the two $S(\textbf{Q})$ are different in the shown plane, despite yielding the same trace over spin correlations, $\sum_\alpha S^{\alpha\alpha} \left(\mathbf{Q}\right)$. 
	%\anj{Ref. \cite{PhysRevB.94.064435} also reports change of spin orientation in the magnetic ground state with change of sign on the $\Gamma$-term.}
	%
	%However, the trace of spin correlations functions looks exactly the same. 
	Some of the diffuse $S(\textbf{Q})$, taken at either phase boundaries or critical points, are also shown. The $S(\textbf{Q})$ labeled 11 and 12 are for the pure AFM and FM Kitaev models.
	}
\label{fig:phasediagram}
\end{figure*}

Experiments have revealed evidence that $\alpha$-RuCl$_3$ is close to the Kitaev QSL \cite{Winter2017, Takagi2019, PhysRevB.103.174417, Suzuki2021}. At low temperatures and magnetic fields it orders magnetically in a zigzag structure \cite{PhysRevB.91.144420, PhysRevB.92.235119, PhysRevB.93.134423, Banerjee2016}, implying the presence of symmetry-allowed interactions additional to the Kitaev Hamiltonian, as is generically predicted by theory \cite{PhysRevLett.102.017205, PhysRevLett.112.077204, PhysRevB.93.214431, PhysRevB.100.075110}. Inelastic neutron scattering (INS) shows scattering dominated by continua at the zone center \cite{Banerjee2017, Do2017, PhysRevLett.118.107203, Banerjee2018, Balz2019}, interpreted as originating from underlying fractional Majorana excitations, or from incoherent excitations due to magnon decay \cite{10.1038/s41467-017-01177-0, PhysRevLett.120.077203, PhysRevResearch.2.033011}; both related to strongly fluctuating quantum states. Similarly, Raman scattering shows a broad scattering continuum at the zone center \cite{PhysRevLett.114.147201, Nasu2016, Du2018, PhysRevB.100.134419, Wang2020a}, and a fermionic temperature dependence, interpreted as indicating fractional excitations. The zigzag order melts in a narrow range of applied in-plane magnetic fields, possibly inducing a QSL state \cite{Banerjee2018,Balz2019,PhysRevLett.119.227208, PhysRevLett.119.037201, PhysRevB.95.180411}. Oscillations of the thermal conductivity were also observed in this field range, suggesting the presence of a Fermi surface \cite{Czajka2021, Inti2021, PhysRevB.104.165133}. Perhaps the most striking reports are those of a half-integer-quantized thermal Hall effect in the same field range \cite{10.1038/s41586-018-0274-0, PhysRevB.102.220404, Yokoi_2021}. Additional experimental evidence for Kitaev interactions in $\alpha$-RuCl$_3$ has been reported using e.g. inelastic X-ray scattering \cite{Li2021Phonon}, thermodynamical \cite{PhysRevB.96.041405, Do2017, PhysRevB.99.094415, PhysRevLett.125.097203, Tanaka2022}, NMR \cite{PhysRevLett.119.037201, Jansa2018}, electron spin resonance \cite{PhysRevLett.125.037202}, microwave absorption \cite{PhysRevB.98.184408}, thermal transport \cite{PhysRevB.95.241112, PhysRevLett.120.217205}, and THz spectroscopy \cite{PhysRevLett.119.227201, PhysRevLett.119.227202, PhysRevB.98.094425, PhysRevB.100.085108, Reschke2019} techniques.

The complexity of magnetic interactions in RuCl$_3$ has hindered determination of an underlying model. Various groups have fit or derived proposed Hamiltonian parameters
%spin Hamiltonians 
for the material \cite{PhysRevB.93.214431, 10.1038/s41467-017-01177-0, PhysRevB.98.060412, PhysRevB.98.094425, PhysRevB.91.241110, PhysRevB.93.155143, Yadav2016, PhysRevB.97.134424, Suzuki2019, PhysRevB.96.054410, PhysRevB.96.115103, PhysRevB.100.075110, PhysRevLett.118.107203, Banerjee2016, PhysRevB.100.085108, PhysRevB.103.174417, Li2021, Suzuki2021, KejingRan:27501}, but these studies disagree significantly about which interactions are present, and on values of specific interaction parameters \cite{PhysRevB.96.064430, Laurell2020, PhysRevResearch.2.033011}. 
Part of the reason for this lack of agreement is that many experimental fits have relied on linear spin wave theory (LSWT), which cannot account for the quantum fluctuations inherent to $\alpha$-RuCl$_3$.
However, the more central issue is that a comparatively large set of weak perturbations are possible, that can significantly modify the magnetic ordering, dynamics and thermal properties of Kitaev materials. With such a high-dimensional parameter space, comparing modeling with experimental data leaves a great deal of uncertainty, unless comprehensive enough to explore the range of possible interactions; an approach which is absent to date.

Scattering data contain considerable information on the magnetic states and interactions in materials. % as well as interactions. 
A difficult step in the quantification of models has been inversion from measured data to a model---the so-called inverse scattering problem, which is usually ill-posed due to loss of phase information. % in scattering. 
In this regard, machine learning (ML) \cite{Mehta2019, doi:10.1080/23746149.2020.1797528} has shown promising results \cite{Doucet_2021, Butler_2021, doi:10.1063/5.0049111, Yu2021, Samarakoon2021}. 
Here we combine ML approaches with large-scale semi-classical simulations (SCSs) \cite{Samarakoon2021}.
ML-SCS techniques have been used to successfully extract couplings from diffuse neutron scattering data and yielded %as well as yielding 
significant insight %from neutron scattering data 
by mapping the physical behavior in %the 
high-dimensional interaction spaces of materials \cite{Samarakoon2020,Samarakoon2021}. We extend these methods to include dynamics data %available
for $\alpha$-RuCl$_3$, allowing a comprehensive fit.

\emph{Experiments.}---%
%\subsubsection{Neutron Diffraction Experiment}
%% Christian's text
Elastic neutron studies were performed at the Spallation Neutron Source (SNS) \cite{mason2006spallation} using the CORELLI beamline \cite{Rosenkranz2008}. A 125 mg $\alpha$-RuCl$_3$ crystal was mounted on an aluminium plate and aligned with the $[h, 0, l]$ plane horizontal. The crystal was rotated through $170$ degrees in $2^\degree$ steps about the vertical axis. The temperature of the measurement was 2 K and the perpendicular wave vector transfer was integrated in the range $l$ = [0.92, 1.08] r.l.u.. The elastic scattering data was previously published as Supplementary Figure S2(a) in Ref.~\cite{Banerjee2018}.

INS was performed on a $0.7$ g single crystal, % with a mass of 0.7 g. 
which was 
%This was 
sealed in a thin-walled aluminium can with 1 atmosphere of Helium gas for thermal contact. Measurements at $4$ K were carried out using the SEQUOIA spectrometer \cite{Granroth2006, Granroth2010} at the SNS. The incident energy was set to $E_i = 22.5$~meV. % and the temperature to 4 K. 
The crystal was mounted with $[h, 0, 0]$ and $[0, 0, l]$ axes in the horizontal plane, and the orthogonal $[0.5k, -k, 0]$ axis pointing vertically upwards. Data were collected by rotating the crystal about the vertical axis over $290^\degree$ in $1^\degree$ steps. 
The data are integrated over the range $[0, 0, l]$= [-3.5, 3.5].
%The data are integrated over ranges $[0.5k, -k, 0]$=[-0.05, 0.05] and (0, 0, L)=[-2, 2].

\emph{Modeling.}---%
%
%Based on current understanding we
We consider a generalized spin-$1/2$ Kitaev-Heisenberg spin (local moment) Hamiltonian,
% \begin{align}
% % \begin{eqnarray}
% H	&=	\sum_{\langle i,j\rangle} \left[ J_1 \mathbf{S}_i \cdot \mathbf{S}_j + KS_i^\gamma S_j^\gamma + \Gamma \left( S_i^\alpha S_j^\beta + S_i^\beta S_j^\alpha \right) \right]	
% &+ J_2 \sum_{\langle\langle i,j\rangle\rangle} \mathbf{S}_i \cdot \mathbf{S}_j + J_3 \sum_{\langle\langle\langle i,j\rangle\rangle\rangle} \mathbf{S}_i \cdot \mathbf{S}_j,	\label{eq:hamiltonian}
% % \end{eqnarray}
% \end{align}
\begin{align}
% \begin{eqnarray}
H	&=	\sum_{\gamma = X,Y,Z} \sum_{\langle i,j\rangle^{\gamma}}\mathbf{S}_i \cdot J_1^{\gamma}  \cdot \mathbf{S}_j \nonumber\\
&+ J_2 \sum_{\langle\langle i,j\rangle\rangle} \mathbf{S}_i \cdot \mathbf{S}_j + J_3 \sum_{\langle\langle\langle i,j\rangle\rangle\rangle} \mathbf{S}_i \cdot \mathbf{S}_j,	\label{eq:hamiltonian}
% \end{eqnarray}
\end{align}
on the honeycomb lattice [Fig.~\ref{fig:lattice}(b)], which is expected to capture relevant interactions in the 2D plane. $\langle \dots \rangle$, $\langle\langle \dots \rangle\rangle$, and $\langle\langle\langle \dots \rangle\rangle\rangle$ represent nearest, next-nearest and third-nearest neighbors, respectively, and
% \begin{align}
% \begin{eqnarray}
${\footnotesize  J_1^{X}	=	\begin{bmatrix}
J_1+K & 0 & 0\\
0 & J_1 & \Gamma\\
0 & \Gamma & J_1\\
\end{bmatrix},     
 J_1^{Y}	=	\begin{bmatrix}
J_1 & 0 & \Gamma\\
0 & J_1+K & 0\\
\Gamma & 0 & J_1\\ 
\end{bmatrix},    \\   
 J_1^{Z}	=	\begin{bmatrix}
J_1 &\Gamma &0\\
\Gamma & J_1 & 0\\
0& 0 & J_1+K\\
\end{bmatrix}	\nonumber\\}$
% \label{eq:hamiltonian}
% \end{eqnarray}
% \end{align}
\newline
\hfill \break
%These exchange
Exchange matrices are defined in the $\{X,Y,Z\}$ coordinate system %in which principal axes are along %the
with principal axes along 
mutually orthogonal normal vectors of three nearest neighbor Ru-Cl-Ru-Cl plaquettes.
% The Greek indices satisfy $(\alpha,\beta,\gamma)=(x,y,z)$ for the $Z$ bond, with other bonds obtained by cyclic permutation. 
Our model includes nearest-neighbor Heisenberg ($J_1$), Kitaev ($K$) and symmetric off-diagonal Gamma ($\Gamma$) interactions, as well as second- ($J_2$) and third-nearest ($J_3$) Heisenberg exchanges.  %In the $J_2=0$ limit 
For $J_2=0$ 
it reduces to a proposed minimal model for $\alpha$-RuCl$_3$ \cite{PhysRevB.93.214431}. %The number of terms in 
Eq.~\eqref{eq:hamiltonian} is, however, restricted compared to some %other
proposed models, notably neglecting e.g. interlayer exchange \cite{Balz2019, PhysRevB.101.174444, PhysRevB.103.174417}, and the $\Gamma'$ term associated with trigonal distortion \cite{PhysRevLett.112.077204}. There are conflicting reports as to the magnitude of $\Gamma'$ \cite{Laurell2020, PhysRevResearch.2.033011, Li2021}, such that neglecting it may not be fully justified. Nevertheless, this choice of Hamiltonian allows us to reduce the computational complexity, and to clearly present our proposed method and its capabilities. We note that ML-SCS techniques have been used to theoretically explore phase diagrams of related %generalized Kitaev models
Hamiltonians \cite{PhysRevResearch.3.023016, PhysRevResearch.3.033223}.

To simulate spin structure and dynamics, %a combination of 
Metropolis (Monte Carlo) sampling \cite{Metropolis1953} and Landau-Lifshitz (LL) dynamics is used \cite{Samarakoon2021}. This incorporates effects beyond LSWT, while %still 
achieving sufficiently good performance to allow generating a sufficient amount of training data. %The spin-$1/2$ 
Spin-$1/2$ 
operators in Eq.~\eqref{eq:hamiltonian} are approximated by classical spin vectors subject to semiclassical normalization, $|\mathbf{S}_i| = \sqrt{S(S+1)}$. Metropolis sampling is carried out at fixed temperature, yielding well-thermalized spin configurations. Spin dynamics is governed by the usual LL equations of motion, see %the 
Supplemental Material (SM) \cite{SuppMat}. % derived from the Poisson bracket of the Hamiltonian \eqref{eq:hamiltonian}. 
The LL equation is solved numerically using a fourth-order Runge-Kutta algorithm with adaptive step size \cite{Huberman_2008}. We use a cluster of 20x20 unit cells (2400 spins) with periodic boundary conditions \cite{SuppMat}. %Spin structure factors are obtained by Fourier transforming real-space correlation functions as described in \cite{Samarakoon2021}.
Neutron magnetic form factor for Ru$^{\rm 3+}$, polarization factors, and instrumental resolution are accounted for, to match with experimental data. Figure \ref{fig:phasediagram}(b) shows sampling of diffuse scattering at different locations in %the high-dimensional 
parameter space. The simulated scattering
 , $S^{\rm sim}(\textbf{Q})$,
% , $S^{\rm sur}(Q)$,
shows complex %and rich 
behavior, reflecting the rich physics of Eq.~\eqref{eq:hamiltonian}.
% The simulated scattering signals show complex and rich behavior, reflecting the rich physics of Eq.~\eqref{eq:hamiltonian}.

\emph{Machine learning method.}---%
\begin{figure}
\centering\includegraphics[width=\columnwidth]{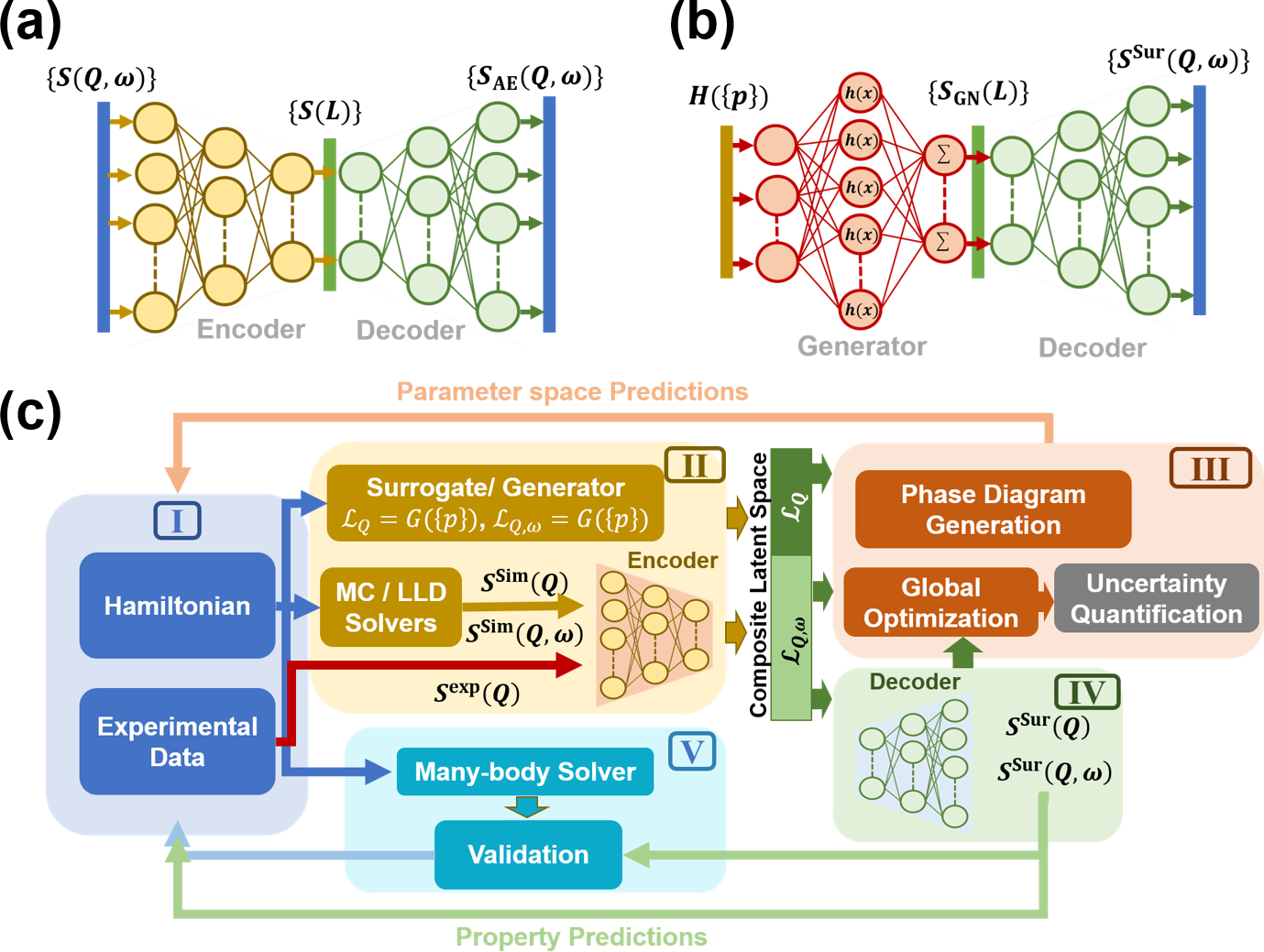}
%\centering\includegraphics[width=\columnwidth]{RuCl3Figure_04.png}
%\centering\includegraphics[width=0.99\textwidth]{sqw_rb2mnf4.jpg}
	\caption{
	%Schematic illustrations of autoencoder, surrogate model, and machine-learning integration into the direct and inverse scattering problem.
	Schematic illustrations of ML methods.
	(a) Autoencoder architecture used to compress 3D/4D $S(\textbf{Q})$/$S(\textbf{Q},\omega)$ volumes into a much lower dimensionality. The autoencoder is trained with simulated data to reproduce the input. The architecture consists of %is composed of 
	two networks: {\it Encoder} and {\it Decoder}. The output of {\it Encoder} contains a compressed (latent space) representation, $S(L)$ of the input $S(Q, \omega)$. {\it Decoder} %takes the $S(L)$ and decompresses the information 
	decompresses $S(L)$ 
	into a filtered structure factor, $S_{\rm AE}(Q, \omega),$ with the original dimensionality. 
	(b) Schematic design of constructing the surrogate for predicting $S(\textbf{Q})$/$S(\textbf{Q}, \omega)$ %for a given 
	given a set of
	model parameters. The surrogate comprises a radial basis network (RBN) as the {\it Generator} mapping parameter space, $H(\{p\}),$ to latent space $S_{\rm GN}(L)$
    and a {\it Decoder} to reconstruct $S(\textbf{Q})$/$S(\textbf{Q}, \omega)$ from latent space representations. The generator network (GN) is trained with simulated data for evaluated parameter sets $\{p\}$ as input and corresponding $S(L)$ as target.
 	(c) The ML workflow implemented here to integrate scattering experiments with theory, and extract model parameters and phase-diagram information. %The workflow is similar to what is reported in Ref. \ref{arXiv} except the extra validation step to compare LLD simulations with Quantum Many-Body Solver (QMBS) results. The workflow is split into five main sections: (I) scattering experiment design and optimization; (II) parameter space exploration and information compression; (III) structure or property predictions; (IV) parameter space predictions; and (V) validation with QMBS. Section II links to both III and IV via a composite latent space, $\mathcal{L}_Q$-$\mathcal{L}_{Q,\omega}$, a compressed version of the large pixel space.
	}
\label{fig:flowchart}
\end{figure}%
%
%Neutron scattering experiments provide rich information about spin-spin correlations. Diffraction data contain the static component while spectroscopy probes the dynamics. 
Our method builds on Ref.~\cite{Samarakoon2020}, which recently demonstrated that an ML-integrated method can be used with the experimental static structure factor, $S^{\rm exp}(\textbf{Q}),$ to extract Hamiltonian parameters from diffuse scattering data on a spin ice. Unlike spin ice, $\alpha$-RuCl3 shows a magnetic diffraction pattern with sharp Bragg peaks associated with long-range order, which does not sufficiently constrain the model parameters. Thus we extend the method to account also for the dynamical structure factor, $S^{\rm exp}(Q,\omega)$ from spectroscopy. %, which contains rich information about magnetic correlations and interactions. 
Although finding a single model to explain the entire 4D scattering is a formidable task, doing so should help avoid fits biased by incomplete information.

%As recently demonstrated in Ref. \cite{Samarakoon2020}, the experimental static structure factor, $S^{\rm exp}(Q),$ can be used to extract exchange energies of the spins from diffuse scattering data. A ML-integrated method was used to solve the inverse scattering problem, which is usually ill-posed due to loss of phase information in scattering. Unlike the magnetic diffuse scattering, a magnetic diffraction pattern with sharp Bragg peaks associated with a long-range order, does not sufficiently constrain the model parameters. Meanwhile, the dynamical structure factor, $S^{\rm exp}(Q, \omega )$ from spectroscopy experiments contains abundant information on the interactions of the magnetic system, although finding a single model to explain the entire 4-dimensional scattering for $S^{\rm exp}(Q, \omega )$, is a formidable task.

%Similarly to Ref. \cite{Samarakoon2020}, a 
A machine-learning-integrated workflow with autoencoder training and global optimization was used to simultaneously fit both $S(\textbf{Q})$ and $S(\textbf{Q}, \omega)$. A four-dimensional hyper-parameter space, $\{p\}$, was explored to learn the uncertainty manifold in the five-dimensional parameter space $\{J_1,K,\Gamma,J_2,J_3\}$ \cite{SuppMat} %(see Appendix \ref{app:parameterization}) 
using a variant of the Efficient Global Optimization algorithm \cite{Jones1998, Samarakoon2020} which we call %refer to as 
the Iterative Mapping algorithm (IMA). % as described in Ref. . 
Autoencoders are unsupervised artificial neural networks with architecture as shown in Fig.~\ref{fig:flowchart}(a). We train two autoencoders \cite{SuppMat} with either $S(\textbf{Q})$ or $S(\textbf{Q}, \omega)$. The {\it Encoder} takes a linearized version of the structure factor $S(\nu)$ [$\nu=Q$ or $\nu=Q,\omega$] and outputs a compressed representation, $S(L)$, reducing the input dimensionality $N_\nu = 10^6-10^8$ pixels down to $N_L = 10^0-10^2$.
%with the input dimensionality of $N_\nu = 10^6 -10^8$ pixels reduced down to $N_L = 10^0-10^2$. 
The {\it Decoder} is a contrary network, %of the {\it Encoder} 
which projects $S(L)$ back to the original dimensionality and predicts $S_{\rm AE}(\nu)$. Our {\it Encoders} and {\it Decoders} are designed to be symmetrical, and the numbers of layers are tuned as described in \cite{SuppMat}.
%The {\it Encoder} and {\it Decoder} are usually designed to be symmetrical and the latent space dimension $N_L$, number of intermediate layers, as well as neurons-per-layer can be tuned for better performance.

%Autoencoders are unsupervised artificial neural networks shown in Fig. \ref{fig:flowchart}(a) (see Appendix \ref{app:autoencoder} for more on autoencoder training). Two autoencoders are trained to take either of $S(Q)$ or $S(Q, \omega)$. The {\it encoder} takes a linearized version of structure factor, $S(\nu)$ and outputs a compressed representation, $S(L)$ with the input dimensionality of  $N_\nu = 10^6 -10^8$ pixels reduced down to $N_L = 10^0-10^2$. The {\it Decoder} is a contrary network of the {\it Encoder} which projects $S(L)$ back to its original dimensionality and predicts $S_{\rm AE}(\nu)$. The {\it Encoder} and {\it Decoder} are usually designed to be symmetrical and the latent space dimension $N_L$, number of intermediate layers, as well as neurons-per-layer can be tuned for better performance.

Two separate Radial Basis Networks (RBN) \cite{Broomhead1988}, shown in Fig.~\ref{fig:flowchart}(b), provide {\it Generator} Networks (GN) to approximately map the Hamiltonian space, $H(\{p\})$ directly to latent space, $\mathcal{L}_\nu$. See \cite{SuppMat} for training details. The GN provides surrogate calculations to bypass the computationally expensive direct solver, allowing exhaustive searches for parameter space mapping as illustrated in Fig.~\ref{fig:flowchart}(c). GN predictions depend on the degree of training of the network, the topography of the parameter space, and the sampling sparsity. They do not fully replace simulations, and should not be used to draw conclusions when detailed information is needed. Complete surrogates predicting structure factors, $S^{\rm sur} (\nu)$, are constructed by linking the GN with corresponding {\rm Decoder}. These surrogates can also be used as low-cost estimators in the IMA as an alternative to the Gaussian Process Regression in Ref. \cite{Samarakoon2020}. 

%Two separate Radial Basis Networks (RBN), (see Appendix for training details), shown in Fig. \ref{fig:flowchart}(b), provide {\it Generator} networks (GN) to map the Hamiltonian space, $H(\{p\})$ directly to latent space, $\mathcal{L}_\nu$. 
% Having Autoencoders are trained with reasonable reconstruction error (squared distance between input and predicted), two separate {\it Generator} networks (GN) are trained to map the Hamiltonian space, $H(\{p\})$ directly to latent space, $\mathcal{L}_\nu$ as shown in Fig. \ref{fig:flowchart}(b). 
%The GN provides surrogate calculations to bypass the computationally expensive direct solver and make exhaustive parameter searches for parameter space mapping as illustrated in  Fig. \ref{fig:flowchart}(c). These predictions depend on the degree of training of the network, and on the topography of the phase space and the sparsity of the sampling. They do not fully replace simulations and should not be used to draw conclusions when detailed information is needed. The complete surrogates to predict structure factors, $S^{\rm sur} (\nu)$ are constructed by linking the GN with the corresponding {\rm Decoder}. These surrogates can also be used as the low-cost estimator in the iterative mapping algorithm work-flow as an alternative to the Gaussian Process Regression as suggested in Ref. \cite{Samarakoon2020}. (see Appendix \ref{app:parameterization} for more information).

As Fig. \ref{fig:flowchart}(c) schematically shows, the workflow can be split into five sections: %. %resolved into five aspects. 
%These are 
I) scattering experiment and hypothesis; II) parameter space exploration and information compression; III) structure or property predictions; IV) parameter space predictions; and V) validation of SCS results using a quantum many-body solver. 
%Classical solver with Quantum. 
The workflow is similar to one proposed in Ref.~\cite{pressure_paper}, but here we add step V and use a composite latent space $\mathcal{L}_Q\cup \mathcal{L}_{Q,\omega}$.
%This workflow is very similar to what is proposed in Ref. \cite{pressure_paper} except the extra Quantum validation step and the composite latent space, $\mathcal{L}_Q-\mathcal{L}_{Q,\omega}$. 
The latent space forms the backbone of the operation, into which experimental data, simulations, and predictions from GN feed, and from which structure, property, and model parameters are predicted.

\emph{ML predictions.}---%
The $S(\textbf{Q})$ (and consequently $S(L)$) provides natural classification of phases, as the correlations of the system are encoded \cite{pressure_paper}. A high-dimensional graphical phase diagram can be constructed easily by projecting $Q$-space into a latent space of $N_L=3$ as suggested in Ref. \cite{pressure_paper}. %Thus, an
An architecture of three intermediate layers with 300-3-300 logistic neurons (activation function as $f(x)=1/(1+e^{-x})$) was empirically found have the highest performance for the $S^{\rm sim}(\textbf{Q})$. 
%The latent vectors are treated as the RGB color components of a phase map (see Ref. \cite{Samarakoon2021}).

\begin{figure}
\centering\includegraphics[width=\columnwidth]{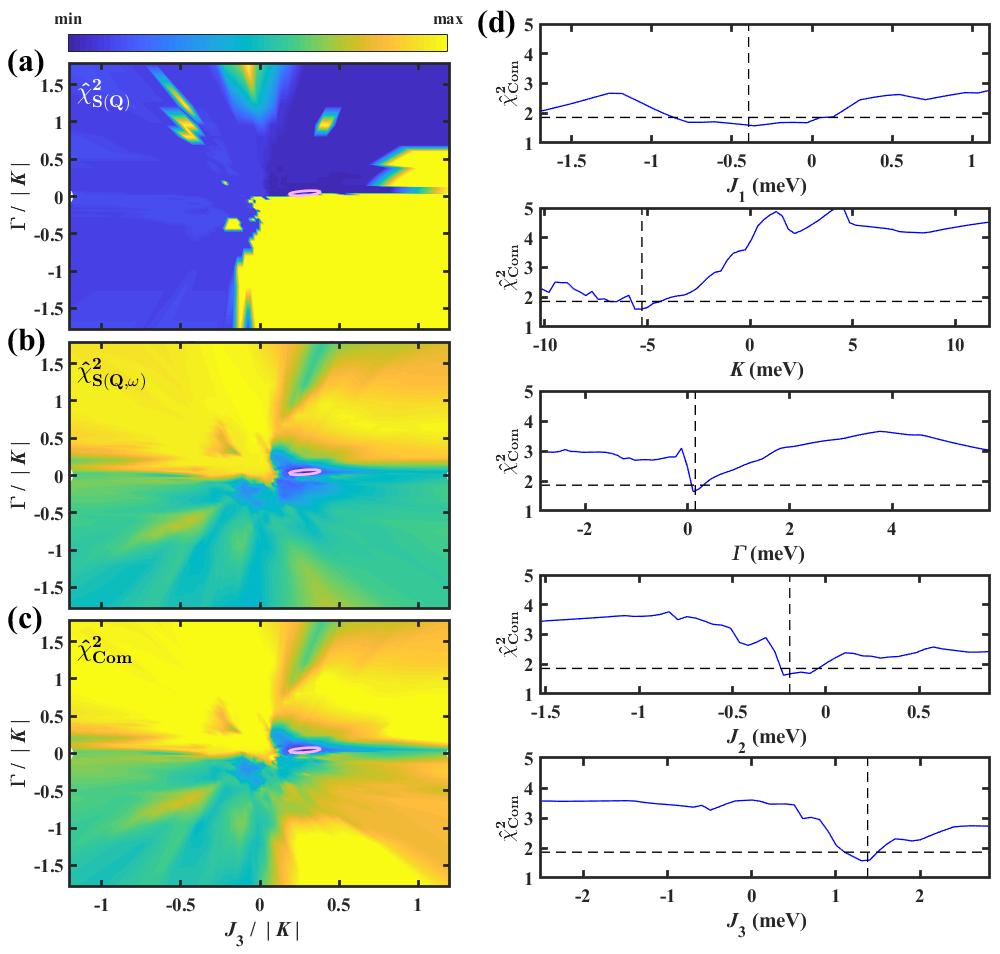}
% \centering\includegraphics[width=0.95\textwidth]{Figure04_Combined2_Valenti2.jpg}
%\centering\includegraphics[width=0.99\textwidth]{sqw_rb2mnf4.jpg}
	\caption{
	Optimal solution for $ \alpha$-RuCl$_3$. (a) shows a slice of $ \hat{\chi}^2_{S(\textbf{Q})}$ as function of $J_3/|K|$ and $\Gamma/|K|$. The problem is strongly underdetermined when only $S(\textbf{Q})$ is taken into account, with a flat fitness landscape indicating many possible solutions. (b) Same slice for $\hat{\chi}^2_{S(Q,\omega)}$. The fitness landscape remains relatively flat when only $S(\mathbf{Q},\omega)$ is accounted for. (c) Slice for the combined $\hat{\chi}^2_{\rm Com}$ function, which accounts for both statics and dynamics. An ellipsoidal approximation to the optimal regions is shown in pink on panels (a), (b) and (c). 
%     We can identify an optimal region, labeled by the pink ellipse [also shown in (a),(b) and ] and characterized by small $\Gamma/K$. 
    (d) shows line-cuts of $\hat{\chi}^2_{\rm Com}$ for individual parameters by fixing other parameters to their optimal values.}% of $J_1 = $ -0.4$\pm$0.4, $K = $-5.3$\pm$0.3, $\Gamma = $0.15$\pm$0.05, $J_2 = $-0.19$\pm$0.15 and $J_3 = $-1.35$\pm$0.15.}
\label{figure_optimal}
\end{figure}

Two phase diagrams are plotted in Figures~\ref{fig:lattice}(c) and \ref{fig:phasediagram}(a). Phases are indicated by color derived by treating latent vectors as RGB color components \cite{Samarakoon2021}.
%The phases are indicated by color, with colors derived using an RGB color scale based on the three most significant latent space components $(S_1 , S_2, S_3)$. 
Fig.~\ref{fig:lattice}(c) corresponds to the $J_2=J_3=0$ hyperplane, and uses the parametrization \cite{PhysRevLett.112.077204} $J_1=\sin \theta \cos \phi$, $K=\sin \theta \sin \phi$, $\Gamma=\cos \theta$, where the energy scale is fixed according to $1=\sqrt{J_1^2+K^2+\Gamma^2}$. Overlaid dashed black lines indicate the theoretically predicted phase diagram. We note that typically our method does not find sharp transitions, so our results are not phase diagrams in a strict sense. Nevertheless, the excellent agreement between Fig.~\ref{fig:lattice}(c) and the phase diagram derived in Ref.~\cite{PhysRevLett.112.077204} shows the merit of the approach. Fig.~\ref{fig:phasediagram}(a) shows the phase diagram in a slice around the optimal solution we find for $\alpha$-RuCl$_3$ (see Fig.~\ref{figure_optimal} and later discussion). Fig.~\ref{fig:phasediagram}(b) indicates the scattering at various points in the two phase diagrams.%, representing a variety of magnetic phases.

To fit experimental scattering data and map uncertainty in the five-dimensional parameter space we employed IMA with cost function ${\hat \chi}_{\rm Com}^2	=	{\hat \chi}_{S(\textbf{Q})}^2	\times {\hat \chi}_{S(Q,\omega)}^2$, where the ${\hat \chi}_{S(\nu)}^2$ are low-cost estimators defined in SM \cite{SuppMat}. IMA samples the parameter space iteratively subject to ${\hat \chi}_{\rm Com}^2 \leq C_{\rm Com}$. The threshold value $C_{\rm Com}$ is iteratively reduced to a final value. The Autoencoders and GN, are retrained at the end of each iteration. Thus, the predictability of the networks becomes reliable towards the minimum of ${\hat \chi}_{\rm Com}^2 $.

\begin{figure}
\centering\includegraphics[ %trim={3cm 2cm, 9cm 2cm},
width=\columnwidth]{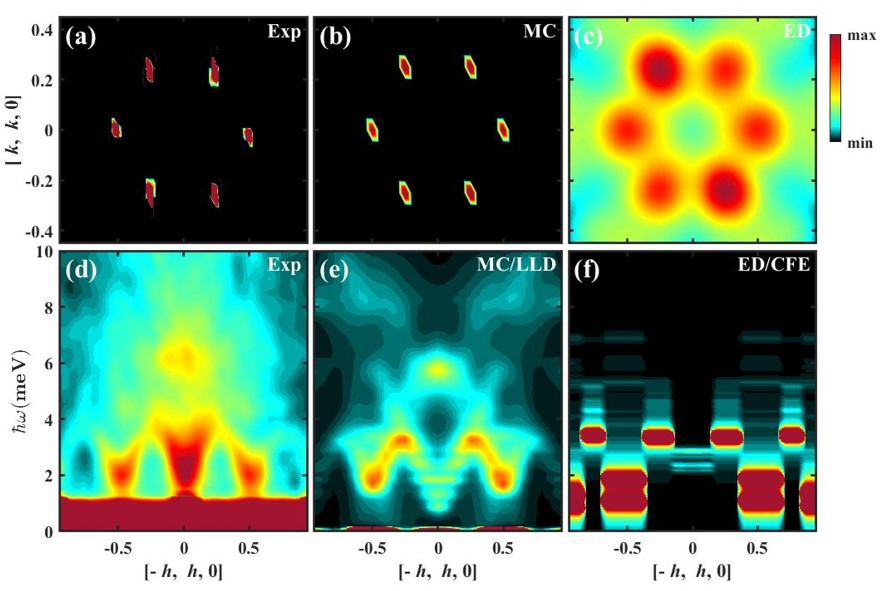}
%\centering\includegraphics[width=0.99\textwidth]{sqw_rb2mnf4.jpg}
	\caption{
	Top row: (a) Experimental 
%     \footnote{The elastic scattering data was previously published as Supplementary Figure S2(a) in Ref. \cite{Banerjee2018}} 
    and (b,c) theoretical static spin structure factors calculated using (b) MC simulation and (c) Lanczos ED for the optimized Hamiltonian parameters. All methods find peaks at the $M$ points, reflecting the zigzag ordering. Bottom row: Inelastic scattering from (d) experiment, (e) LLD, and (f) Lanczos for the same parameter set. 
	}
\label{fig:sqw}
\end{figure}

% \emph{Incorporating dynamics.}---%
\emph{Results and Discussion.}---%
% Discuss around Fig. \ref{figure_optimal} and Fig. \label{fig:sqw}
% Fig.~\ref{figure_optimal} shows the obtained optimal fit. %fit is shown in Fig.~\ref{figure_optimal}. 
Figure~\ref{figure_optimal} shows slices and cuts of the final ${\hat \chi}_{\rm Com}^2 $in parameter space. 
Due to uncertainties in the data, %minimization of
minimizing ${\hat \chi}_{\rm Com}^2 $ leads to a region of potential fits, indicated by the ellipsoid in panels (a)-(c). %in Fig. \ref{figure_optimal}.  
Additional Hamiltonian terms may need to be included in the modeling to capture all relevant interactions and achieve higher fitting certainty. This restriction aside, we have identified several parameter sets with particularly low ${\hat \chi}_{\rm Com}^2 $, and these were investigated more closely. Fig.~\ref{fig:sqw} shows $S(\textbf{Q})$ and $S(\textbf{Q},\omega)$ from experiment, LL simulation, and Lanczos exact diagonalization (ED) for the optimized parameter set: 
$J_1 =  -0.4\pm0.4$ meV, $K = -5.3\pm0.3$ meV, $\Gamma = 0.15\pm0.05$ meV, $J_2 = -0.19\pm0.15$ meV and $J_3 = 1.35\pm0.15$ meV.
The LL-simulated spectrum shows intensity at both the $M$ and $\Gamma$ points, although the intensity at $\Gamma$ is lower than in the experiment. In addition, the simulation captures the curvature of the spin wave dispersion along the $\Gamma\rightarrow M$ path, as well as the feature at $\omega=6$ meV. ED is subject to finite-size restrictions and low momentum resolution, but captures the magnetic order and energy scale of the low-energy scattering ($\omega<5$ meV).
%We also show the DSF calculated using ED for the same parameters in Fig.~\ref{fig:sqw} (f).

How does our optimized solution compare to other proposed models for $\alpha$-RuCl$_3$? 
Using 
the surrogates %generative model 
we can easily calculate $\hat{\chi}^2$ values for proposed models described by Eq.~\eqref{eq:hamiltonian}. By this metric our fit outperforms other models in the literature at describing the neutron data, see SM \cite{SuppMat}.
Our fit has a Kitaev interaction strength comparable to a previous INS fit \cite{10.1038/s41467-017-01177-0}, but lower $\Gamma/K$ and higher $J_3/K$. However, the energy scale is generally smaller than for models predicted by band structure calculations, and for models that seek to explain the experimental magnetic specific heat $C(T)$ \cite{Laurell2020}. Consequently, thermal pure quantum state \cite{PhysRevLett.108.240401, Kawamura2017} $C(T)$ results for the optimized solution fail to capture the experimentally observed high-temperature peak \cite{PhysRevB.99.094415, SuppMat}. One of our identified near-optimized parameter sets performs better in this regard, but worse at reproducing subtle spectral features \cite{SuppMat}. This reinforces the point that Eq.~\eqref{eq:hamiltonian} may miss some important term.

%\textcolor{red}{
One limitation of our approach is the use of SCS. This was necessary %in order 
to generate large amounts of training data, and allowed us to generate phase diagrams. However, quantum effects can be significant close to phase boundaries, thus locally diminishing the reliability of our networks and requiring many-body verification. % with a many-body solver. %. Thus reliability of our networks is diminished near transitions, requiring verification with a quantum many-body solver. 
This is particularly important in $\alpha$-RuCl$_3$, which is close to a phase transition under magnetic fields, and where many Hamiltonian parameters matter. 
Our optimized parameters are close to a transition between the Z.Z. and Z.Z. (2D) orders, but using ED we fortunately find the SCS results %for our optimized parameter set 
are physical, and correctly identifies the ground state. In contrast, the recently proposed Hamiltonian of Ref.~\cite{Li2021} is close to a transition between FM and Z.Z. orders, and our SCS predict FM, while ED finds Z.Z. %an FM groundstate, while ED finds zigzag order. 
This suggests it may be useful to retrain the networks using many-body simulations in regions close to phase boundaries to increase physical predictability.
%}

Our analysis shows that subtle changes in parameters affect the spectra and ordering. This means that other Hamiltonian terms could also account for the results. This implies that zero field neutron scattering is probably insufficient to constrain the model beyond the treatment here. For a more definitive understanding of $\alpha$-RuCl$_3$ additional data is needed. Simulations of field dependence suggest that high field spectroscopy measurements should be helpful here in disentangling the contributions of competing terms. Neutron scattering with its ability to capture wavevector and energy effects would be particularly valuable. Co-analysis of high field data along with zero field measurements used here, as well as other observable properties, can then be undertaken using the machine-learning-based approach.

%\emph{Discussion}
%\anj{
%Discuss the inconsistencies between best calculation with data and realistic models being complex
%Reason as the model considered is incomplete and possible extensions will improve the results 
%}

\emph{Conclusion.}---%
We have demonstrated unsupervised ML-SCS methods can be used to solve the inverse scattering problem inherent to INS experiments, thereby extending previous methods to also account for dynamics. Our approach can be applied to a wide range of magnetic systems, to obtain phase diagrams and fit the full 4D experimental scattering, as long as sufficient amounts of training data can be generated.
For $\alpha$-RuCl$_3$ we find a relatively flat fitness landscape, producing an uncertain fit. 
It does not fully explain the experimental scattering, likely due to interactions not considered here. 
%Some inconsistency with experimental scattering is found, likely due to interactions not considered here. 
Nevertheless, the optimal parameters reproduce many smaller scattering features, not captured by other proposed models. 
Improved algorithms are needed to extend the method further to even higher-dimensional parameter spaces and to fully constrain Hamiltonians. This can be done iteratively, building on previously simulated data. With future advances in computing power, we hope such methods may be used to rapidly and reliably identify the crucial physics of new materials.

%We have demonstrated that unsupervised ML-SCS methods can be used to solve the inverse scattering problem inherent in INS experiments, Our methods can be applied to a wide range of systems... For $\alpha$-RuCl$_3$ we find a spin Hamiltonian that...
%New more optimal algorithms are needed to extend the method further to more higher-dimensional parameter spaces, to fully constrain the Hamiltonians.

\section{Acknowledgements}
We thank J. Q. Yan for valuable discussions and for providing detailed crystal growth instructions. 
D.A.T,  A.B., S.E.N., and S.O. have been supported by the U.S. Department of Energy, Office of Science, National Quantum Information Science Research Centers, Quantum Science Center. %The research by D. A. T. was sponsored by the Quantum Science Center. 
A.M.S. was supported by the U.S. Department of Energy, Office of Science, Materials Sciences and Engineering Division and Scientific User Facilities Division. The research by P.L. and the early stage of S.O.'s effort were supported by the scientific Discovery through Advanced Computing (SciDAC) program funded by U.S. Department of Energy, Office of Science, Advanced Scientific Computing Research and Basic Energy Sciences, Division of Materials Sciences and Engineering. A portion of this research used resources at the Spallation Neutron Source, a DOE Office of Science User Facility operated by the Oak Ridge National Laboratory. The computer modeling used resources of the Oak Ridge Leadership Computing Facility, which is a DOE Office of Science User Facility supported under Contract DE-AC05-00OR22725, and of the Compute and Data Environment for Science (CADES) at the Oak Ridge National Laboratory, which is supported by the Office of Science of the U.S. Department of Energy under Contract No. DE-AC05-00OR22725.

%\appendix{\subsection{Experiment Details}\label{app:experiment}
%
%\subsubsection{Neutron Diffraction Experiment}
% Christian's text
%Elastic neutron studies were performed at Spallation Neutron Source (SNS) using the CORELLI beamline. A 125 mg $\alpha$-RuCl$_3$ crystal was mounted on an aluminium plate and aligned with the (H,0,L) plane horizontal. The crystal was rotated through $170$ degrees in $2^\degree$ steps about the vertical axis. The temperature of the measurement was 2 K and the perpendicular wave vector transfer was integrated in the range L = [0.92, 1.08] r.l.u..
%
%
%\subsubsection{Neutron Spectroscopy Experiment}
% Christian's text
%Inelastic neutron scattering was performed on a single crystal with a mass of 0.7 g. This was sealed in a thin-walled aluminium can with 1 atmosphere of Helium gas for thermal contact. Measurements were carried out using the SEQUOIA spectrometer at the Spallation Neutron Source (SNS). The incident energy was set to $E_i = 22.5$~meV and the temperature to 4 K. The crystal was mounted with the (H, 0, 0) and (0, 0, L) axes in the horizontal plane, and the orthogonal (0.5K, -K, 0) axis pointing vertically upwards. Data were collected by rotating the crystal about the vertical axis over $290^\degree$ in $1^\degree$ steps. The data are integrated over ranges (0.5K, -K, 0)=[-0.05, 0.05] and (0, 0, L)=[-2, 2].}

\bibliography{kitaev, MachineLearning, otherRef}

\begin{thebibliography}{95}%
\makeatletter
\providecommand \@ifxundefined [1]{%
 \@ifx{#1\undefined}
}%
\providecommand \@ifnum [1]{%
 \ifnum #1\expandafter \@firstoftwo
 \else \expandafter \@secondoftwo
 \fi
}%
\providecommand \@ifx [1]{%
 \ifx #1\expandafter \@firstoftwo
 \else \expandafter \@secondoftwo
 \fi
}%
\providecommand \natexlab [1]{#1}%
\providecommand \enquote  [1]{``#1''}%
\providecommand \bibnamefont  [1]{#1}%
\providecommand \bibfnamefont [1]{#1}%
\providecommand \citenamefont [1]{#1}%
\providecommand \href@noop [0]{\@secondoftwo}%
\providecommand \href [0]{\begingroup \@sanitize@url \@href}%
\providecommand \@href[1]{\@@startlink{#1}\@@href}%
\providecommand \@@href[1]{\endgroup#1\@@endlink}%
\providecommand \@sanitize@url [0]{\catcode `\\12\catcode `\$12\catcode
  `\&12\catcode `\#12\catcode `\^12\catcode `\_12\catcode `\%12\relax}%
\providecommand \@@startlink[1]{}%
\providecommand \@@endlink[0]{}%
\providecommand \url  [0]{\begingroup\@sanitize@url \@url }%
\providecommand \@url [1]{\endgroup\@href {#1}{\urlprefix }}%
\providecommand \urlprefix  [0]{URL }%
\providecommand \Eprint [0]{\href }%
\providecommand \doibase [0]{https://doi.org/}%
\providecommand \selectlanguage [0]{\@gobble}%
\providecommand \bibinfo  [0]{\@secondoftwo}%
\providecommand \bibfield  [0]{\@secondoftwo}%
\providecommand \translation [1]{[#1]}%
\providecommand \BibitemOpen [0]{}%
\providecommand \bibitemStop [0]{}%
\providecommand \bibitemNoStop [0]{.\EOS\space}%
\providecommand \EOS [0]{\spacefactor3000\relax}%
\providecommand \BibitemShut  [1]{\csname bibitem#1\endcsname}%
\let\auto@bib@innerbib\@empty
%</preamble>
\bibitem [{\citenamefont {Balents}(2010)}]{Balents2010}%
  \BibitemOpen
  \bibfield  {author} {\bibinfo {author} {\bibfnamefont {L.}~\bibnamefont
  {Balents}},\ }\href {https://doi.org/10.1038/nature08917} {\bibfield
  {journal} {\bibinfo  {journal} {Nature (London)}\ }\textbf {\bibinfo {volume}
  {464}},\ \bibinfo {pages} {199} (\bibinfo {year} {2010})}\BibitemShut
  {NoStop}%
\bibitem [{\citenamefont {Lacroix}\ \emph {et~al.}(2011)\citenamefont
  {Lacroix}, \citenamefont {Mendels},\ and\ \citenamefont
  {Mila}}]{Lacroix2011}%
  \BibitemOpen
  \bibinfo {editor} {\bibfnamefont {C.}~\bibnamefont {Lacroix}}, \bibinfo
  {editor} {\bibfnamefont {P.}~\bibnamefont {Mendels}},\ and\ \bibinfo {editor}
  {\bibfnamefont {F.}~\bibnamefont {Mila}},\ eds.,\ \href@noop {} {\emph
  {\bibinfo {title} {Introduction to Frustrated Magnetism: Materials,
  Experiments, Theory}}},\ Springer Series in Solid-State Sciences\ (\bibinfo
  {publisher} {Springer, Berlin},\ \bibinfo {year} {2011})\BibitemShut
  {NoStop}%
\bibitem [{\citenamefont {Savary}\ and\ \citenamefont
  {Balents}(2017)}]{Savary2017}%
  \BibitemOpen
  \bibfield  {author} {\bibinfo {author} {\bibfnamefont {L.}~\bibnamefont
  {Savary}}\ and\ \bibinfo {author} {\bibfnamefont {L.}~\bibnamefont
  {Balents}},\ }\href {https://doi.org/10.1088/0034-4885/80/1/016502}
  {\bibfield  {journal} {\bibinfo  {journal} {Rep. Prog. Phys.}\ }\textbf
  {\bibinfo {volume} {80}},\ \bibinfo {pages} {016502} (\bibinfo {year}
  {2017})}\BibitemShut {NoStop}%
\bibitem [{\citenamefont {Plumb}\ \emph {et~al.}(2014)\citenamefont {Plumb},
  \citenamefont {Clancy}, \citenamefont {Sandilands}, \citenamefont {Shankar},
  \citenamefont {Hu}, \citenamefont {Burch}, \citenamefont {Kee},\ and\
  \citenamefont {Kim}}]{PhysRevB.90.041112}%
  \BibitemOpen
  \bibfield  {author} {\bibinfo {author} {\bibfnamefont {K.~W.}\ \bibnamefont
  {Plumb}}, \bibinfo {author} {\bibfnamefont {J.~P.}\ \bibnamefont {Clancy}},
  \bibinfo {author} {\bibfnamefont {L.~J.}\ \bibnamefont {Sandilands}},
  \bibinfo {author} {\bibfnamefont {V.~V.}\ \bibnamefont {Shankar}}, \bibinfo
  {author} {\bibfnamefont {Y.~F.}\ \bibnamefont {Hu}}, \bibinfo {author}
  {\bibfnamefont {K.~S.}\ \bibnamefont {Burch}}, \bibinfo {author}
  {\bibfnamefont {H.-Y.}\ \bibnamefont {Kee}},\ and\ \bibinfo {author}
  {\bibfnamefont {Y.-J.}\ \bibnamefont {Kim}},\ }\href
  {https://doi.org/10.1103/PhysRevB.90.041112} {\bibfield  {journal} {\bibinfo
  {journal} {Phys. Rev. B}\ }\textbf {\bibinfo {volume} {90}},\ \bibinfo
  {pages} {041112} (\bibinfo {year} {2014})}\BibitemShut {NoStop}%
\bibitem [{\citenamefont {Sandilands}\ \emph {et~al.}(2015)\citenamefont
  {Sandilands}, \citenamefont {Tian}, \citenamefont {Plumb}, \citenamefont
  {Kim},\ and\ \citenamefont {Burch}}]{PhysRevLett.114.147201}%
  \BibitemOpen
  \bibfield  {author} {\bibinfo {author} {\bibfnamefont {L.~J.}\ \bibnamefont
  {Sandilands}}, \bibinfo {author} {\bibfnamefont {Y.}~\bibnamefont {Tian}},
  \bibinfo {author} {\bibfnamefont {K.~W.}\ \bibnamefont {Plumb}}, \bibinfo
  {author} {\bibfnamefont {Y.-J.}\ \bibnamefont {Kim}},\ and\ \bibinfo {author}
  {\bibfnamefont {K.~S.}\ \bibnamefont {Burch}},\ }\href
  {https://doi.org/10.1103/PhysRevLett.114.147201} {\bibfield  {journal}
  {\bibinfo  {journal} {Phys. Rev. Lett.}\ }\textbf {\bibinfo {volume} {114}},\
  \bibinfo {pages} {147201} (\bibinfo {year} {2015})}\BibitemShut {NoStop}%
\bibitem [{\citenamefont {Banerjee}\ \emph {et~al.}(2016)\citenamefont
  {Banerjee}, \citenamefont {Bridges}, \citenamefont {Yan}, \citenamefont
  {Aczel}, \citenamefont {Li}, \citenamefont {Stone}, \citenamefont {Granroth},
  \citenamefont {Lumsden}, \citenamefont {Yiu}, \citenamefont {Knolle},
  \citenamefont {Bhattacharjee}, \citenamefont {Kovrizhin}, \citenamefont
  {Moessner}, \citenamefont {Tennant}, \citenamefont {Mandrus},\ and\
  \citenamefont {Nagler}}]{Banerjee2016}%
  \BibitemOpen
  \bibfield  {author} {\bibinfo {author} {\bibfnamefont {A.}~\bibnamefont
  {Banerjee}}, \bibinfo {author} {\bibfnamefont {C.~A.}\ \bibnamefont
  {Bridges}}, \bibinfo {author} {\bibfnamefont {J.-Q.}\ \bibnamefont {Yan}},
  \bibinfo {author} {\bibfnamefont {A.~A.}\ \bibnamefont {Aczel}}, \bibinfo
  {author} {\bibfnamefont {L.}~\bibnamefont {Li}}, \bibinfo {author}
  {\bibfnamefont {M.~B.}\ \bibnamefont {Stone}}, \bibinfo {author}
  {\bibfnamefont {G.~E.}\ \bibnamefont {Granroth}}, \bibinfo {author}
  {\bibfnamefont {M.~D.}\ \bibnamefont {Lumsden}}, \bibinfo {author}
  {\bibfnamefont {Y.}~\bibnamefont {Yiu}}, \bibinfo {author} {\bibfnamefont
  {J.}~\bibnamefont {Knolle}}, \bibinfo {author} {\bibfnamefont
  {S.}~\bibnamefont {Bhattacharjee}}, \bibinfo {author} {\bibfnamefont {D.~L.}\
  \bibnamefont {Kovrizhin}}, \bibinfo {author} {\bibfnamefont {R.}~\bibnamefont
  {Moessner}}, \bibinfo {author} {\bibfnamefont {D.~A.}\ \bibnamefont
  {Tennant}}, \bibinfo {author} {\bibfnamefont {D.~G.}\ \bibnamefont
  {Mandrus}},\ and\ \bibinfo {author} {\bibfnamefont {S.~E.}\ \bibnamefont
  {Nagler}},\ }\href {https://doi.org/10.1038/nmat4604} {\bibfield  {journal}
  {\bibinfo  {journal} {Nat. Mat.}\ }\textbf {\bibinfo {volume} {15}},\
  \bibinfo {pages} {733} (\bibinfo {year} {2016})}\BibitemShut {NoStop}%
\bibitem [{\citenamefont {Banerjee}\ \emph {et~al.}(2017)\citenamefont
  {Banerjee}, \citenamefont {Yan}, \citenamefont {Knolle}, \citenamefont
  {Bridges}, \citenamefont {Stone}, \citenamefont {Lumsden}, \citenamefont
  {Mandrus}, \citenamefont {Tennant}, \citenamefont {Moessner},\ and\
  \citenamefont {Nagler}}]{Banerjee2017}%
  \BibitemOpen
  \bibfield  {author} {\bibinfo {author} {\bibfnamefont {A.}~\bibnamefont
  {Banerjee}}, \bibinfo {author} {\bibfnamefont {J.}~\bibnamefont {Yan}},
  \bibinfo {author} {\bibfnamefont {J.}~\bibnamefont {Knolle}}, \bibinfo
  {author} {\bibfnamefont {C.~A.}\ \bibnamefont {Bridges}}, \bibinfo {author}
  {\bibfnamefont {M.~B.}\ \bibnamefont {Stone}}, \bibinfo {author}
  {\bibfnamefont {M.~D.}\ \bibnamefont {Lumsden}}, \bibinfo {author}
  {\bibfnamefont {D.~G.}\ \bibnamefont {Mandrus}}, \bibinfo {author}
  {\bibfnamefont {D.~A.}\ \bibnamefont {Tennant}}, \bibinfo {author}
  {\bibfnamefont {R.}~\bibnamefont {Moessner}},\ and\ \bibinfo {author}
  {\bibfnamefont {S.~E.}\ \bibnamefont {Nagler}},\ }\href
  {https://doi.org/10.1126/science.aah6015} {\bibfield  {journal} {\bibinfo
  {journal} {Science}\ }\textbf {\bibinfo {volume} {356}},\ \bibinfo {pages}
  {1055} (\bibinfo {year} {2017})}\BibitemShut {NoStop}%
\bibitem [{\citenamefont {Do}\ \emph {et~al.}(2017)\citenamefont {Do},
  \citenamefont {Park}, \citenamefont {Yoshitake}, \citenamefont {Nasu},
  \citenamefont {Motome}, \citenamefont {Kwon}, \citenamefont {Adroja},
  \citenamefont {Voneshen}, \citenamefont {Kim}, \citenamefont {Jang},
  \citenamefont {Park}, \citenamefont {Choi},\ and\ \citenamefont
  {Ji}}]{Do2017}%
  \BibitemOpen
  \bibfield  {author} {\bibinfo {author} {\bibfnamefont {S.-H.}\ \bibnamefont
  {Do}}, \bibinfo {author} {\bibfnamefont {S.-Y.}\ \bibnamefont {Park}},
  \bibinfo {author} {\bibfnamefont {J.}~\bibnamefont {Yoshitake}}, \bibinfo
  {author} {\bibfnamefont {J.}~\bibnamefont {Nasu}}, \bibinfo {author}
  {\bibfnamefont {Y.}~\bibnamefont {Motome}}, \bibinfo {author} {\bibfnamefont
  {Y.~S.}\ \bibnamefont {Kwon}}, \bibinfo {author} {\bibfnamefont {D.~T.}\
  \bibnamefont {Adroja}}, \bibinfo {author} {\bibfnamefont {D.~J.}\
  \bibnamefont {Voneshen}}, \bibinfo {author} {\bibfnamefont {K.}~\bibnamefont
  {Kim}}, \bibinfo {author} {\bibfnamefont {T.-H.}\ \bibnamefont {Jang}},
  \bibinfo {author} {\bibfnamefont {J.-H.}\ \bibnamefont {Park}}, \bibinfo
  {author} {\bibfnamefont {K.-Y.}\ \bibnamefont {Choi}},\ and\ \bibinfo
  {author} {\bibfnamefont {S.}~\bibnamefont {Ji}},\ }\href
  {https://doi.org/10.1038/nphys4264} {\bibfield  {journal} {\bibinfo
  {journal} {Nat. Phys.}\ }\textbf {\bibinfo {volume} {13}},\ \bibinfo {pages}
  {1079} (\bibinfo {year} {2017})}\BibitemShut {NoStop}%
\bibitem [{\citenamefont {Ran}\ \emph {et~al.}(2017)\citenamefont {Ran},
  \citenamefont {Wang}, \citenamefont {Wang}, \citenamefont {Dong},
  \citenamefont {Ren}, \citenamefont {Bao}, \citenamefont {Li}, \citenamefont
  {Ma}, \citenamefont {Gan}, \citenamefont {Zhang}, \citenamefont {Park},
  \citenamefont {Deng}, \citenamefont {Danilkin}, \citenamefont {Yu},
  \citenamefont {Li},\ and\ \citenamefont {Wen}}]{PhysRevLett.118.107203}%
  \BibitemOpen
  \bibfield  {author} {\bibinfo {author} {\bibfnamefont {K.}~\bibnamefont
  {Ran}}, \bibinfo {author} {\bibfnamefont {J.}~\bibnamefont {Wang}}, \bibinfo
  {author} {\bibfnamefont {W.}~\bibnamefont {Wang}}, \bibinfo {author}
  {\bibfnamefont {Z.-Y.}\ \bibnamefont {Dong}}, \bibinfo {author}
  {\bibfnamefont {X.}~\bibnamefont {Ren}}, \bibinfo {author} {\bibfnamefont
  {S.}~\bibnamefont {Bao}}, \bibinfo {author} {\bibfnamefont {S.}~\bibnamefont
  {Li}}, \bibinfo {author} {\bibfnamefont {Z.}~\bibnamefont {Ma}}, \bibinfo
  {author} {\bibfnamefont {Y.}~\bibnamefont {Gan}}, \bibinfo {author}
  {\bibfnamefont {Y.}~\bibnamefont {Zhang}}, \bibinfo {author} {\bibfnamefont
  {J.~T.}\ \bibnamefont {Park}}, \bibinfo {author} {\bibfnamefont
  {G.}~\bibnamefont {Deng}}, \bibinfo {author} {\bibfnamefont {S.}~\bibnamefont
  {Danilkin}}, \bibinfo {author} {\bibfnamefont {S.-L.}\ \bibnamefont {Yu}},
  \bibinfo {author} {\bibfnamefont {J.-X.}\ \bibnamefont {Li}},\ and\ \bibinfo
  {author} {\bibfnamefont {J.}~\bibnamefont {Wen}},\ }\href
  {https://doi.org/10.1103/PhysRevLett.118.107203} {\bibfield  {journal}
  {\bibinfo  {journal} {Phys. Rev. Lett.}\ }\textbf {\bibinfo {volume} {118}},\
  \bibinfo {pages} {107203} (\bibinfo {year} {2017})}\BibitemShut {NoStop}%
\bibitem [{\citenamefont {Banerjee}\ \emph {et~al.}(2018)\citenamefont
  {Banerjee}, \citenamefont {Lampen-Kelley}, \citenamefont {Knolle},
  \citenamefont {Balz}, \citenamefont {Aczel}, \citenamefont {Winn},
  \citenamefont {Liu}, \citenamefont {Pajerowski}, \citenamefont {Yang},
  \citenamefont {Bridges}, \citenamefont {Savici}, \citenamefont {Chakoumakos},
  \citenamefont {Lumsden}, \citenamefont {Tennant}, \citenamefont {Moessner},
  \citenamefont {Mandrus},\ and\ \citenamefont {Nagler}}]{Banerjee2018}%
  \BibitemOpen
  \bibfield  {author} {\bibinfo {author} {\bibfnamefont {A.}~\bibnamefont
  {Banerjee}}, \bibinfo {author} {\bibfnamefont {P.}~\bibnamefont
  {Lampen-Kelley}}, \bibinfo {author} {\bibfnamefont {J.}~\bibnamefont
  {Knolle}}, \bibinfo {author} {\bibfnamefont {C.}~\bibnamefont {Balz}},
  \bibinfo {author} {\bibfnamefont {A.~A.}\ \bibnamefont {Aczel}}, \bibinfo
  {author} {\bibfnamefont {B.}~\bibnamefont {Winn}}, \bibinfo {author}
  {\bibfnamefont {Y.}~\bibnamefont {Liu}}, \bibinfo {author} {\bibfnamefont
  {D.}~\bibnamefont {Pajerowski}}, \bibinfo {author} {\bibfnamefont
  {J.}~\bibnamefont {Yang}}, \bibinfo {author} {\bibfnamefont {C.~A.}\
  \bibnamefont {Bridges}}, \bibinfo {author} {\bibfnamefont {A.~T.}\
  \bibnamefont {Savici}}, \bibinfo {author} {\bibfnamefont {B.~C.}\
  \bibnamefont {Chakoumakos}}, \bibinfo {author} {\bibfnamefont {M.~D.}\
  \bibnamefont {Lumsden}}, \bibinfo {author} {\bibfnamefont {D.~A.}\
  \bibnamefont {Tennant}}, \bibinfo {author} {\bibfnamefont {R.}~\bibnamefont
  {Moessner}}, \bibinfo {author} {\bibfnamefont {D.~G.}\ \bibnamefont
  {Mandrus}},\ and\ \bibinfo {author} {\bibfnamefont {S.~E.}\ \bibnamefont
  {Nagler}},\ }\href {https://doi.org/10.1038/s41535-018-0079-2} {\bibfield
  {journal} {\bibinfo  {journal} {npj Quantum Mater.}\ }\textbf {\bibinfo
  {volume} {3}},\ \bibinfo {pages} {8} (\bibinfo {year} {2018})}\BibitemShut
  {NoStop}%
\bibitem [{\citenamefont {Balz}\ \emph {et~al.}(2019)\citenamefont {Balz},
  \citenamefont {Lampen-Kelley}, \citenamefont {Banerjee}, \citenamefont {Yan},
  \citenamefont {Lu}, \citenamefont {Hu}, \citenamefont {Yadav}, \citenamefont
  {Takano}, \citenamefont {Liu}, \citenamefont {Tennant}, \citenamefont
  {Lumsden}, \citenamefont {Mandrus},\ and\ \citenamefont {Nagler}}]{Balz2019}%
  \BibitemOpen
  \bibfield  {author} {\bibinfo {author} {\bibfnamefont {C.}~\bibnamefont
  {Balz}}, \bibinfo {author} {\bibfnamefont {P.}~\bibnamefont {Lampen-Kelley}},
  \bibinfo {author} {\bibfnamefont {A.}~\bibnamefont {Banerjee}}, \bibinfo
  {author} {\bibfnamefont {J.}~\bibnamefont {Yan}}, \bibinfo {author}
  {\bibfnamefont {Z.}~\bibnamefont {Lu}}, \bibinfo {author} {\bibfnamefont
  {X.}~\bibnamefont {Hu}}, \bibinfo {author} {\bibfnamefont {S.~M.}\
  \bibnamefont {Yadav}}, \bibinfo {author} {\bibfnamefont {Y.}~\bibnamefont
  {Takano}}, \bibinfo {author} {\bibfnamefont {Y.}~\bibnamefont {Liu}},
  \bibinfo {author} {\bibfnamefont {D.~A.}\ \bibnamefont {Tennant}}, \bibinfo
  {author} {\bibfnamefont {M.~D.}\ \bibnamefont {Lumsden}}, \bibinfo {author}
  {\bibfnamefont {D.}~\bibnamefont {Mandrus}},\ and\ \bibinfo {author}
  {\bibfnamefont {S.~E.}\ \bibnamefont {Nagler}},\ }\href
  {https://doi.org/10.1103/PhysRevB.100.060405} {\bibfield  {journal} {\bibinfo
   {journal} {Phys. Rev. B}\ }\textbf {\bibinfo {volume} {100}},\ \bibinfo
  {pages} {060405} (\bibinfo {year} {2019})}\BibitemShut {NoStop}%
\bibitem [{\citenamefont {Winter}\ \emph
  {et~al.}(2017{\natexlab{a}})\citenamefont {Winter}, \citenamefont {Tsirlin},
  \citenamefont {Daghofer}, \citenamefont {van~den Brink}, \citenamefont
  {Singh}, \citenamefont {Gegenwart},\ and\ \citenamefont
  {Valent\'i}}]{Winter2017}%
  \BibitemOpen
  \bibfield  {author} {\bibinfo {author} {\bibfnamefont {S.~M.}\ \bibnamefont
  {Winter}}, \bibinfo {author} {\bibfnamefont {A.~A.}\ \bibnamefont {Tsirlin}},
  \bibinfo {author} {\bibfnamefont {M.}~\bibnamefont {Daghofer}}, \bibinfo
  {author} {\bibfnamefont {J.}~\bibnamefont {van~den Brink}}, \bibinfo {author}
  {\bibfnamefont {Y.}~\bibnamefont {Singh}}, \bibinfo {author} {\bibfnamefont
  {P.}~\bibnamefont {Gegenwart}},\ and\ \bibinfo {author} {\bibfnamefont
  {R.}~\bibnamefont {Valent\'i}},\ }\href
  {https://doi.org/10.1088/1361-648X/aa8cf5} {\bibfield  {journal} {\bibinfo
  {journal} {J. Phys. Condens. Matter}\ }\textbf {\bibinfo {volume} {29}},\
  \bibinfo {pages} {493002} (\bibinfo {year} {2017}{\natexlab{a}})}\BibitemShut
  {NoStop}%
\bibitem [{\citenamefont {Takagi}\ \emph {et~al.}(2019)\citenamefont {Takagi},
  \citenamefont {Takayama}, \citenamefont {Jackeli}, \citenamefont
  {Khaliullin},\ and\ \citenamefont {Nagler}}]{Takagi2019}%
  \BibitemOpen
  \bibfield  {author} {\bibinfo {author} {\bibfnamefont {H.}~\bibnamefont
  {Takagi}}, \bibinfo {author} {\bibfnamefont {T.}~\bibnamefont {Takayama}},
  \bibinfo {author} {\bibfnamefont {G.}~\bibnamefont {Jackeli}}, \bibinfo
  {author} {\bibfnamefont {G.}~\bibnamefont {Khaliullin}},\ and\ \bibinfo
  {author} {\bibfnamefont {S.~E.}\ \bibnamefont {Nagler}},\ }\href
  {https://doi.org/10.1038/s42254-019-0038-2} {\bibfield  {journal} {\bibinfo
  {journal} {Nat. Rev. Phys.}\ }\textbf {\bibinfo {volume} {1}},\ \bibinfo
  {pages} {264} (\bibinfo {year} {2019})}\BibitemShut {NoStop}%
\bibitem [{\citenamefont {Motome}\ and\ \citenamefont
  {Nasu}(2020)}]{doi:10.7566/JPSJ.89.012002}%
  \BibitemOpen
  \bibfield  {author} {\bibinfo {author} {\bibfnamefont {Y.}~\bibnamefont
  {Motome}}\ and\ \bibinfo {author} {\bibfnamefont {J.}~\bibnamefont {Nasu}},\
  }\href {https://doi.org/https://doi.org/10.7566/JPSJ.89.012002} {\bibfield
  {journal} {\bibinfo  {journal} {J. Phys. Soc. Jpn.}\ }\textbf {\bibinfo
  {volume} {89}},\ \bibinfo {pages} {012002} (\bibinfo {year}
  {2020})}\BibitemShut {NoStop}%
\bibitem [{\citenamefont {Kitaev}(2006)}]{Kitaev2006}%
  \BibitemOpen
  \bibfield  {author} {\bibinfo {author} {\bibfnamefont {A.}~\bibnamefont
  {Kitaev}},\ }\href {https://doi.org/10.1016/j.aop.2005.10.005} {\bibfield
  {journal} {\bibinfo  {journal} {Ann. Phys. (N.Y.)}\ }\textbf {\bibinfo
  {volume} {321}},\ \bibinfo {pages} {2} (\bibinfo {year} {2006})}\BibitemShut
  {NoStop}%
\bibitem [{\citenamefont {Hermanns}\ \emph {et~al.}(2018)\citenamefont
  {Hermanns}, \citenamefont {Kimchi},\ and\ \citenamefont
  {Knolle}}]{Hermanns2018}%
  \BibitemOpen
  \bibfield  {author} {\bibinfo {author} {\bibfnamefont {M.}~\bibnamefont
  {Hermanns}}, \bibinfo {author} {\bibfnamefont {I.}~\bibnamefont {Kimchi}},\
  and\ \bibinfo {author} {\bibfnamefont {J.}~\bibnamefont {Knolle}},\ }\href
  {https://doi.org/10.1146/annurev-conmatphys-033117-053934} {\bibfield
  {journal} {\bibinfo  {journal} {Annu. Rev. Condens. Matter Phys.}\ }\textbf
  {\bibinfo {volume} {9}},\ \bibinfo {pages} {17} (\bibinfo {year}
  {2018})}\BibitemShut {NoStop}%
\bibitem [{\citenamefont {Nayak}\ \emph {et~al.}(2008)\citenamefont {Nayak},
  \citenamefont {Simon}, \citenamefont {Stern}, \citenamefont {Freedman},\ and\
  \citenamefont {Das~Sarma}}]{RevModPhys.80.1083}%
  \BibitemOpen
  \bibfield  {author} {\bibinfo {author} {\bibfnamefont {C.}~\bibnamefont
  {Nayak}}, \bibinfo {author} {\bibfnamefont {S.~H.}\ \bibnamefont {Simon}},
  \bibinfo {author} {\bibfnamefont {A.}~\bibnamefont {Stern}}, \bibinfo
  {author} {\bibfnamefont {M.}~\bibnamefont {Freedman}},\ and\ \bibinfo
  {author} {\bibfnamefont {S.}~\bibnamefont {Das~Sarma}},\ }\href
  {https://doi.org/10.1103/RevModPhys.80.1083} {\bibfield  {journal} {\bibinfo
  {journal} {Rev. Mod. Phys.}\ }\textbf {\bibinfo {volume} {80}},\ \bibinfo
  {pages} {1083} (\bibinfo {year} {2008})}\BibitemShut {NoStop}%
\bibitem [{\citenamefont {Aasen}\ \emph {et~al.}(2020)\citenamefont {Aasen},
  \citenamefont {Mong}, \citenamefont {Hunt}, \citenamefont {Mandrus},\ and\
  \citenamefont {Alicea}}]{PhysRevX.10.031014}%
  \BibitemOpen
  \bibfield  {author} {\bibinfo {author} {\bibfnamefont {D.}~\bibnamefont
  {Aasen}}, \bibinfo {author} {\bibfnamefont {R.~S.~K.}\ \bibnamefont {Mong}},
  \bibinfo {author} {\bibfnamefont {B.~M.}\ \bibnamefont {Hunt}}, \bibinfo
  {author} {\bibfnamefont {D.}~\bibnamefont {Mandrus}},\ and\ \bibinfo {author}
  {\bibfnamefont {J.}~\bibnamefont {Alicea}},\ }\href
  {https://doi.org/10.1103/PhysRevX.10.031014} {\bibfield  {journal} {\bibinfo
  {journal} {Phys. Rev. X}\ }\textbf {\bibinfo {volume} {10}},\ \bibinfo
  {pages} {031014} (\bibinfo {year} {2020})}\BibitemShut {NoStop}%
\bibitem [{\citenamefont {Klocke}\ \emph {et~al.}(2021)\citenamefont {Klocke},
  \citenamefont {Aasen}, \citenamefont {Mong}, \citenamefont {Demler},\ and\
  \citenamefont {Alicea}}]{PhysRevLett.126.177204}%
  \BibitemOpen
  \bibfield  {author} {\bibinfo {author} {\bibfnamefont {K.}~\bibnamefont
  {Klocke}}, \bibinfo {author} {\bibfnamefont {D.}~\bibnamefont {Aasen}},
  \bibinfo {author} {\bibfnamefont {R.~S.~K.}\ \bibnamefont {Mong}}, \bibinfo
  {author} {\bibfnamefont {E.~A.}\ \bibnamefont {Demler}},\ and\ \bibinfo
  {author} {\bibfnamefont {J.}~\bibnamefont {Alicea}},\ }\href
  {https://doi.org/10.1103/PhysRevLett.126.177204} {\bibfield  {journal}
  {\bibinfo  {journal} {Phys. Rev. Lett.}\ }\textbf {\bibinfo {volume} {126}},\
  \bibinfo {pages} {177204} (\bibinfo {year} {2021})}\BibitemShut {NoStop}%
\bibitem [{\citenamefont {Rau}\ \emph {et~al.}(2014)\citenamefont {Rau},
  \citenamefont {Lee},\ and\ \citenamefont {Kee}}]{PhysRevLett.112.077204}%
  \BibitemOpen
  \bibfield  {author} {\bibinfo {author} {\bibfnamefont {J.~G.}\ \bibnamefont
  {Rau}}, \bibinfo {author} {\bibfnamefont {E.~K.-H.}\ \bibnamefont {Lee}},\
  and\ \bibinfo {author} {\bibfnamefont {H.-Y.}\ \bibnamefont {Kee}},\ }\href
  {https://doi.org/10.1103/PhysRevLett.112.077204} {\bibfield  {journal}
  {\bibinfo  {journal} {Phys. Rev. Lett.}\ }\textbf {\bibinfo {volume} {112}},\
  \bibinfo {pages} {077204} (\bibinfo {year} {2014})}\BibitemShut {NoStop}%
\bibitem [{Sup()}]{SuppMat}%
  \BibitemOpen
  \href@noop {} {}\bibinfo {note} {See Supplemental Material, which includes
  Refs.~\cite{neutron_scattering, PhysRevB.95.174429, Mu2021, chollet2015keras,
  RevModPhys.66.763, moshtagh2005minimum}, at [URL will be inserted by
  publisher] for more details of the experiments and calculations.}\BibitemShut
  {Stop}%
\bibitem [{\citenamefont {Chaloupka}\ and\ \citenamefont
  {Khaliullin}(2016)}]{PhysRevB.94.064435}%
  \BibitemOpen
  \bibfield  {author} {\bibinfo {author} {\bibfnamefont {J.}~\bibnamefont
  {Chaloupka}}\ and\ \bibinfo {author} {\bibfnamefont {G.}~\bibnamefont
  {Khaliullin}},\ }\href {https://doi.org/10.1103/PhysRevB.94.064435}
  {\bibfield  {journal} {\bibinfo  {journal} {Phys. Rev. B}\ }\textbf {\bibinfo
  {volume} {94}},\ \bibinfo {pages} {064435} (\bibinfo {year}
  {2016})}\BibitemShut {NoStop}%
\bibitem [{\citenamefont {Balz}\ \emph {et~al.}(2021)\citenamefont {Balz},
  \citenamefont {Janssen}, \citenamefont {Lampen-Kelley}, \citenamefont
  {Banerjee}, \citenamefont {Liu}, \citenamefont {Yan}, \citenamefont
  {Mandrus}, \citenamefont {Vojta},\ and\ \citenamefont
  {Nagler}}]{PhysRevB.103.174417}%
  \BibitemOpen
  \bibfield  {author} {\bibinfo {author} {\bibfnamefont {C.}~\bibnamefont
  {Balz}}, \bibinfo {author} {\bibfnamefont {L.}~\bibnamefont {Janssen}},
  \bibinfo {author} {\bibfnamefont {P.}~\bibnamefont {Lampen-Kelley}}, \bibinfo
  {author} {\bibfnamefont {A.}~\bibnamefont {Banerjee}}, \bibinfo {author}
  {\bibfnamefont {Y.~H.}\ \bibnamefont {Liu}}, \bibinfo {author} {\bibfnamefont
  {J.-Q.}\ \bibnamefont {Yan}}, \bibinfo {author} {\bibfnamefont {D.~G.}\
  \bibnamefont {Mandrus}}, \bibinfo {author} {\bibfnamefont {M.}~\bibnamefont
  {Vojta}},\ and\ \bibinfo {author} {\bibfnamefont {S.~E.}\ \bibnamefont
  {Nagler}},\ }\href {https://doi.org/10.1103/PhysRevB.103.174417} {\bibfield
  {journal} {\bibinfo  {journal} {Phys. Rev. B}\ }\textbf {\bibinfo {volume}
  {103}},\ \bibinfo {pages} {174417} (\bibinfo {year} {2021})}\BibitemShut
  {NoStop}%
\bibitem [{\citenamefont {Suzuki}\ \emph {et~al.}(2021)\citenamefont {Suzuki},
  \citenamefont {Liu}, \citenamefont {Bertinshaw}, \citenamefont {Ueda},
  \citenamefont {Kim}, \citenamefont {Laha}, \citenamefont {Weber},
  \citenamefont {Yang}, \citenamefont {Wang}, \citenamefont {H.~Takahash~and},
  \citenamefont {Minola}, \citenamefont {Lotsch}, \citenamefont {Kim},
  \citenamefont {Yavaş}, \citenamefont {Daghofer}, \citenamefont {Chaloupka},
  \citenamefont {Khaliullin}, \citenamefont {Gretarsson},\ and\ \citenamefont
  {Keimer}}]{Suzuki2021}%
  \BibitemOpen
  \bibfield  {author} {\bibinfo {author} {\bibfnamefont {H.}~\bibnamefont
  {Suzuki}}, \bibinfo {author} {\bibfnamefont {H.}~\bibnamefont {Liu}},
  \bibinfo {author} {\bibfnamefont {J.}~\bibnamefont {Bertinshaw}}, \bibinfo
  {author} {\bibfnamefont {K.}~\bibnamefont {Ueda}}, \bibinfo {author}
  {\bibfnamefont {H.}~\bibnamefont {Kim}}, \bibinfo {author} {\bibfnamefont
  {S.}~\bibnamefont {Laha}}, \bibinfo {author} {\bibfnamefont {D.}~\bibnamefont
  {Weber}}, \bibinfo {author} {\bibfnamefont {Z.}~\bibnamefont {Yang}},
  \bibinfo {author} {\bibfnamefont {L.}~\bibnamefont {Wang}}, \bibinfo {author}
  {\bibfnamefont {K.~F.}\ \bibnamefont {H.~Takahash~and}}, \bibinfo {author}
  {\bibfnamefont {M.}~\bibnamefont {Minola}}, \bibinfo {author} {\bibfnamefont
  {B.~V.}\ \bibnamefont {Lotsch}}, \bibinfo {author} {\bibfnamefont {B.~J.}\
  \bibnamefont {Kim}}, \bibinfo {author} {\bibfnamefont {H.}~\bibnamefont
  {Yavaş}}, \bibinfo {author} {\bibfnamefont {M.}~\bibnamefont {Daghofer}},
  \bibinfo {author} {\bibfnamefont {J.}~\bibnamefont {Chaloupka}}, \bibinfo
  {author} {\bibfnamefont {G.}~\bibnamefont {Khaliullin}}, \bibinfo {author}
  {\bibfnamefont {H.}~\bibnamefont {Gretarsson}},\ and\ \bibinfo {author}
  {\bibfnamefont {B.}~\bibnamefont {Keimer}},\ }\href
  {https://doi.org/https://doi.org/10.1038/s41467-021-24722-4} {\bibfield
  {journal} {\bibinfo  {journal} {Nat. Commun.}\ }\textbf {\bibinfo {volume}
  {12}},\ \bibinfo {pages} {4512} (\bibinfo {year} {2021})}\BibitemShut
  {NoStop}%
\bibitem [{\citenamefont {Sears}\ \emph {et~al.}(2015)\citenamefont {Sears},
  \citenamefont {Songvilay}, \citenamefont {Plumb}, \citenamefont {Clancy},
  \citenamefont {Qiu}, \citenamefont {Zhao}, \citenamefont {Parshall},\ and\
  \citenamefont {Kim}}]{PhysRevB.91.144420}%
  \BibitemOpen
  \bibfield  {author} {\bibinfo {author} {\bibfnamefont {J.~A.}\ \bibnamefont
  {Sears}}, \bibinfo {author} {\bibfnamefont {M.}~\bibnamefont {Songvilay}},
  \bibinfo {author} {\bibfnamefont {K.~W.}\ \bibnamefont {Plumb}}, \bibinfo
  {author} {\bibfnamefont {J.~P.}\ \bibnamefont {Clancy}}, \bibinfo {author}
  {\bibfnamefont {Y.}~\bibnamefont {Qiu}}, \bibinfo {author} {\bibfnamefont
  {Y.}~\bibnamefont {Zhao}}, \bibinfo {author} {\bibfnamefont {D.}~\bibnamefont
  {Parshall}},\ and\ \bibinfo {author} {\bibfnamefont {Y.-J.}\ \bibnamefont
  {Kim}},\ }\href {https://doi.org/10.1103/PhysRevB.91.144420} {\bibfield
  {journal} {\bibinfo  {journal} {Phys. Rev. B}\ }\textbf {\bibinfo {volume}
  {91}},\ \bibinfo {pages} {144420} (\bibinfo {year} {2015})}\BibitemShut
  {NoStop}%
\bibitem [{\citenamefont {Johnson}\ \emph {et~al.}(2015)\citenamefont
  {Johnson}, \citenamefont {Williams}, \citenamefont {Haghighirad},
  \citenamefont {Singleton}, \citenamefont {Zapf}, \citenamefont {Manuel},
  \citenamefont {Mazin}, \citenamefont {Li}, \citenamefont {Jeschke},
  \citenamefont {Valent\'{\i}},\ and\ \citenamefont
  {Coldea}}]{PhysRevB.92.235119}%
  \BibitemOpen
  \bibfield  {author} {\bibinfo {author} {\bibfnamefont {R.~D.}\ \bibnamefont
  {Johnson}}, \bibinfo {author} {\bibfnamefont {S.~C.}\ \bibnamefont
  {Williams}}, \bibinfo {author} {\bibfnamefont {A.~A.}\ \bibnamefont
  {Haghighirad}}, \bibinfo {author} {\bibfnamefont {J.}~\bibnamefont
  {Singleton}}, \bibinfo {author} {\bibfnamefont {V.}~\bibnamefont {Zapf}},
  \bibinfo {author} {\bibfnamefont {P.}~\bibnamefont {Manuel}}, \bibinfo
  {author} {\bibfnamefont {I.~I.}\ \bibnamefont {Mazin}}, \bibinfo {author}
  {\bibfnamefont {Y.}~\bibnamefont {Li}}, \bibinfo {author} {\bibfnamefont
  {H.~O.}\ \bibnamefont {Jeschke}}, \bibinfo {author} {\bibfnamefont
  {R.}~\bibnamefont {Valent\'{\i}}},\ and\ \bibinfo {author} {\bibfnamefont
  {R.}~\bibnamefont {Coldea}},\ }\href
  {https://doi.org/10.1103/PhysRevB.92.235119} {\bibfield  {journal} {\bibinfo
  {journal} {Phys. Rev. B}\ }\textbf {\bibinfo {volume} {92}},\ \bibinfo
  {pages} {235119} (\bibinfo {year} {2015})}\BibitemShut {NoStop}%
\bibitem [{\citenamefont {Cao}\ \emph {et~al.}(2016)\citenamefont {Cao},
  \citenamefont {Banerjee}, \citenamefont {Yan}, \citenamefont {Bridges},
  \citenamefont {Lumsden}, \citenamefont {Mandrus}, \citenamefont {Tennant},
  \citenamefont {Chakoumakos},\ and\ \citenamefont
  {Nagler}}]{PhysRevB.93.134423}%
  \BibitemOpen
  \bibfield  {author} {\bibinfo {author} {\bibfnamefont {H.~B.}\ \bibnamefont
  {Cao}}, \bibinfo {author} {\bibfnamefont {A.}~\bibnamefont {Banerjee}},
  \bibinfo {author} {\bibfnamefont {J.-Q.}\ \bibnamefont {Yan}}, \bibinfo
  {author} {\bibfnamefont {C.~A.}\ \bibnamefont {Bridges}}, \bibinfo {author}
  {\bibfnamefont {M.~D.}\ \bibnamefont {Lumsden}}, \bibinfo {author}
  {\bibfnamefont {D.~G.}\ \bibnamefont {Mandrus}}, \bibinfo {author}
  {\bibfnamefont {D.~A.}\ \bibnamefont {Tennant}}, \bibinfo {author}
  {\bibfnamefont {B.~C.}\ \bibnamefont {Chakoumakos}},\ and\ \bibinfo {author}
  {\bibfnamefont {S.~E.}\ \bibnamefont {Nagler}},\ }\href
  {https://doi.org/10.1103/PhysRevB.93.134423} {\bibfield  {journal} {\bibinfo
  {journal} {Phys. Rev. B}\ }\textbf {\bibinfo {volume} {93}},\ \bibinfo
  {pages} {134423} (\bibinfo {year} {2016})}\BibitemShut {NoStop}%
\bibitem [{\citenamefont {Jackeli}\ and\ \citenamefont
  {Khaliullin}(2009)}]{PhysRevLett.102.017205}%
  \BibitemOpen
  \bibfield  {author} {\bibinfo {author} {\bibfnamefont {G.}~\bibnamefont
  {Jackeli}}\ and\ \bibinfo {author} {\bibfnamefont {G.}~\bibnamefont
  {Khaliullin}},\ }\href {https://doi.org/10.1103/PhysRevLett.102.017205}
  {\bibfield  {journal} {\bibinfo  {journal} {Phys. Rev. Lett.}\ }\textbf
  {\bibinfo {volume} {102}},\ \bibinfo {pages} {017205} (\bibinfo {year}
  {2009})}\BibitemShut {NoStop}%
\bibitem [{\citenamefont {Winter}\ \emph {et~al.}(2016)\citenamefont {Winter},
  \citenamefont {Li}, \citenamefont {Jeschke},\ and\ \citenamefont
  {Valent\'{\i}}}]{PhysRevB.93.214431}%
  \BibitemOpen
  \bibfield  {author} {\bibinfo {author} {\bibfnamefont {S.~M.}\ \bibnamefont
  {Winter}}, \bibinfo {author} {\bibfnamefont {Y.}~\bibnamefont {Li}}, \bibinfo
  {author} {\bibfnamefont {H.~O.}\ \bibnamefont {Jeschke}},\ and\ \bibinfo
  {author} {\bibfnamefont {R.}~\bibnamefont {Valent\'{\i}}},\ }\href
  {https://doi.org/10.1103/PhysRevB.93.214431} {\bibfield  {journal} {\bibinfo
  {journal} {Phys. Rev. B}\ }\textbf {\bibinfo {volume} {93}},\ \bibinfo
  {pages} {214431} (\bibinfo {year} {2016})}\BibitemShut {NoStop}%
\bibitem [{\citenamefont {Eichstaedt}\ \emph {et~al.}(2019)\citenamefont
  {Eichstaedt}, \citenamefont {Zhang}, \citenamefont {Laurell}, \citenamefont
  {Okamoto}, \citenamefont {Eguiluz},\ and\ \citenamefont
  {Berlijn}}]{PhysRevB.100.075110}%
  \BibitemOpen
  \bibfield  {author} {\bibinfo {author} {\bibfnamefont {C.}~\bibnamefont
  {Eichstaedt}}, \bibinfo {author} {\bibfnamefont {Y.}~\bibnamefont {Zhang}},
  \bibinfo {author} {\bibfnamefont {P.}~\bibnamefont {Laurell}}, \bibinfo
  {author} {\bibfnamefont {S.}~\bibnamefont {Okamoto}}, \bibinfo {author}
  {\bibfnamefont {A.~G.}\ \bibnamefont {Eguiluz}},\ and\ \bibinfo {author}
  {\bibfnamefont {T.}~\bibnamefont {Berlijn}},\ }\href
  {https://doi.org/10.1103/PhysRevB.100.075110} {\bibfield  {journal} {\bibinfo
   {journal} {Phys. Rev. B}\ }\textbf {\bibinfo {volume} {100}},\ \bibinfo
  {pages} {075110} (\bibinfo {year} {2019})}\BibitemShut {NoStop}%
\bibitem [{\citenamefont {Winter}\ \emph
  {et~al.}(2017{\natexlab{b}})\citenamefont {Winter}, \citenamefont {Riedl},
  \citenamefont {Maksimov}, \citenamefont {Chernyshev}, \citenamefont
  {Honecker},\ and\ \citenamefont {Valent\'i}}]{10.1038/s41467-017-01177-0}%
  \BibitemOpen
  \bibfield  {author} {\bibinfo {author} {\bibfnamefont {S.~M.}\ \bibnamefont
  {Winter}}, \bibinfo {author} {\bibfnamefont {K.}~\bibnamefont {Riedl}},
  \bibinfo {author} {\bibfnamefont {P.~A.}\ \bibnamefont {Maksimov}}, \bibinfo
  {author} {\bibfnamefont {A.~L.}\ \bibnamefont {Chernyshev}}, \bibinfo
  {author} {\bibfnamefont {A.}~\bibnamefont {Honecker}},\ and\ \bibinfo
  {author} {\bibfnamefont {R.}~\bibnamefont {Valent\'i}},\ }\href
  {https://doi.org/10.1038/s41467-017-01177-0} {\bibfield  {journal} {\bibinfo
  {journal} {Nat. Commun.}\ }\textbf {\bibinfo {volume} {8}},\ \bibinfo {pages}
  {1152} (\bibinfo {year} {2017}{\natexlab{b}})}\BibitemShut {NoStop}%
\bibitem [{\citenamefont {Winter}\ \emph {et~al.}(2018)\citenamefont {Winter},
  \citenamefont {Riedl}, \citenamefont {Kaib}, \citenamefont {Coldea},\ and\
  \citenamefont {Valent\'{\i}}}]{PhysRevLett.120.077203}%
  \BibitemOpen
  \bibfield  {author} {\bibinfo {author} {\bibfnamefont {S.~M.}\ \bibnamefont
  {Winter}}, \bibinfo {author} {\bibfnamefont {K.}~\bibnamefont {Riedl}},
  \bibinfo {author} {\bibfnamefont {D.}~\bibnamefont {Kaib}}, \bibinfo {author}
  {\bibfnamefont {R.}~\bibnamefont {Coldea}},\ and\ \bibinfo {author}
  {\bibfnamefont {R.}~\bibnamefont {Valent\'{\i}}},\ }\href
  {https://doi.org/10.1103/PhysRevLett.120.077203} {\bibfield  {journal}
  {\bibinfo  {journal} {Phys. Rev. Lett.}\ }\textbf {\bibinfo {volume} {120}},\
  \bibinfo {pages} {077203} (\bibinfo {year} {2018})}\BibitemShut {NoStop}%
\bibitem [{\citenamefont {Maksimov}\ and\ \citenamefont
  {Chernyshev}(2020)}]{PhysRevResearch.2.033011}%
  \BibitemOpen
  \bibfield  {author} {\bibinfo {author} {\bibfnamefont {P.~A.}\ \bibnamefont
  {Maksimov}}\ and\ \bibinfo {author} {\bibfnamefont {A.~L.}\ \bibnamefont
  {Chernyshev}},\ }\href {https://doi.org/10.1103/PhysRevResearch.2.033011}
  {\bibfield  {journal} {\bibinfo  {journal} {Phys. Rev. Research}\ }\textbf
  {\bibinfo {volume} {2}},\ \bibinfo {pages} {033011} (\bibinfo {year}
  {2020})}\BibitemShut {NoStop}%
\bibitem [{\citenamefont {Nasu}\ \emph {et~al.}(2016)\citenamefont {Nasu},
  \citenamefont {Knolle}, \citenamefont {Kovrizhin}, \citenamefont {Motome},\
  and\ \citenamefont {Moessner}}]{Nasu2016}%
  \BibitemOpen
  \bibfield  {author} {\bibinfo {author} {\bibfnamefont {J.}~\bibnamefont
  {Nasu}}, \bibinfo {author} {\bibfnamefont {J.}~\bibnamefont {Knolle}},
  \bibinfo {author} {\bibfnamefont {D.~L.}\ \bibnamefont {Kovrizhin}}, \bibinfo
  {author} {\bibfnamefont {Y.}~\bibnamefont {Motome}},\ and\ \bibinfo {author}
  {\bibfnamefont {R.}~\bibnamefont {Moessner}},\ }\href
  {https://doi.org/10.1038/nphys3809} {\bibfield  {journal} {\bibinfo
  {journal} {Nat. Phys.}\ }\textbf {\bibinfo {volume} {12}},\ \bibinfo {pages}
  {912} (\bibinfo {year} {2016})}\BibitemShut {NoStop}%
\bibitem [{\citenamefont {Du}\ \emph {et~al.}(2018)\citenamefont {Du},
  \citenamefont {Huang}, \citenamefont {Wang}, \citenamefont {Wang},
  \citenamefont {Yang}, \citenamefont {Tang}, \citenamefont {Liao},
  \citenamefont {Shi}, \citenamefont {Shi},\ and\ \citenamefont
  {Zhou}}]{Du2018}%
  \BibitemOpen
  \bibfield  {author} {\bibinfo {author} {\bibfnamefont {L.}~\bibnamefont
  {Du}}, \bibinfo {author} {\bibfnamefont {Y.}~\bibnamefont {Huang}}, \bibinfo
  {author} {\bibfnamefont {Y.}~\bibnamefont {Wang}}, \bibinfo {author}
  {\bibfnamefont {Q.}~\bibnamefont {Wang}}, \bibinfo {author} {\bibfnamefont
  {R.}~\bibnamefont {Yang}}, \bibinfo {author} {\bibfnamefont {J.}~\bibnamefont
  {Tang}}, \bibinfo {author} {\bibfnamefont {M.}~\bibnamefont {Liao}}, \bibinfo
  {author} {\bibfnamefont {D.}~\bibnamefont {Shi}}, \bibinfo {author}
  {\bibfnamefont {Y.}~\bibnamefont {Shi}},\ and\ \bibinfo {author}
  {\bibfnamefont {X.}~\bibnamefont {Zhou}},\ }\href
  {https://doi.org/10.1088/2053-1583/aaee29} {\bibfield  {journal} {\bibinfo
  {journal} {2D Mater.}\ }\textbf {\bibinfo {volume} {6}},\ \bibinfo {pages}
  {015014} (\bibinfo {year} {2018})}\BibitemShut {NoStop}%
\bibitem [{\citenamefont {Mai}\ \emph {et~al.}(2019)\citenamefont {Mai},
  \citenamefont {McCreary}, \citenamefont {Lampen-Kelley}, \citenamefont
  {Butch}, \citenamefont {Simpson}, \citenamefont {Yan}, \citenamefont
  {Nagler}, \citenamefont {Mandrus}, \citenamefont {Walker},\ and\
  \citenamefont {Aguilar}}]{PhysRevB.100.134419}%
  \BibitemOpen
  \bibfield  {author} {\bibinfo {author} {\bibfnamefont {T.~T.}\ \bibnamefont
  {Mai}}, \bibinfo {author} {\bibfnamefont {A.}~\bibnamefont {McCreary}},
  \bibinfo {author} {\bibfnamefont {P.}~\bibnamefont {Lampen-Kelley}}, \bibinfo
  {author} {\bibfnamefont {N.}~\bibnamefont {Butch}}, \bibinfo {author}
  {\bibfnamefont {J.~R.}\ \bibnamefont {Simpson}}, \bibinfo {author}
  {\bibfnamefont {J.-Q.}\ \bibnamefont {Yan}}, \bibinfo {author} {\bibfnamefont
  {S.~E.}\ \bibnamefont {Nagler}}, \bibinfo {author} {\bibfnamefont
  {D.}~\bibnamefont {Mandrus}}, \bibinfo {author} {\bibfnamefont {A.~R.~H.}\
  \bibnamefont {Walker}},\ and\ \bibinfo {author} {\bibfnamefont {R.~V.}\
  \bibnamefont {Aguilar}},\ }\href
  {https://doi.org/10.1103/PhysRevB.100.134419} {\bibfield  {journal} {\bibinfo
   {journal} {Phys. Rev. B}\ }\textbf {\bibinfo {volume} {100}},\ \bibinfo
  {pages} {134419} (\bibinfo {year} {2019})}\BibitemShut {NoStop}%
\bibitem [{\citenamefont {Wang}\ \emph {et~al.}(2020)\citenamefont {Wang},
  \citenamefont {Osterhoudt}, \citenamefont {Tian}, \citenamefont
  {Lampen-Kelley}, \citenamefont {Banerjee}, \citenamefont {Goldstein},
  \citenamefont {Yan}, \citenamefont {Knolle}, \citenamefont {Ji},
  \citenamefont {Cava}, \citenamefont {Nasu}, \citenamefont {Motome},
  \citenamefont {Nagler}, \citenamefont {Mandrus},\ and\ \citenamefont
  {Burch}}]{Wang2020a}%
  \BibitemOpen
  \bibfield  {author} {\bibinfo {author} {\bibfnamefont {Y.}~\bibnamefont
  {Wang}}, \bibinfo {author} {\bibfnamefont {G.~B.}\ \bibnamefont
  {Osterhoudt}}, \bibinfo {author} {\bibfnamefont {Y.}~\bibnamefont {Tian}},
  \bibinfo {author} {\bibfnamefont {P.}~\bibnamefont {Lampen-Kelley}}, \bibinfo
  {author} {\bibfnamefont {A.}~\bibnamefont {Banerjee}}, \bibinfo {author}
  {\bibfnamefont {T.}~\bibnamefont {Goldstein}}, \bibinfo {author}
  {\bibfnamefont {J.}~\bibnamefont {Yan}}, \bibinfo {author} {\bibfnamefont
  {J.}~\bibnamefont {Knolle}}, \bibinfo {author} {\bibfnamefont
  {H.}~\bibnamefont {Ji}}, \bibinfo {author} {\bibfnamefont {R.~J.}\
  \bibnamefont {Cava}}, \bibinfo {author} {\bibfnamefont {J.}~\bibnamefont
  {Nasu}}, \bibinfo {author} {\bibfnamefont {Y.}~\bibnamefont {Motome}},
  \bibinfo {author} {\bibfnamefont {S.~E.}\ \bibnamefont {Nagler}}, \bibinfo
  {author} {\bibfnamefont {D.}~\bibnamefont {Mandrus}},\ and\ \bibinfo {author}
  {\bibfnamefont {K.~S.}\ \bibnamefont {Burch}},\ }\href
  {https://doi.org/https://doi.org/10.1038/s41535-020-0216-6} {\bibfield
  {journal} {\bibinfo  {journal} {npj Quantum Mater.}\ }\textbf {\bibinfo
  {volume} {5}},\ \bibinfo {pages} {14} (\bibinfo {year} {2020})}\BibitemShut
  {NoStop}%
\bibitem [{\citenamefont {Zheng}\ \emph {et~al.}(2017)\citenamefont {Zheng},
  \citenamefont {Ran}, \citenamefont {Li}, \citenamefont {Wang}, \citenamefont
  {Wang}, \citenamefont {Liu}, \citenamefont {Liu}, \citenamefont {Normand},
  \citenamefont {Wen},\ and\ \citenamefont {Yu}}]{PhysRevLett.119.227208}%
  \BibitemOpen
  \bibfield  {author} {\bibinfo {author} {\bibfnamefont {J.}~\bibnamefont
  {Zheng}}, \bibinfo {author} {\bibfnamefont {K.}~\bibnamefont {Ran}}, \bibinfo
  {author} {\bibfnamefont {T.}~\bibnamefont {Li}}, \bibinfo {author}
  {\bibfnamefont {J.}~\bibnamefont {Wang}}, \bibinfo {author} {\bibfnamefont
  {P.}~\bibnamefont {Wang}}, \bibinfo {author} {\bibfnamefont {B.}~\bibnamefont
  {Liu}}, \bibinfo {author} {\bibfnamefont {Z.-X.}\ \bibnamefont {Liu}},
  \bibinfo {author} {\bibfnamefont {B.}~\bibnamefont {Normand}}, \bibinfo
  {author} {\bibfnamefont {J.}~\bibnamefont {Wen}},\ and\ \bibinfo {author}
  {\bibfnamefont {W.}~\bibnamefont {Yu}},\ }\href
  {https://doi.org/10.1103/PhysRevLett.119.227208} {\bibfield  {journal}
  {\bibinfo  {journal} {Phys. Rev. Lett.}\ }\textbf {\bibinfo {volume} {119}},\
  \bibinfo {pages} {227208} (\bibinfo {year} {2017})}\BibitemShut {NoStop}%
\bibitem [{\citenamefont {Baek}\ \emph {et~al.}(2017)\citenamefont {Baek},
  \citenamefont {Do}, \citenamefont {Choi}, \citenamefont {Kwon}, \citenamefont
  {Wolter}, \citenamefont {Nishimoto}, \citenamefont {van~den Brink},\ and\
  \citenamefont {B\"uchner}}]{PhysRevLett.119.037201}%
  \BibitemOpen
  \bibfield  {author} {\bibinfo {author} {\bibfnamefont {S.-H.}\ \bibnamefont
  {Baek}}, \bibinfo {author} {\bibfnamefont {S.-H.}\ \bibnamefont {Do}},
  \bibinfo {author} {\bibfnamefont {K.-Y.}\ \bibnamefont {Choi}}, \bibinfo
  {author} {\bibfnamefont {Y.~S.}\ \bibnamefont {Kwon}}, \bibinfo {author}
  {\bibfnamefont {A.~U.~B.}\ \bibnamefont {Wolter}}, \bibinfo {author}
  {\bibfnamefont {S.}~\bibnamefont {Nishimoto}}, \bibinfo {author}
  {\bibfnamefont {J.}~\bibnamefont {van~den Brink}},\ and\ \bibinfo {author}
  {\bibfnamefont {B.}~\bibnamefont {B\"uchner}},\ }\href
  {https://doi.org/10.1103/PhysRevLett.119.037201} {\bibfield  {journal}
  {\bibinfo  {journal} {Phys. Rev. Lett.}\ }\textbf {\bibinfo {volume} {119}},\
  \bibinfo {pages} {037201} (\bibinfo {year} {2017})}\BibitemShut {NoStop}%
\bibitem [{\citenamefont {Sears}\ \emph {et~al.}(2017)\citenamefont {Sears},
  \citenamefont {Zhao}, \citenamefont {Xu}, \citenamefont {Lynn},\ and\
  \citenamefont {Kim}}]{PhysRevB.95.180411}%
  \BibitemOpen
  \bibfield  {author} {\bibinfo {author} {\bibfnamefont {J.~A.}\ \bibnamefont
  {Sears}}, \bibinfo {author} {\bibfnamefont {Y.}~\bibnamefont {Zhao}},
  \bibinfo {author} {\bibfnamefont {Z.}~\bibnamefont {Xu}}, \bibinfo {author}
  {\bibfnamefont {J.~W.}\ \bibnamefont {Lynn}},\ and\ \bibinfo {author}
  {\bibfnamefont {Y.-J.}\ \bibnamefont {Kim}},\ }\href
  {https://doi.org/10.1103/PhysRevB.95.180411} {\bibfield  {journal} {\bibinfo
  {journal} {Phys. Rev. B}\ }\textbf {\bibinfo {volume} {95}},\ \bibinfo
  {pages} {180411} (\bibinfo {year} {2017})}\BibitemShut {NoStop}%
\bibitem [{\citenamefont {Czajka}\ \emph {et~al.}(2021)\citenamefont {Czajka},
  \citenamefont {Gao}, \citenamefont {Hirschberger}, \citenamefont
  {Lampen-Kelley}, \citenamefont {Banerjee}, \citenamefont {Yan}, \citenamefont
  {Mandrus}, \citenamefont {Nagler},\ and\ \citenamefont {Ong}}]{Czajka2021}%
  \BibitemOpen
  \bibfield  {author} {\bibinfo {author} {\bibfnamefont {P.}~\bibnamefont
  {Czajka}}, \bibinfo {author} {\bibfnamefont {T.}~\bibnamefont {Gao}},
  \bibinfo {author} {\bibfnamefont {M.}~\bibnamefont {Hirschberger}}, \bibinfo
  {author} {\bibfnamefont {P.}~\bibnamefont {Lampen-Kelley}}, \bibinfo {author}
  {\bibfnamefont {A.}~\bibnamefont {Banerjee}}, \bibinfo {author}
  {\bibfnamefont {J.}~\bibnamefont {Yan}}, \bibinfo {author} {\bibfnamefont
  {D.~G.}\ \bibnamefont {Mandrus}}, \bibinfo {author} {\bibfnamefont {S.~E.}\
  \bibnamefont {Nagler}},\ and\ \bibinfo {author} {\bibfnamefont {N.~P.}\
  \bibnamefont {Ong}},\ }\href
  {https://doi.org/https://doi.org/10.1038/s41567-021-01243-x} {\bibfield
  {journal} {\bibinfo  {journal} {Nat. Phys.}\ }\textbf {\bibinfo {volume}
  {17}},\ \bibinfo {pages} {915} (\bibinfo {year} {2021})}\BibitemShut
  {NoStop}%
\bibitem [{\citenamefont {Villadiego}(2021)}]{Inti2021}%
  \BibitemOpen
  \bibfield  {author} {\bibinfo {author} {\bibfnamefont {I.~S.}\ \bibnamefont
  {Villadiego}},\ }\href {https://doi.org/10.1103/PhysRevB.104.195149}
  {\bibfield  {journal} {\bibinfo  {journal} {Phys. Rev. B}\ }\textbf {\bibinfo
  {volume} {104}},\ \bibinfo {pages} {195149} (\bibinfo {year}
  {2021})}\BibitemShut {NoStop}%
\bibitem [{\citenamefont {Kr\"uger}\ and\ \citenamefont
  {Janssen}(2021)}]{PhysRevB.104.165133}%
  \BibitemOpen
  \bibfield  {author} {\bibinfo {author} {\bibfnamefont {W.~G.~F.}\
  \bibnamefont {Kr\"uger}}\ and\ \bibinfo {author} {\bibfnamefont
  {L.}~\bibnamefont {Janssen}},\ }\href
  {https://doi.org/10.1103/PhysRevB.104.165133} {\bibfield  {journal} {\bibinfo
   {journal} {Phys. Rev. B}\ }\textbf {\bibinfo {volume} {104}},\ \bibinfo
  {pages} {165133} (\bibinfo {year} {2021})}\BibitemShut {NoStop}%
\bibitem [{\citenamefont {Kasahara}\ \emph
  {et~al.}(2018{\natexlab{a}})\citenamefont {Kasahara}, \citenamefont
  {Ohnishi}, \citenamefont {Mizukami}, \citenamefont {Tanaka}, \citenamefont
  {Ma}, \citenamefont {Sugii}, \citenamefont {Kurita}, \citenamefont {Tanaka},
  \citenamefont {Nasu}, \citenamefont {Motome}, \citenamefont {Shibauchi},\
  and\ \citenamefont {Matsuda}}]{10.1038/s41586-018-0274-0}%
  \BibitemOpen
  \bibfield  {author} {\bibinfo {author} {\bibfnamefont {Y.}~\bibnamefont
  {Kasahara}}, \bibinfo {author} {\bibfnamefont {T.}~\bibnamefont {Ohnishi}},
  \bibinfo {author} {\bibfnamefont {Y.}~\bibnamefont {Mizukami}}, \bibinfo
  {author} {\bibfnamefont {O.}~\bibnamefont {Tanaka}}, \bibinfo {author}
  {\bibfnamefont {S.}~\bibnamefont {Ma}}, \bibinfo {author} {\bibfnamefont
  {K.}~\bibnamefont {Sugii}}, \bibinfo {author} {\bibfnamefont
  {N.}~\bibnamefont {Kurita}}, \bibinfo {author} {\bibfnamefont
  {H.}~\bibnamefont {Tanaka}}, \bibinfo {author} {\bibfnamefont
  {J.}~\bibnamefont {Nasu}}, \bibinfo {author} {\bibfnamefont {Y.}~\bibnamefont
  {Motome}}, \bibinfo {author} {\bibfnamefont {T.}~\bibnamefont {Shibauchi}},\
  and\ \bibinfo {author} {\bibfnamefont {Y.}~\bibnamefont {Matsuda}},\ }\href
  {https://doi.org/10.1038/s41586-018-0274-0} {\bibfield  {journal} {\bibinfo
  {journal} {Nature (London)}\ }\textbf {\bibinfo {volume} {559}},\ \bibinfo
  {pages} {227} (\bibinfo {year} {2018}{\natexlab{a}})}\BibitemShut {NoStop}%
\bibitem [{\citenamefont {Yamashita}\ \emph {et~al.}(2020)\citenamefont
  {Yamashita}, \citenamefont {Gouchi}, \citenamefont {Uwatoko}, \citenamefont
  {Kurita},\ and\ \citenamefont {Tanaka}}]{PhysRevB.102.220404}%
  \BibitemOpen
  \bibfield  {author} {\bibinfo {author} {\bibfnamefont {M.}~\bibnamefont
  {Yamashita}}, \bibinfo {author} {\bibfnamefont {J.}~\bibnamefont {Gouchi}},
  \bibinfo {author} {\bibfnamefont {Y.}~\bibnamefont {Uwatoko}}, \bibinfo
  {author} {\bibfnamefont {N.}~\bibnamefont {Kurita}},\ and\ \bibinfo {author}
  {\bibfnamefont {H.}~\bibnamefont {Tanaka}},\ }\href
  {https://doi.org/10.1103/PhysRevB.102.220404} {\bibfield  {journal} {\bibinfo
   {journal} {Phys. Rev. B}\ }\textbf {\bibinfo {volume} {102}},\ \bibinfo
  {pages} {220404} (\bibinfo {year} {2020})}\BibitemShut {NoStop}%
\bibitem [{\citenamefont {Yokoi}\ \emph {et~al.}(2021)\citenamefont {Yokoi},
  \citenamefont {Ma}, \citenamefont {Kasahara}, \citenamefont {Kasahara},
  \citenamefont {Shibauchi}, \citenamefont {Kurita}, \citenamefont {Tanaka},
  \citenamefont {Nasu}, \citenamefont {Motome}, \citenamefont {Hickey},
  \citenamefont {Trebst},\ and\ \citenamefont {Matsuda}}]{Yokoi_2021}%
  \BibitemOpen
  \bibfield  {author} {\bibinfo {author} {\bibfnamefont {T.}~\bibnamefont
  {Yokoi}}, \bibinfo {author} {\bibfnamefont {S.}~\bibnamefont {Ma}}, \bibinfo
  {author} {\bibfnamefont {Y.}~\bibnamefont {Kasahara}}, \bibinfo {author}
  {\bibfnamefont {S.}~\bibnamefont {Kasahara}}, \bibinfo {author}
  {\bibfnamefont {T.}~\bibnamefont {Shibauchi}}, \bibinfo {author}
  {\bibfnamefont {N.}~\bibnamefont {Kurita}}, \bibinfo {author} {\bibfnamefont
  {H.}~\bibnamefont {Tanaka}}, \bibinfo {author} {\bibfnamefont
  {J.}~\bibnamefont {Nasu}}, \bibinfo {author} {\bibfnamefont {Y.}~\bibnamefont
  {Motome}}, \bibinfo {author} {\bibfnamefont {C.}~\bibnamefont {Hickey}},
  \bibinfo {author} {\bibfnamefont {S.}~\bibnamefont {Trebst}},\ and\ \bibinfo
  {author} {\bibfnamefont {Y.}~\bibnamefont {Matsuda}},\ }\href
  {https://doi.org/10.1126/science.aay5551} {\bibfield  {journal} {\bibinfo
  {journal} {Science}\ }\textbf {\bibinfo {volume} {373}},\ \bibinfo {pages}
  {568} (\bibinfo {year} {2021})}\BibitemShut {NoStop}%
\bibitem [{\citenamefont {Li}\ \emph {et~al.}(2021{\natexlab{a}})\citenamefont
  {Li}, \citenamefont {Zhang}, \citenamefont {Said}, \citenamefont {Fabbris},
  \citenamefont {Mazzone}, \citenamefont {Yan}, \citenamefont {Mandrus},
  \citenamefont {Hal\'asz}, \citenamefont {Okamoto}, \citenamefont {Murakami},
  \citenamefont {Dean}, \citenamefont {Lee},\ and\ \citenamefont
  {Miao}}]{Li2021Phonon}%
  \BibitemOpen
  \bibfield  {author} {\bibinfo {author} {\bibfnamefont {H.}~\bibnamefont
  {Li}}, \bibinfo {author} {\bibfnamefont {T.~T.}\ \bibnamefont {Zhang}},
  \bibinfo {author} {\bibfnamefont {A.}~\bibnamefont {Said}}, \bibinfo {author}
  {\bibfnamefont {G.}~\bibnamefont {Fabbris}}, \bibinfo {author} {\bibfnamefont
  {D.~G.}\ \bibnamefont {Mazzone}}, \bibinfo {author} {\bibfnamefont {J.~Q.}\
  \bibnamefont {Yan}}, \bibinfo {author} {\bibfnamefont {D.}~\bibnamefont
  {Mandrus}}, \bibinfo {author} {\bibfnamefont {G.~B.}\ \bibnamefont
  {Hal\'asz}}, \bibinfo {author} {\bibfnamefont {S.}~\bibnamefont {Okamoto}},
  \bibinfo {author} {\bibfnamefont {S.}~\bibnamefont {Murakami}}, \bibinfo
  {author} {\bibfnamefont {M.~P.~M.}\ \bibnamefont {Dean}}, \bibinfo {author}
  {\bibfnamefont {H.~N.}\ \bibnamefont {Lee}},\ and\ \bibinfo {author}
  {\bibfnamefont {H.}~\bibnamefont {Miao}},\ }\href
  {https://doi.org/https://doi.org/10.1038/s41467-021-23826-1} {\bibfield
  {journal} {\bibinfo  {journal} {Nat. Commun.}\ }\textbf {\bibinfo {volume}
  {12}},\ \bibinfo {pages} {3513} (\bibinfo {year}
  {2021}{\natexlab{a}})}\BibitemShut {NoStop}%
\bibitem [{\citenamefont {Wolter}\ \emph {et~al.}(2017)\citenamefont {Wolter},
  \citenamefont {Corredor}, \citenamefont {Janssen}, \citenamefont {Nenkov},
  \citenamefont {Sch\"onecker}, \citenamefont {Do}, \citenamefont {Choi},
  \citenamefont {Albrecht}, \citenamefont {Hunger}, \citenamefont {Doert},
  \citenamefont {Vojta},\ and\ \citenamefont {B\"uchner}}]{PhysRevB.96.041405}%
  \BibitemOpen
  \bibfield  {author} {\bibinfo {author} {\bibfnamefont {A.~U.~B.}\
  \bibnamefont {Wolter}}, \bibinfo {author} {\bibfnamefont {L.~T.}\
  \bibnamefont {Corredor}}, \bibinfo {author} {\bibfnamefont {L.}~\bibnamefont
  {Janssen}}, \bibinfo {author} {\bibfnamefont {K.}~\bibnamefont {Nenkov}},
  \bibinfo {author} {\bibfnamefont {S.}~\bibnamefont {Sch\"onecker}}, \bibinfo
  {author} {\bibfnamefont {S.-H.}\ \bibnamefont {Do}}, \bibinfo {author}
  {\bibfnamefont {K.-Y.}\ \bibnamefont {Choi}}, \bibinfo {author}
  {\bibfnamefont {R.}~\bibnamefont {Albrecht}}, \bibinfo {author}
  {\bibfnamefont {J.}~\bibnamefont {Hunger}}, \bibinfo {author} {\bibfnamefont
  {T.}~\bibnamefont {Doert}}, \bibinfo {author} {\bibfnamefont
  {M.}~\bibnamefont {Vojta}},\ and\ \bibinfo {author} {\bibfnamefont
  {B.}~\bibnamefont {B\"uchner}},\ }\href
  {https://doi.org/10.1103/PhysRevB.96.041405} {\bibfield  {journal} {\bibinfo
  {journal} {Phys. Rev. B}\ }\textbf {\bibinfo {volume} {96}},\ \bibinfo
  {pages} {041405} (\bibinfo {year} {2017})}\BibitemShut {NoStop}%
\bibitem [{\citenamefont {Widmann}\ \emph {et~al.}(2019)\citenamefont
  {Widmann}, \citenamefont {Tsurkan}, \citenamefont {Prishchenko},
  \citenamefont {Mazurenko}, \citenamefont {Tsirlin},\ and\ \citenamefont
  {Loidl}}]{PhysRevB.99.094415}%
  \BibitemOpen
  \bibfield  {author} {\bibinfo {author} {\bibfnamefont {S.}~\bibnamefont
  {Widmann}}, \bibinfo {author} {\bibfnamefont {V.}~\bibnamefont {Tsurkan}},
  \bibinfo {author} {\bibfnamefont {D.~A.}\ \bibnamefont {Prishchenko}},
  \bibinfo {author} {\bibfnamefont {V.~G.}\ \bibnamefont {Mazurenko}}, \bibinfo
  {author} {\bibfnamefont {A.~A.}\ \bibnamefont {Tsirlin}},\ and\ \bibinfo
  {author} {\bibfnamefont {A.}~\bibnamefont {Loidl}},\ }\href
  {https://doi.org/10.1103/PhysRevB.99.094415} {\bibfield  {journal} {\bibinfo
  {journal} {Phys. Rev. B}\ }\textbf {\bibinfo {volume} {99}},\ \bibinfo
  {pages} {094415} (\bibinfo {year} {2019})}\BibitemShut {NoStop}%
\bibitem [{\citenamefont {Bachus}\ \emph {et~al.}(2020)\citenamefont {Bachus},
  \citenamefont {Kaib}, \citenamefont {Tokiwa}, \citenamefont {Jesche},
  \citenamefont {Tsurkan}, \citenamefont {Loidl}, \citenamefont {Winter},
  \citenamefont {Tsirlin}, \citenamefont {Valent\'{\i}},\ and\ \citenamefont
  {Gegenwart}}]{PhysRevLett.125.097203}%
  \BibitemOpen
  \bibfield  {author} {\bibinfo {author} {\bibfnamefont {S.}~\bibnamefont
  {Bachus}}, \bibinfo {author} {\bibfnamefont {D.~A.~S.}\ \bibnamefont {Kaib}},
  \bibinfo {author} {\bibfnamefont {Y.}~\bibnamefont {Tokiwa}}, \bibinfo
  {author} {\bibfnamefont {A.}~\bibnamefont {Jesche}}, \bibinfo {author}
  {\bibfnamefont {V.}~\bibnamefont {Tsurkan}}, \bibinfo {author} {\bibfnamefont
  {A.}~\bibnamefont {Loidl}}, \bibinfo {author} {\bibfnamefont {S.~M.}\
  \bibnamefont {Winter}}, \bibinfo {author} {\bibfnamefont {A.~A.}\
  \bibnamefont {Tsirlin}}, \bibinfo {author} {\bibfnamefont {R.}~\bibnamefont
  {Valent\'{\i}}},\ and\ \bibinfo {author} {\bibfnamefont {P.}~\bibnamefont
  {Gegenwart}},\ }\href {https://doi.org/10.1103/PhysRevLett.125.097203}
  {\bibfield  {journal} {\bibinfo  {journal} {Phys. Rev. Lett.}\ }\textbf
  {\bibinfo {volume} {125}},\ \bibinfo {pages} {097203} (\bibinfo {year}
  {2020})}\BibitemShut {NoStop}%
\bibitem [{\citenamefont {Tanaka}\ \emph {et~al.}(2022)\citenamefont {Tanaka},
  \citenamefont {Mizukami}, \citenamefont {Harasawa}, \citenamefont
  {Hashimoto}, \citenamefont {Hwang}, \citenamefont {Kurita}, \citenamefont
  {Tanaka}, \citenamefont {Fujimoto}, \citenamefont {Matsuda}, \citenamefont
  {Moon},\ and\ \citenamefont {Shibauchi}}]{Tanaka2022}%
  \BibitemOpen
  \bibfield  {author} {\bibinfo {author} {\bibfnamefont {O.}~\bibnamefont
  {Tanaka}}, \bibinfo {author} {\bibfnamefont {Y.}~\bibnamefont {Mizukami}},
  \bibinfo {author} {\bibfnamefont {R.}~\bibnamefont {Harasawa}}, \bibinfo
  {author} {\bibfnamefont {K.}~\bibnamefont {Hashimoto}}, \bibinfo {author}
  {\bibfnamefont {K.}~\bibnamefont {Hwang}}, \bibinfo {author} {\bibfnamefont
  {N.}~\bibnamefont {Kurita}}, \bibinfo {author} {\bibfnamefont
  {H.}~\bibnamefont {Tanaka}}, \bibinfo {author} {\bibfnamefont
  {S.}~\bibnamefont {Fujimoto}}, \bibinfo {author} {\bibfnamefont
  {Y.}~\bibnamefont {Matsuda}}, \bibinfo {author} {\bibfnamefont {E.-G.}\
  \bibnamefont {Moon}},\ and\ \bibinfo {author} {\bibfnamefont
  {T.}~\bibnamefont {Shibauchi}},\ }\bibfield  {journal} {\bibinfo  {journal}
  {Nat. Phys.}\ }\href
  {https://doi.org/https://doi.org/10.1038/s41567-021-01488-6}
  {https://doi.org/10.1038/s41567-021-01488-6} (\bibinfo {year}
  {2022})\BibitemShut {NoStop}%
\bibitem [{\citenamefont {Jan\v{s}a}\ \emph {et~al.}(2018)\citenamefont
  {Jan\v{s}a}, \citenamefont {Zorko}, \citenamefont {Gomil\v{s}ek},
  \citenamefont {Pregelj}, \citenamefont {Kr\"amer}, \citenamefont {Biner},
  \citenamefont {Biffin}, \citenamefont {R\"uegg},\ and\ \citenamefont
  {Klanj\v{s}ek}}]{Jansa2018}%
  \BibitemOpen
  \bibfield  {author} {\bibinfo {author} {\bibfnamefont {N.}~\bibnamefont
  {Jan\v{s}a}}, \bibinfo {author} {\bibfnamefont {A.}~\bibnamefont {Zorko}},
  \bibinfo {author} {\bibfnamefont {M.}~\bibnamefont {Gomil\v{s}ek}}, \bibinfo
  {author} {\bibfnamefont {M.}~\bibnamefont {Pregelj}}, \bibinfo {author}
  {\bibfnamefont {K.~W.}\ \bibnamefont {Kr\"amer}}, \bibinfo {author}
  {\bibfnamefont {D.}~\bibnamefont {Biner}}, \bibinfo {author} {\bibfnamefont
  {A.}~\bibnamefont {Biffin}}, \bibinfo {author} {\bibfnamefont
  {C.}~\bibnamefont {R\"uegg}},\ and\ \bibinfo {author} {\bibfnamefont
  {M.}~\bibnamefont {Klanj\v{s}ek}},\ }\href
  {https://doi.org/10.1038/s41567-018-0129-5} {\bibfield  {journal} {\bibinfo
  {journal} {Nat. Phys.}\ }\textbf {\bibinfo {volume} {14}},\ \bibinfo {pages}
  {786} (\bibinfo {year} {2018})}\BibitemShut {NoStop}%
\bibitem [{\citenamefont {Ponomaryov}\ \emph {et~al.}(2020)\citenamefont
  {Ponomaryov}, \citenamefont {Zviagina}, \citenamefont {Wosnitza},
  \citenamefont {Lampen-Kelley}, \citenamefont {Banerjee}, \citenamefont {Yan},
  \citenamefont {Bridges}, \citenamefont {Mandrus}, \citenamefont {Nagler},\
  and\ \citenamefont {Zvyagin}}]{PhysRevLett.125.037202}%
  \BibitemOpen
  \bibfield  {author} {\bibinfo {author} {\bibfnamefont {A.~N.}\ \bibnamefont
  {Ponomaryov}}, \bibinfo {author} {\bibfnamefont {L.}~\bibnamefont
  {Zviagina}}, \bibinfo {author} {\bibfnamefont {J.}~\bibnamefont {Wosnitza}},
  \bibinfo {author} {\bibfnamefont {P.}~\bibnamefont {Lampen-Kelley}}, \bibinfo
  {author} {\bibfnamefont {A.}~\bibnamefont {Banerjee}}, \bibinfo {author}
  {\bibfnamefont {J.-Q.}\ \bibnamefont {Yan}}, \bibinfo {author} {\bibfnamefont
  {C.~A.}\ \bibnamefont {Bridges}}, \bibinfo {author} {\bibfnamefont {D.~G.}\
  \bibnamefont {Mandrus}}, \bibinfo {author} {\bibfnamefont {S.~E.}\
  \bibnamefont {Nagler}},\ and\ \bibinfo {author} {\bibfnamefont {S.~A.}\
  \bibnamefont {Zvyagin}},\ }\href
  {https://doi.org/10.1103/PhysRevLett.125.037202} {\bibfield  {journal}
  {\bibinfo  {journal} {Phys. Rev. Lett.}\ }\textbf {\bibinfo {volume} {125}},\
  \bibinfo {pages} {037202} (\bibinfo {year} {2020})}\BibitemShut {NoStop}%
\bibitem [{\citenamefont {Wellm}\ \emph {et~al.}(2018)\citenamefont {Wellm},
  \citenamefont {Zeisner}, \citenamefont {Alfonsov}, \citenamefont {Wolter},
  \citenamefont {Roslova}, \citenamefont {Isaeva}, \citenamefont {Doert},
  \citenamefont {Vojta}, \citenamefont {B\"uchner},\ and\ \citenamefont
  {Kataev}}]{PhysRevB.98.184408}%
  \BibitemOpen
  \bibfield  {author} {\bibinfo {author} {\bibfnamefont {C.}~\bibnamefont
  {Wellm}}, \bibinfo {author} {\bibfnamefont {J.}~\bibnamefont {Zeisner}},
  \bibinfo {author} {\bibfnamefont {A.}~\bibnamefont {Alfonsov}}, \bibinfo
  {author} {\bibfnamefont {A.~U.~B.}\ \bibnamefont {Wolter}}, \bibinfo {author}
  {\bibfnamefont {M.}~\bibnamefont {Roslova}}, \bibinfo {author} {\bibfnamefont
  {A.}~\bibnamefont {Isaeva}}, \bibinfo {author} {\bibfnamefont
  {T.}~\bibnamefont {Doert}}, \bibinfo {author} {\bibfnamefont
  {M.}~\bibnamefont {Vojta}}, \bibinfo {author} {\bibfnamefont
  {B.}~\bibnamefont {B\"uchner}},\ and\ \bibinfo {author} {\bibfnamefont
  {V.}~\bibnamefont {Kataev}},\ }\href
  {https://doi.org/10.1103/PhysRevB.98.184408} {\bibfield  {journal} {\bibinfo
  {journal} {Phys. Rev. B}\ }\textbf {\bibinfo {volume} {98}},\ \bibinfo
  {pages} {184408} (\bibinfo {year} {2018})}\BibitemShut {NoStop}%
\bibitem [{\citenamefont {Hirobe}\ \emph {et~al.}(2017)\citenamefont {Hirobe},
  \citenamefont {Sato}, \citenamefont {Shiomi}, \citenamefont {Tanaka},\ and\
  \citenamefont {Saitoh}}]{PhysRevB.95.241112}%
  \BibitemOpen
  \bibfield  {author} {\bibinfo {author} {\bibfnamefont {D.}~\bibnamefont
  {Hirobe}}, \bibinfo {author} {\bibfnamefont {M.}~\bibnamefont {Sato}},
  \bibinfo {author} {\bibfnamefont {Y.}~\bibnamefont {Shiomi}}, \bibinfo
  {author} {\bibfnamefont {H.}~\bibnamefont {Tanaka}},\ and\ \bibinfo {author}
  {\bibfnamefont {E.}~\bibnamefont {Saitoh}},\ }\href
  {https://doi.org/10.1103/PhysRevB.95.241112} {\bibfield  {journal} {\bibinfo
  {journal} {Phys. Rev. B}\ }\textbf {\bibinfo {volume} {95}},\ \bibinfo
  {pages} {241112} (\bibinfo {year} {2017})}\BibitemShut {NoStop}%
\bibitem [{\citenamefont {Kasahara}\ \emph
  {et~al.}(2018{\natexlab{b}})\citenamefont {Kasahara}, \citenamefont {Sugii},
  \citenamefont {Ohnishi}, \citenamefont {Shimozawa}, \citenamefont
  {Yamashita}, \citenamefont {Kurita}, \citenamefont {Tanaka}, \citenamefont
  {Nasu}, \citenamefont {Motome}, \citenamefont {Shibauchi},\ and\
  \citenamefont {Matsuda}}]{PhysRevLett.120.217205}%
  \BibitemOpen
  \bibfield  {author} {\bibinfo {author} {\bibfnamefont {Y.}~\bibnamefont
  {Kasahara}}, \bibinfo {author} {\bibfnamefont {K.}~\bibnamefont {Sugii}},
  \bibinfo {author} {\bibfnamefont {T.}~\bibnamefont {Ohnishi}}, \bibinfo
  {author} {\bibfnamefont {M.}~\bibnamefont {Shimozawa}}, \bibinfo {author}
  {\bibfnamefont {M.}~\bibnamefont {Yamashita}}, \bibinfo {author}
  {\bibfnamefont {N.}~\bibnamefont {Kurita}}, \bibinfo {author} {\bibfnamefont
  {H.}~\bibnamefont {Tanaka}}, \bibinfo {author} {\bibfnamefont
  {J.}~\bibnamefont {Nasu}}, \bibinfo {author} {\bibfnamefont {Y.}~\bibnamefont
  {Motome}}, \bibinfo {author} {\bibfnamefont {T.}~\bibnamefont {Shibauchi}},\
  and\ \bibinfo {author} {\bibfnamefont {Y.}~\bibnamefont {Matsuda}},\ }\href
  {https://doi.org/10.1103/PhysRevLett.120.217205} {\bibfield  {journal}
  {\bibinfo  {journal} {Phys. Rev. Lett.}\ }\textbf {\bibinfo {volume} {120}},\
  \bibinfo {pages} {217205} (\bibinfo {year} {2018}{\natexlab{b}})}\BibitemShut
  {NoStop}%
\bibitem [{\citenamefont {Little}\ \emph {et~al.}(2017)\citenamefont {Little},
  \citenamefont {Wu}, \citenamefont {Lampen-Kelley}, \citenamefont {Banerjee},
  \citenamefont {Patankar}, \citenamefont {Rees}, \citenamefont {Bridges},
  \citenamefont {Yan}, \citenamefont {Mandrus}, \citenamefont {Nagler},\ and\
  \citenamefont {Orenstein}}]{PhysRevLett.119.227201}%
  \BibitemOpen
  \bibfield  {author} {\bibinfo {author} {\bibfnamefont {A.}~\bibnamefont
  {Little}}, \bibinfo {author} {\bibfnamefont {L.}~\bibnamefont {Wu}}, \bibinfo
  {author} {\bibfnamefont {P.}~\bibnamefont {Lampen-Kelley}}, \bibinfo {author}
  {\bibfnamefont {A.}~\bibnamefont {Banerjee}}, \bibinfo {author}
  {\bibfnamefont {S.}~\bibnamefont {Patankar}}, \bibinfo {author}
  {\bibfnamefont {D.}~\bibnamefont {Rees}}, \bibinfo {author} {\bibfnamefont
  {C.~A.}\ \bibnamefont {Bridges}}, \bibinfo {author} {\bibfnamefont {J.-Q.}\
  \bibnamefont {Yan}}, \bibinfo {author} {\bibfnamefont {D.}~\bibnamefont
  {Mandrus}}, \bibinfo {author} {\bibfnamefont {S.~E.}\ \bibnamefont
  {Nagler}},\ and\ \bibinfo {author} {\bibfnamefont {J.}~\bibnamefont
  {Orenstein}},\ }\href {https://doi.org/10.1103/PhysRevLett.119.227201}
  {\bibfield  {journal} {\bibinfo  {journal} {Phys. Rev. Lett.}\ }\textbf
  {\bibinfo {volume} {119}},\ \bibinfo {pages} {227201} (\bibinfo {year}
  {2017})}\BibitemShut {NoStop}%
\bibitem [{\citenamefont {Wang}\ \emph
  {et~al.}(2017{\natexlab{a}})\citenamefont {Wang}, \citenamefont {Reschke},
  \citenamefont {H\"uvonen}, \citenamefont {Do}, \citenamefont {Choi},
  \citenamefont {Gensch}, \citenamefont {Nagel}, \citenamefont {R{\~o}{\~o}m},\
  and\ \citenamefont {Loidl}}]{PhysRevLett.119.227202}%
  \BibitemOpen
  \bibfield  {author} {\bibinfo {author} {\bibfnamefont {Z.}~\bibnamefont
  {Wang}}, \bibinfo {author} {\bibfnamefont {S.}~\bibnamefont {Reschke}},
  \bibinfo {author} {\bibfnamefont {D.}~\bibnamefont {H\"uvonen}}, \bibinfo
  {author} {\bibfnamefont {S.-H.}\ \bibnamefont {Do}}, \bibinfo {author}
  {\bibfnamefont {K.-Y.}\ \bibnamefont {Choi}}, \bibinfo {author}
  {\bibfnamefont {M.}~\bibnamefont {Gensch}}, \bibinfo {author} {\bibfnamefont
  {U.}~\bibnamefont {Nagel}}, \bibinfo {author} {\bibfnamefont
  {T.}~\bibnamefont {R{\~o}{\~o}m}},\ and\ \bibinfo {author} {\bibfnamefont
  {A.}~\bibnamefont {Loidl}},\ }\href
  {https://doi.org/10.1103/PhysRevLett.119.227202} {\bibfield  {journal}
  {\bibinfo  {journal} {Phys. Rev. Lett.}\ }\textbf {\bibinfo {volume} {119}},\
  \bibinfo {pages} {227202} (\bibinfo {year} {2017}{\natexlab{a}})}\BibitemShut
  {NoStop}%
\bibitem [{\citenamefont {Wu}\ \emph {et~al.}(2018)\citenamefont {Wu},
  \citenamefont {Little}, \citenamefont {Aldape}, \citenamefont {Rees},
  \citenamefont {Thewalt}, \citenamefont {Lampen-Kelley}, \citenamefont
  {Banerjee}, \citenamefont {Bridges}, \citenamefont {Yan}, \citenamefont
  {Boone}, \citenamefont {Patankar}, \citenamefont {Goldhaber-Gordon},
  \citenamefont {Mandrus}, \citenamefont {Nagler}, \citenamefont {Altman},\
  and\ \citenamefont {Orenstein}}]{PhysRevB.98.094425}%
  \BibitemOpen
  \bibfield  {author} {\bibinfo {author} {\bibfnamefont {L.}~\bibnamefont
  {Wu}}, \bibinfo {author} {\bibfnamefont {A.}~\bibnamefont {Little}}, \bibinfo
  {author} {\bibfnamefont {E.~E.}\ \bibnamefont {Aldape}}, \bibinfo {author}
  {\bibfnamefont {D.}~\bibnamefont {Rees}}, \bibinfo {author} {\bibfnamefont
  {E.}~\bibnamefont {Thewalt}}, \bibinfo {author} {\bibfnamefont
  {P.}~\bibnamefont {Lampen-Kelley}}, \bibinfo {author} {\bibfnamefont
  {A.}~\bibnamefont {Banerjee}}, \bibinfo {author} {\bibfnamefont {C.~A.}\
  \bibnamefont {Bridges}}, \bibinfo {author} {\bibfnamefont {J.-Q.}\
  \bibnamefont {Yan}}, \bibinfo {author} {\bibfnamefont {D.}~\bibnamefont
  {Boone}}, \bibinfo {author} {\bibfnamefont {S.}~\bibnamefont {Patankar}},
  \bibinfo {author} {\bibfnamefont {D.}~\bibnamefont {Goldhaber-Gordon}},
  \bibinfo {author} {\bibfnamefont {D.}~\bibnamefont {Mandrus}}, \bibinfo
  {author} {\bibfnamefont {S.~E.}\ \bibnamefont {Nagler}}, \bibinfo {author}
  {\bibfnamefont {E.}~\bibnamefont {Altman}},\ and\ \bibinfo {author}
  {\bibfnamefont {J.}~\bibnamefont {Orenstein}},\ }\href
  {https://doi.org/10.1103/PhysRevB.98.094425} {\bibfield  {journal} {\bibinfo
  {journal} {Phys. Rev. B}\ }\textbf {\bibinfo {volume} {98}},\ \bibinfo
  {pages} {094425} (\bibinfo {year} {2018})}\BibitemShut {NoStop}%
\bibitem [{\citenamefont {Ozel}\ \emph {et~al.}(2019)\citenamefont {Ozel},
  \citenamefont {Belvin}, \citenamefont {Baldini}, \citenamefont {Kimchi},
  \citenamefont {Do}, \citenamefont {Choi},\ and\ \citenamefont
  {Gedik}}]{PhysRevB.100.085108}%
  \BibitemOpen
  \bibfield  {author} {\bibinfo {author} {\bibfnamefont {I.~O.}\ \bibnamefont
  {Ozel}}, \bibinfo {author} {\bibfnamefont {C.~A.}\ \bibnamefont {Belvin}},
  \bibinfo {author} {\bibfnamefont {E.}~\bibnamefont {Baldini}}, \bibinfo
  {author} {\bibfnamefont {I.}~\bibnamefont {Kimchi}}, \bibinfo {author}
  {\bibfnamefont {S.}~\bibnamefont {Do}}, \bibinfo {author} {\bibfnamefont
  {K.-Y.}\ \bibnamefont {Choi}},\ and\ \bibinfo {author} {\bibfnamefont
  {N.}~\bibnamefont {Gedik}},\ }\href
  {https://doi.org/10.1103/PhysRevB.100.085108} {\bibfield  {journal} {\bibinfo
   {journal} {Phys. Rev. B}\ }\textbf {\bibinfo {volume} {100}},\ \bibinfo
  {pages} {085108} (\bibinfo {year} {2019})}\BibitemShut {NoStop}%
\bibitem [{\citenamefont {Reschke}\ \emph {et~al.}(2019)\citenamefont
  {Reschke}, \citenamefont {Tsurkan}, \citenamefont {Do}, \citenamefont {Choi},
  \citenamefont {Lunkenheimer}, \citenamefont {Wang},\ and\ \citenamefont
  {Loidl}}]{Reschke2019}%
  \BibitemOpen
  \bibfield  {author} {\bibinfo {author} {\bibfnamefont {S.}~\bibnamefont
  {Reschke}}, \bibinfo {author} {\bibfnamefont {V.}~\bibnamefont {Tsurkan}},
  \bibinfo {author} {\bibfnamefont {S.-H.}\ \bibnamefont {Do}}, \bibinfo
  {author} {\bibfnamefont {K.-Y.}\ \bibnamefont {Choi}}, \bibinfo {author}
  {\bibfnamefont {P.}~\bibnamefont {Lunkenheimer}}, \bibinfo {author}
  {\bibfnamefont {Z.}~\bibnamefont {Wang}},\ and\ \bibinfo {author}
  {\bibfnamefont {A.}~\bibnamefont {Loidl}},\ }\href
  {https://doi.org/10.1103/PhysRevB.100.100403} {\bibfield  {journal} {\bibinfo
   {journal} {Phys. Rev. B}\ }\textbf {\bibinfo {volume} {100}},\ \bibinfo
  {pages} {100403} (\bibinfo {year} {2019})}\BibitemShut {NoStop}%
\bibitem [{\citenamefont {Cookmeyer}\ and\ \citenamefont
  {Moore}(2018)}]{PhysRevB.98.060412}%
  \BibitemOpen
  \bibfield  {author} {\bibinfo {author} {\bibfnamefont {J.}~\bibnamefont
  {Cookmeyer}}\ and\ \bibinfo {author} {\bibfnamefont {J.~E.}\ \bibnamefont
  {Moore}},\ }\href {https://doi.org/10.1103/PhysRevB.98.060412} {\bibfield
  {journal} {\bibinfo  {journal} {Phys. Rev. B}\ }\textbf {\bibinfo {volume}
  {98}},\ \bibinfo {pages} {060412} (\bibinfo {year} {2018})}\BibitemShut
  {NoStop}%
\bibitem [{\citenamefont {Kim}\ \emph {et~al.}(2015)\citenamefont {Kim},
  \citenamefont {V.}, \citenamefont {Catuneanu},\ and\ \citenamefont
  {Kee}}]{PhysRevB.91.241110}%
  \BibitemOpen
  \bibfield  {author} {\bibinfo {author} {\bibfnamefont {H.-S.}\ \bibnamefont
  {Kim}}, \bibinfo {author} {\bibfnamefont {V.~S.}\ \bibnamefont {V.}},
  \bibinfo {author} {\bibfnamefont {A.}~\bibnamefont {Catuneanu}},\ and\
  \bibinfo {author} {\bibfnamefont {H.-Y.}\ \bibnamefont {Kee}},\ }\href
  {https://doi.org/10.1103/PhysRevB.91.241110} {\bibfield  {journal} {\bibinfo
  {journal} {Phys. Rev. B}\ }\textbf {\bibinfo {volume} {91}},\ \bibinfo
  {pages} {241110} (\bibinfo {year} {2015})}\BibitemShut {NoStop}%
\bibitem [{\citenamefont {Kim}\ and\ \citenamefont
  {Kee}(2016)}]{PhysRevB.93.155143}%
  \BibitemOpen
  \bibfield  {author} {\bibinfo {author} {\bibfnamefont {H.-S.}\ \bibnamefont
  {Kim}}\ and\ \bibinfo {author} {\bibfnamefont {H.-Y.}\ \bibnamefont {Kee}},\
  }\href {https://doi.org/10.1103/PhysRevB.93.155143} {\bibfield  {journal}
  {\bibinfo  {journal} {Phys. Rev. B}\ }\textbf {\bibinfo {volume} {93}},\
  \bibinfo {pages} {155143} (\bibinfo {year} {2016})}\BibitemShut {NoStop}%
\bibitem [{\citenamefont {Yadav}\ \emph {et~al.}(2016)\citenamefont {Yadav},
  \citenamefont {Bogdanov}, \citenamefont {Katukuri}, \citenamefont
  {Nishimoto}, \citenamefont {van~den Brink},\ and\ \citenamefont
  {Hozoi}}]{Yadav2016}%
  \BibitemOpen
  \bibfield  {author} {\bibinfo {author} {\bibfnamefont {R.}~\bibnamefont
  {Yadav}}, \bibinfo {author} {\bibfnamefont {N.~A.}\ \bibnamefont {Bogdanov}},
  \bibinfo {author} {\bibfnamefont {V.~M.}\ \bibnamefont {Katukuri}}, \bibinfo
  {author} {\bibfnamefont {S.}~\bibnamefont {Nishimoto}}, \bibinfo {author}
  {\bibfnamefont {J.}~\bibnamefont {van~den Brink}},\ and\ \bibinfo {author}
  {\bibfnamefont {L.}~\bibnamefont {Hozoi}},\ }\href
  {https://doi.org/10.1038/srep37925} {\bibfield  {journal} {\bibinfo
  {journal} {Sci. Rep.}\ }\textbf {\bibinfo {volume} {6}},\ \bibinfo {pages}
  {37925} (\bibinfo {year} {2016})}\BibitemShut {NoStop}%
\bibitem [{\citenamefont {Suzuki}\ and\ \citenamefont
  {Suga}(2018)}]{PhysRevB.97.134424}%
  \BibitemOpen
  \bibfield  {author} {\bibinfo {author} {\bibfnamefont {T.}~\bibnamefont
  {Suzuki}}\ and\ \bibinfo {author} {\bibfnamefont {S.-i.}\ \bibnamefont
  {Suga}},\ }\href {https://doi.org/10.1103/PhysRevB.97.134424} {\bibfield
  {journal} {\bibinfo  {journal} {Phys. Rev. B}\ }\textbf {\bibinfo {volume}
  {97}},\ \bibinfo {pages} {134424} (\bibinfo {year} {2018})}\BibitemShut
  {NoStop}%
\bibitem [{\citenamefont {Suzuki}\ and\ \citenamefont
  {Suga}(2019)}]{Suzuki2019}%
  \BibitemOpen
  \bibfield  {author} {\bibinfo {author} {\bibfnamefont {T.}~\bibnamefont
  {Suzuki}}\ and\ \bibinfo {author} {\bibfnamefont {S.-i.}\ \bibnamefont
  {Suga}},\ }\href {https://doi.org/10.1103/PhysRevB.99.249902} {\bibfield
  {journal} {\bibinfo  {journal} {Phys. Rev. B}\ }\textbf {\bibinfo {volume}
  {99}},\ \bibinfo {pages} {249902} (\bibinfo {year} {2019})}\BibitemShut
  {NoStop}%
\bibitem [{\citenamefont {Hou}\ \emph {et~al.}(2017)\citenamefont {Hou},
  \citenamefont {Xiang},\ and\ \citenamefont {Gong}}]{PhysRevB.96.054410}%
  \BibitemOpen
  \bibfield  {author} {\bibinfo {author} {\bibfnamefont {Y.~S.}\ \bibnamefont
  {Hou}}, \bibinfo {author} {\bibfnamefont {H.~J.}\ \bibnamefont {Xiang}},\
  and\ \bibinfo {author} {\bibfnamefont {X.~G.}\ \bibnamefont {Gong}},\ }\href
  {https://doi.org/10.1103/PhysRevB.96.054410} {\bibfield  {journal} {\bibinfo
  {journal} {Phys. Rev. B}\ }\textbf {\bibinfo {volume} {96}},\ \bibinfo
  {pages} {054410} (\bibinfo {year} {2017})}\BibitemShut {NoStop}%
\bibitem [{\citenamefont {Wang}\ \emph
  {et~al.}(2017{\natexlab{b}})\citenamefont {Wang}, \citenamefont {Dong},
  \citenamefont {Yu},\ and\ \citenamefont {Li}}]{PhysRevB.96.115103}%
  \BibitemOpen
  \bibfield  {author} {\bibinfo {author} {\bibfnamefont {W.}~\bibnamefont
  {Wang}}, \bibinfo {author} {\bibfnamefont {Z.-Y.}\ \bibnamefont {Dong}},
  \bibinfo {author} {\bibfnamefont {S.-L.}\ \bibnamefont {Yu}},\ and\ \bibinfo
  {author} {\bibfnamefont {J.-X.}\ \bibnamefont {Li}},\ }\href
  {https://doi.org/10.1103/PhysRevB.96.115103} {\bibfield  {journal} {\bibinfo
  {journal} {Phys. Rev. B}\ }\textbf {\bibinfo {volume} {96}},\ \bibinfo
  {pages} {115103} (\bibinfo {year} {2017}{\natexlab{b}})}\BibitemShut
  {NoStop}%
\bibitem [{\citenamefont {Li}\ \emph {et~al.}(2021{\natexlab{b}})\citenamefont
  {Li}, \citenamefont {Zhang}, \citenamefont {Wang}, \citenamefont {Wu},
  \citenamefont {Gao}, \citenamefont {Qu}, \citenamefont {Liu}, \citenamefont
  {Gong},\ and\ \citenamefont {Li}}]{Li2021}%
  \BibitemOpen
  \bibfield  {author} {\bibinfo {author} {\bibfnamefont {H.}~\bibnamefont
  {Li}}, \bibinfo {author} {\bibfnamefont {H.-K.}\ \bibnamefont {Zhang}},
  \bibinfo {author} {\bibfnamefont {J.}~\bibnamefont {Wang}}, \bibinfo {author}
  {\bibfnamefont {H.-Q.}\ \bibnamefont {Wu}}, \bibinfo {author} {\bibfnamefont
  {Y.}~\bibnamefont {Gao}}, \bibinfo {author} {\bibfnamefont {D.-W.}\
  \bibnamefont {Qu}}, \bibinfo {author} {\bibfnamefont {Z.-X.}\ \bibnamefont
  {Liu}}, \bibinfo {author} {\bibfnamefont {S.-S.}\ \bibnamefont {Gong}},\ and\
  \bibinfo {author} {\bibfnamefont {W.}~\bibnamefont {Li}},\ }\href
  {https://doi.org/https://doi.org/10.1038/s41467-021-24257-8} {\bibfield
  {journal} {\bibinfo  {journal} {Nat. Commun.}\ }\textbf {\bibinfo {volume}
  {12}},\ \bibinfo {pages} {4007} (\bibinfo {year}
  {2021}{\natexlab{b}})}\BibitemShut {NoStop}%
\bibitem [{\citenamefont {Ran}\ \emph {et~al.}(2022)\citenamefont {Ran},
  \citenamefont {Wang}, \citenamefont {Bao}, \citenamefont {Cai}, \citenamefont
  {Shangguan}, \citenamefont {Ma}, \citenamefont {Wang}, \citenamefont {Dong},
  \citenamefont {Čermák}, \citenamefont {Schneidewind}, \citenamefont {Meng},
  \citenamefont {Lu}, \citenamefont {Yu}, \citenamefont {Li},\ and\
  \citenamefont {Wen}}]{KejingRan:27501}%
  \BibitemOpen
  \bibfield  {author} {\bibinfo {author} {\bibfnamefont {K.}~\bibnamefont
  {Ran}}, \bibinfo {author} {\bibfnamefont {J.}~\bibnamefont {Wang}}, \bibinfo
  {author} {\bibfnamefont {S.}~\bibnamefont {Bao}}, \bibinfo {author}
  {\bibfnamefont {Z.}~\bibnamefont {Cai}}, \bibinfo {author} {\bibfnamefont
  {Y.}~\bibnamefont {Shangguan}}, \bibinfo {author} {\bibfnamefont
  {Z.}~\bibnamefont {Ma}}, \bibinfo {author} {\bibfnamefont {W.}~\bibnamefont
  {Wang}}, \bibinfo {author} {\bibfnamefont {Z.-Y.}\ \bibnamefont {Dong}},
  \bibinfo {author} {\bibfnamefont {P.}~\bibnamefont {Čermák}}, \bibinfo
  {author} {\bibfnamefont {A.}~\bibnamefont {Schneidewind}}, \bibinfo {author}
  {\bibfnamefont {S.}~\bibnamefont {Meng}}, \bibinfo {author} {\bibfnamefont
  {Z.}~\bibnamefont {Lu}}, \bibinfo {author} {\bibfnamefont {S.-L.}\
  \bibnamefont {Yu}}, \bibinfo {author} {\bibfnamefont {J.-X.}\ \bibnamefont
  {Li}},\ and\ \bibinfo {author} {\bibfnamefont {J.}~\bibnamefont {Wen}},\
  }\href {https://doi.org/10.1088/0256-307X/39/2/027501} {\bibfield  {journal}
  {\bibinfo  {journal} {Chin. Phys. Lett.}\ }\textbf {\bibinfo {volume} {39}},\
  \bibinfo {eid} {027501} (\bibinfo {year} {2022})}\BibitemShut {NoStop}%
\bibitem [{\citenamefont {Janssen}\ \emph {et~al.}(2017)\citenamefont
  {Janssen}, \citenamefont {Andrade},\ and\ \citenamefont
  {Vojta}}]{PhysRevB.96.064430}%
  \BibitemOpen
  \bibfield  {author} {\bibinfo {author} {\bibfnamefont {L.}~\bibnamefont
  {Janssen}}, \bibinfo {author} {\bibfnamefont {E.~C.}\ \bibnamefont
  {Andrade}},\ and\ \bibinfo {author} {\bibfnamefont {M.}~\bibnamefont
  {Vojta}},\ }\href {https://doi.org/10.1103/PhysRevB.96.064430} {\bibfield
  {journal} {\bibinfo  {journal} {Phys. Rev. B}\ }\textbf {\bibinfo {volume}
  {96}},\ \bibinfo {pages} {064430} (\bibinfo {year} {2017})}\BibitemShut
  {NoStop}%
\bibitem [{\citenamefont {Laurell}\ and\ \citenamefont
  {Okamoto}(2020)}]{Laurell2020}%
  \BibitemOpen
  \bibfield  {author} {\bibinfo {author} {\bibfnamefont {P.}~\bibnamefont
  {Laurell}}\ and\ \bibinfo {author} {\bibfnamefont {S.}~\bibnamefont
  {Okamoto}},\ }\href {https://doi.org/10.1038/s41535-019-0203-y} {\bibfield
  {journal} {\bibinfo  {journal} {npj Quantum Mater.}\ }\textbf {\bibinfo
  {volume} {5}},\ \bibinfo {pages} {2} (\bibinfo {year} {2020})}\BibitemShut
  {NoStop}%
\bibitem [{\citenamefont {Mehta}\ \emph {et~al.}(2019)\citenamefont {Mehta},
  \citenamefont {Bukov}, \citenamefont {Wang}, \citenamefont {Day},
  \citenamefont {Richardson}, \citenamefont {Fisher},\ and\ \citenamefont
  {Schwab}}]{Mehta2019}%
  \BibitemOpen
  \bibfield  {author} {\bibinfo {author} {\bibfnamefont {P.}~\bibnamefont
  {Mehta}}, \bibinfo {author} {\bibfnamefont {M.}~\bibnamefont {Bukov}},
  \bibinfo {author} {\bibfnamefont {C.-H.}\ \bibnamefont {Wang}}, \bibinfo
  {author} {\bibfnamefont {A.~G.}\ \bibnamefont {Day}}, \bibinfo {author}
  {\bibfnamefont {C.}~\bibnamefont {Richardson}}, \bibinfo {author}
  {\bibfnamefont {C.~K.}\ \bibnamefont {Fisher}},\ and\ \bibinfo {author}
  {\bibfnamefont {D.~J.}\ \bibnamefont {Schwab}},\ }\href
  {https://doi.org/https://doi.org/10.1016/j.physrep.2019.03.001} {\bibfield
  {journal} {\bibinfo  {journal} {Phys. Rep.}\ }\textbf {\bibinfo {volume}
  {810}},\ \bibinfo {pages} {1 } (\bibinfo {year} {2019})}\BibitemShut
  {NoStop}%
\bibitem [{\citenamefont
  {Carrasquilla}(2020)}]{doi:10.1080/23746149.2020.1797528}%
  \BibitemOpen
  \bibfield  {author} {\bibinfo {author} {\bibfnamefont {J.}~\bibnamefont
  {Carrasquilla}},\ }\href {https://doi.org/10.1080/23746149.2020.1797528}
  {\bibfield  {journal} {\bibinfo  {journal} {Adv. Phys. X}\ }\textbf {\bibinfo
  {volume} {5}},\ \bibinfo {pages} {1797528} (\bibinfo {year}
  {2020})}\BibitemShut {NoStop}%
\bibitem [{\citenamefont {Doucet}\ \emph {et~al.}(2021)\citenamefont {Doucet},
  \citenamefont {Samarakoon}, \citenamefont {Do}, \citenamefont {Heller},
  \citenamefont {Archibald}, \citenamefont {Tennant}, \citenamefont {Proffen},\
  and\ \citenamefont {Granroth}}]{Doucet_2021}%
  \BibitemOpen
  \bibfield  {author} {\bibinfo {author} {\bibfnamefont {M.}~\bibnamefont
  {Doucet}}, \bibinfo {author} {\bibfnamefont {A.~M.}\ \bibnamefont
  {Samarakoon}}, \bibinfo {author} {\bibfnamefont {C.}~\bibnamefont {Do}},
  \bibinfo {author} {\bibfnamefont {W.~T.}\ \bibnamefont {Heller}}, \bibinfo
  {author} {\bibfnamefont {R.}~\bibnamefont {Archibald}}, \bibinfo {author}
  {\bibfnamefont {D.~A.}\ \bibnamefont {Tennant}}, \bibinfo {author}
  {\bibfnamefont {T.}~\bibnamefont {Proffen}},\ and\ \bibinfo {author}
  {\bibfnamefont {G.~E.}\ \bibnamefont {Granroth}},\ }\href
  {https://doi.org/10.1088/2632-2153/abcf88} {\bibfield  {journal} {\bibinfo
  {journal} {Mach. Learn. Sci. Technol.}\ }\textbf {\bibinfo {volume} {2}},\
  \bibinfo {pages} {023001} (\bibinfo {year} {2021})}\BibitemShut {NoStop}%
\bibitem [{\citenamefont {Butler}\ \emph {et~al.}(2021)\citenamefont {Butler},
  \citenamefont {Le}, \citenamefont {Thiyagalingam},\ and\ \citenamefont
  {Perring}}]{Butler_2021}%
  \BibitemOpen
  \bibfield  {author} {\bibinfo {author} {\bibfnamefont {K.~T.}\ \bibnamefont
  {Butler}}, \bibinfo {author} {\bibfnamefont {M.~D.}\ \bibnamefont {Le}},
  \bibinfo {author} {\bibfnamefont {J.}~\bibnamefont {Thiyagalingam}},\ and\
  \bibinfo {author} {\bibfnamefont {T.~G.}\ \bibnamefont {Perring}},\ }\href
  {https://doi.org/10.1088/1361-648x/abea1c} {\bibfield  {journal} {\bibinfo
  {journal} {J. Phys. Condens. Matter}\ }\textbf {\bibinfo {volume} {33}},\
  \bibinfo {pages} {194006} (\bibinfo {year} {2021})}\BibitemShut {NoStop}%
\bibitem [{\citenamefont {Chen}\ \emph {et~al.}(2021)\citenamefont {Chen},
  \citenamefont {Andrejevic}, \citenamefont {Drucker}, \citenamefont {Nguyen},
  \citenamefont {Xian}, \citenamefont {Smidt}, \citenamefont {Wang},
  \citenamefont {Ernstorfer}, \citenamefont {Tennant}, \citenamefont {Chan},\
  and\ \citenamefont {Li}}]{doi:10.1063/5.0049111}%
  \BibitemOpen
  \bibfield  {author} {\bibinfo {author} {\bibfnamefont {Z.}~\bibnamefont
  {Chen}}, \bibinfo {author} {\bibfnamefont {N.}~\bibnamefont {Andrejevic}},
  \bibinfo {author} {\bibfnamefont {N.~C.}\ \bibnamefont {Drucker}}, \bibinfo
  {author} {\bibfnamefont {T.}~\bibnamefont {Nguyen}}, \bibinfo {author}
  {\bibfnamefont {R.~P.}\ \bibnamefont {Xian}}, \bibinfo {author}
  {\bibfnamefont {T.}~\bibnamefont {Smidt}}, \bibinfo {author} {\bibfnamefont
  {Y.}~\bibnamefont {Wang}}, \bibinfo {author} {\bibfnamefont {R.}~\bibnamefont
  {Ernstorfer}}, \bibinfo {author} {\bibfnamefont {D.~A.}\ \bibnamefont
  {Tennant}}, \bibinfo {author} {\bibfnamefont {M.}~\bibnamefont {Chan}},\ and\
  \bibinfo {author} {\bibfnamefont {M.}~\bibnamefont {Li}},\ }\href
  {https://doi.org/10.1063/5.0049111} {\bibfield  {journal} {\bibinfo
  {journal} {Chem. Phys. Rev.}\ }\textbf {\bibinfo {volume} {2}},\ \bibinfo
  {pages} {031301} (\bibinfo {year} {2021})}\BibitemShut {NoStop}%
\bibitem [{\citenamefont {Yu}\ \emph {et~al.}(2021)\citenamefont {Yu},
  \citenamefont {Gao}, \citenamefont {Chen},\ and\ \citenamefont
  {Li}}]{Yu2021}%
  \BibitemOpen
  \bibfield  {author} {\bibinfo {author} {\bibfnamefont {S.}~\bibnamefont
  {Yu}}, \bibinfo {author} {\bibfnamefont {Y.}~\bibnamefont {Gao}}, \bibinfo
  {author} {\bibfnamefont {B.-B.}\ \bibnamefont {Chen}},\ and\ \bibinfo
  {author} {\bibfnamefont {W.}~\bibnamefont {Li}},\ }\href
  {https://doi.org/https://doi.org/10.1088/0256-307x/38/9/097502} {\bibfield
  {journal} {\bibinfo  {journal} {Chin. Phys. Lett.}\ }\textbf {\bibinfo
  {volume} {38}},\ \bibinfo {pages} {097502} (\bibinfo {year}
  {2021})}\BibitemShut {NoStop}%
\bibitem [{\citenamefont {Tennant}\ and\ \citenamefont
  {Samarakoon}(2021)}]{Samarakoon2021}%
  \BibitemOpen
  \bibfield  {author} {\bibinfo {author} {\bibfnamefont {A.}~\bibnamefont
  {Tennant}}\ and\ \bibinfo {author} {\bibfnamefont {A.}~\bibnamefont
  {Samarakoon}},\ }\href
  {http://iopscience.iop.org/article/10.1088/1361-648X/abe818} {\bibfield
  {journal} {\bibinfo  {journal} {J. Phys. Condens. Matter}\ }\textbf {\bibinfo
  {volume} {34}},\ \bibinfo {pages} {044002} (\bibinfo {year}
  {2021})}\BibitemShut {NoStop}%
\bibitem [{\citenamefont {Samarakoon}\ \emph {et~al.}(2020)\citenamefont
  {Samarakoon}, \citenamefont {Barros}, \citenamefont {Li}, \citenamefont
  {Eisenbach}, \citenamefont {Zhang}, \citenamefont {Ye}, \citenamefont
  {Sharma}, \citenamefont {Dun}, \citenamefont {Zhou}, \citenamefont {Grigera},
  \citenamefont {Batista},\ and\ \citenamefont {Tennant}}]{Samarakoon2020}%
  \BibitemOpen
  \bibfield  {author} {\bibinfo {author} {\bibfnamefont {A.~M.}\ \bibnamefont
  {Samarakoon}}, \bibinfo {author} {\bibfnamefont {K.}~\bibnamefont {Barros}},
  \bibinfo {author} {\bibfnamefont {Y.~W.}\ \bibnamefont {Li}}, \bibinfo
  {author} {\bibfnamefont {M.}~\bibnamefont {Eisenbach}}, \bibinfo {author}
  {\bibfnamefont {Q.}~\bibnamefont {Zhang}}, \bibinfo {author} {\bibfnamefont
  {F.}~\bibnamefont {Ye}}, \bibinfo {author} {\bibfnamefont {V.}~\bibnamefont
  {Sharma}}, \bibinfo {author} {\bibfnamefont {Z.~L.}\ \bibnamefont {Dun}},
  \bibinfo {author} {\bibfnamefont {H.}~\bibnamefont {Zhou}}, \bibinfo {author}
  {\bibfnamefont {S.~A.}\ \bibnamefont {Grigera}}, \bibinfo {author}
  {\bibfnamefont {C.~D.}\ \bibnamefont {Batista}},\ and\ \bibinfo {author}
  {\bibfnamefont {D.~A.}\ \bibnamefont {Tennant}},\ }\href
  {https://doi.org/10.1038/s41467-020-14660-y} {\bibfield  {journal} {\bibinfo
  {journal} {Nat. Commun.}\ }\textbf {\bibinfo {volume} {11}},\ \bibinfo
  {pages} {892} (\bibinfo {year} {2020})}\BibitemShut {NoStop}%
\bibitem [{\citenamefont {Mason}\ \emph {et~al.}(2006)\citenamefont {Mason},
  \citenamefont {Abernathy}, \citenamefont {Anderson}, \citenamefont {Ankner},
  \citenamefont {Egami}, \citenamefont {Ehlers}, \citenamefont {Ekkebus},
  \citenamefont {Granroth}, \citenamefont {Hagen}, \citenamefont {Herwig},
  \citenamefont {Hodges}, \citenamefont {Hoffmann}, \citenamefont {Horak},
  \citenamefont {Horton}, \citenamefont {Klose}, \citenamefont {Larese},
  \citenamefont {Mesecar}, \citenamefont {Myles}, \citenamefont {Neuefeind},
  \citenamefont {Ohl}, \citenamefont {Tulk}, \citenamefont {Wang},\ and\
  \citenamefont {Zhao}}]{mason2006spallation}%
  \BibitemOpen
  \bibfield  {author} {\bibinfo {author} {\bibfnamefont {T.~E.}\ \bibnamefont
  {Mason}}, \bibinfo {author} {\bibfnamefont {D.}~\bibnamefont {Abernathy}},
  \bibinfo {author} {\bibfnamefont {I.}~\bibnamefont {Anderson}}, \bibinfo
  {author} {\bibfnamefont {J.}~\bibnamefont {Ankner}}, \bibinfo {author}
  {\bibfnamefont {T.}~\bibnamefont {Egami}}, \bibinfo {author} {\bibfnamefont
  {G.}~\bibnamefont {Ehlers}}, \bibinfo {author} {\bibfnamefont
  {A.}~\bibnamefont {Ekkebus}}, \bibinfo {author} {\bibfnamefont
  {G.}~\bibnamefont {Granroth}}, \bibinfo {author} {\bibfnamefont
  {M.}~\bibnamefont {Hagen}}, \bibinfo {author} {\bibfnamefont
  {K.}~\bibnamefont {Herwig}}, \bibinfo {author} {\bibfnamefont
  {J.}~\bibnamefont {Hodges}}, \bibinfo {author} {\bibfnamefont
  {C.}~\bibnamefont {Hoffmann}}, \bibinfo {author} {\bibfnamefont
  {C.}~\bibnamefont {Horak}}, \bibinfo {author} {\bibfnamefont
  {L.}~\bibnamefont {Horton}}, \bibinfo {author} {\bibfnamefont
  {F.}~\bibnamefont {Klose}}, \bibinfo {author} {\bibfnamefont
  {J.}~\bibnamefont {Larese}}, \bibinfo {author} {\bibfnamefont
  {A.}~\bibnamefont {Mesecar}}, \bibinfo {author} {\bibfnamefont
  {D.}~\bibnamefont {Myles}}, \bibinfo {author} {\bibfnamefont
  {J.}~\bibnamefont {Neuefeind}}, \bibinfo {author} {\bibfnamefont
  {M.}~\bibnamefont {Ohl}}, \bibinfo {author} {\bibfnamefont {C.}~\bibnamefont
  {Tulk}}, \bibinfo {author} {\bibfnamefont {X.-L.}\ \bibnamefont {Wang}},\
  and\ \bibinfo {author} {\bibfnamefont {J.}~\bibnamefont {Zhao}},\ }\href
  {https://doi.org/https://doi.org/10.1016/j.physb.2006.05.281} {\bibfield
  {journal} {\bibinfo  {journal} {Physica B: Condensed Matter}\ }\textbf
  {\bibinfo {volume} {385-386}},\ \bibinfo {pages} {955} (\bibinfo {year}
  {2006})}\BibitemShut {NoStop}%
\bibitem [{\citenamefont {Rosenkranz}\ and\ \citenamefont
  {Osborn}(2008)}]{Rosenkranz2008}%
  \BibitemOpen
  \bibfield  {author} {\bibinfo {author} {\bibfnamefont {S.}~\bibnamefont
  {Rosenkranz}}\ and\ \bibinfo {author} {\bibfnamefont {R.}~\bibnamefont
  {Osborn}},\ }\href {https://doi.org/10.1007/s12043-008-0259-x} {\bibfield
  {journal} {\bibinfo  {journal} {Pramana-J. Phys.}\ }\textbf {\bibinfo
  {volume} {71}},\ \bibinfo {pages} {705} (\bibinfo {year} {2008})}\BibitemShut
  {NoStop}%
\bibitem [{\citenamefont {Granroth}\ \emph {et~al.}(2006)\citenamefont
  {Granroth}, \citenamefont {Vandergriff},\ and\ \citenamefont
  {Nagler}}]{Granroth2006}%
  \BibitemOpen
  \bibfield  {author} {\bibinfo {author} {\bibfnamefont {G.~E.}\ \bibnamefont
  {Granroth}}, \bibinfo {author} {\bibfnamefont {D.~H.}\ \bibnamefont
  {Vandergriff}},\ and\ \bibinfo {author} {\bibfnamefont {S.~E.}\ \bibnamefont
  {Nagler}},\ }\href {https://doi.org/10.1016/j.physb.2006.05.379} {\bibfield
  {journal} {\bibinfo  {journal} {Physica B: Condensed Matter}\ }\textbf
  {\bibinfo {volume} {385-86}},\ \bibinfo {pages} {1104} (\bibinfo {year}
  {2006})}\BibitemShut {NoStop}%
\bibitem [{\citenamefont {Granroth}\ \emph {et~al.}(2010)\citenamefont
  {Granroth}, \citenamefont {Kolesnikov}, \citenamefont {Sherline},
  \citenamefont {Clancy}, \citenamefont {Ross}, \citenamefont {Ruff},
  \citenamefont {Gaulin},\ and\ \citenamefont {Nagler}}]{Granroth2010}%
  \BibitemOpen
  \bibfield  {author} {\bibinfo {author} {\bibfnamefont {G.~E.}\ \bibnamefont
  {Granroth}}, \bibinfo {author} {\bibfnamefont {A.~I.}\ \bibnamefont
  {Kolesnikov}}, \bibinfo {author} {\bibfnamefont {T.~E.}\ \bibnamefont
  {Sherline}}, \bibinfo {author} {\bibfnamefont {J.~P.}\ \bibnamefont
  {Clancy}}, \bibinfo {author} {\bibfnamefont {K.~A.}\ \bibnamefont {Ross}},
  \bibinfo {author} {\bibfnamefont {J.~P.~C.}\ \bibnamefont {Ruff}}, \bibinfo
  {author} {\bibfnamefont {B.~D.}\ \bibnamefont {Gaulin}},\ and\ \bibinfo
  {author} {\bibfnamefont {S.~E.}\ \bibnamefont {Nagler}},\ }\href
  {http://stacks.iop.org/1742-6596/251/i=1/a=012058} {\bibfield  {journal}
  {\bibinfo  {journal} {J. Phys.: Conf. Ser.}\ }\textbf {\bibinfo {volume}
  {251}},\ \bibinfo {pages} {012058} (\bibinfo {year} {2010})}\BibitemShut
  {NoStop}%
\bibitem [{\citenamefont {Janssen}\ \emph {et~al.}(2020)\citenamefont
  {Janssen}, \citenamefont {Koch},\ and\ \citenamefont
  {Vojta}}]{PhysRevB.101.174444}%
  \BibitemOpen
  \bibfield  {author} {\bibinfo {author} {\bibfnamefont {L.}~\bibnamefont
  {Janssen}}, \bibinfo {author} {\bibfnamefont {S.}~\bibnamefont {Koch}},\ and\
  \bibinfo {author} {\bibfnamefont {M.}~\bibnamefont {Vojta}},\ }\href
  {https://doi.org/10.1103/PhysRevB.101.174444} {\bibfield  {journal} {\bibinfo
   {journal} {Phys. Rev. B}\ }\textbf {\bibinfo {volume} {101}},\ \bibinfo
  {pages} {174444} (\bibinfo {year} {2020})}\BibitemShut {NoStop}%
\bibitem [{\citenamefont {Liu}\ \emph {et~al.}(2021)\citenamefont {Liu},
  \citenamefont {Sadoune}, \citenamefont {Rao}, \citenamefont {Greitemann},\
  and\ \citenamefont {Pollet}}]{PhysRevResearch.3.023016}%
  \BibitemOpen
  \bibfield  {author} {\bibinfo {author} {\bibfnamefont {K.}~\bibnamefont
  {Liu}}, \bibinfo {author} {\bibfnamefont {N.}~\bibnamefont {Sadoune}},
  \bibinfo {author} {\bibfnamefont {N.}~\bibnamefont {Rao}}, \bibinfo {author}
  {\bibfnamefont {J.}~\bibnamefont {Greitemann}},\ and\ \bibinfo {author}
  {\bibfnamefont {L.}~\bibnamefont {Pollet}},\ }\href
  {https://doi.org/10.1103/PhysRevResearch.3.023016} {\bibfield  {journal}
  {\bibinfo  {journal} {Phys. Rev. Research}\ }\textbf {\bibinfo {volume}
  {3}},\ \bibinfo {pages} {023016} (\bibinfo {year} {2021})}\BibitemShut
  {NoStop}%
\bibitem [{\citenamefont {Rao}\ \emph {et~al.}(2021)\citenamefont {Rao},
  \citenamefont {Liu}, \citenamefont {Machaczek},\ and\ \citenamefont
  {Pollet}}]{PhysRevResearch.3.033223}%
  \BibitemOpen
  \bibfield  {author} {\bibinfo {author} {\bibfnamefont {N.}~\bibnamefont
  {Rao}}, \bibinfo {author} {\bibfnamefont {K.}~\bibnamefont {Liu}}, \bibinfo
  {author} {\bibfnamefont {M.}~\bibnamefont {Machaczek}},\ and\ \bibinfo
  {author} {\bibfnamefont {L.}~\bibnamefont {Pollet}},\ }\href
  {https://doi.org/10.1103/PhysRevResearch.3.033223} {\bibfield  {journal}
  {\bibinfo  {journal} {Phys. Rev. Research}\ }\textbf {\bibinfo {volume}
  {3}},\ \bibinfo {pages} {033223} (\bibinfo {year} {2021})}\BibitemShut
  {NoStop}%
\bibitem [{\citenamefont {Metropolis}\ \emph {et~al.}(1953)\citenamefont
  {Metropolis}, \citenamefont {Rosenbluth}, \citenamefont {Rosenbluth},
  \citenamefont {Teller},\ and\ \citenamefont {Teller}}]{Metropolis1953}%
  \BibitemOpen
  \bibfield  {author} {\bibinfo {author} {\bibfnamefont {N.}~\bibnamefont
  {Metropolis}}, \bibinfo {author} {\bibfnamefont {A.~W.}\ \bibnamefont
  {Rosenbluth}}, \bibinfo {author} {\bibfnamefont {M.~N.}\ \bibnamefont
  {Rosenbluth}}, \bibinfo {author} {\bibfnamefont {A.~H.}\ \bibnamefont
  {Teller}},\ and\ \bibinfo {author} {\bibfnamefont {E.}~\bibnamefont
  {Teller}},\ }\href {https://doi.org/10.1063/1.1699114} {\bibfield  {journal}
  {\bibinfo  {journal} {J. Chem. Phys.}\ }\textbf {\bibinfo {volume} {21}},\
  \bibinfo {pages} {1087} (\bibinfo {year} {1953})}\BibitemShut {NoStop}%
\bibitem [{\citenamefont {Huberman}\ \emph {et~al.}(2008)\citenamefont
  {Huberman}, \citenamefont {Tennant}, \citenamefont {Cowley}, \citenamefont
  {Coldea},\ and\ \citenamefont {Frost}}]{Huberman_2008}%
  \BibitemOpen
  \bibfield  {author} {\bibinfo {author} {\bibfnamefont {T.}~\bibnamefont
  {Huberman}}, \bibinfo {author} {\bibfnamefont {D.~A.}\ \bibnamefont
  {Tennant}}, \bibinfo {author} {\bibfnamefont {R.~A.}\ \bibnamefont {Cowley}},
  \bibinfo {author} {\bibfnamefont {R.}~\bibnamefont {Coldea}},\ and\ \bibinfo
  {author} {\bibfnamefont {C.~D.}\ \bibnamefont {Frost}},\ }\href
  {https://doi.org/10.1088/1742-5468/2008/05/p05017} {\bibfield  {journal}
  {\bibinfo  {journal} {J. Stat. Mech.: Theory Exp.}\ }\textbf {\bibinfo
  {volume} {2008}}\bibinfo  {number} { (05)},\ \bibinfo {pages}
  {P05017}}\BibitemShut {NoStop}%
\bibitem [{\citenamefont {Jones}\ \emph {et~al.}(1998)\citenamefont {Jones},
  \citenamefont {Schonlau},\ and\ \citenamefont {Welch}}]{Jones1998}%
  \BibitemOpen
\bibfield  {number} {  }\bibfield  {author} {\bibinfo {author} {\bibfnamefont
  {D.~R.}\ \bibnamefont {Jones}}, \bibinfo {author} {\bibfnamefont
  {M.}~\bibnamefont {Schonlau}},\ and\ \bibinfo {author} {\bibfnamefont
  {W.~J.}\ \bibnamefont {Welch}},\ }\href
  {https://doi.org/10.1023/A:1008306431147} {\bibfield  {journal} {\bibinfo
  {journal} {J. Glob. Optim.}\ }\textbf {\bibinfo {volume} {13}},\ \bibinfo
  {pages} {455} (\bibinfo {year} {1998})}\BibitemShut {NoStop}%
\bibitem [{\citenamefont {Broomhead}\ and\ \citenamefont
  {Lowe}(1988)}]{Broomhead1988}%
  \BibitemOpen
  \bibfield  {author} {\bibinfo {author} {\bibfnamefont {D.~S.}\ \bibnamefont
  {Broomhead}}\ and\ \bibinfo {author} {\bibfnamefont {D.}~\bibnamefont
  {Lowe}},\ }\href {https://apps.dtic.mil/sti/citations/ADA196234} {\emph
  {\bibinfo {title} {Radial Basis Functions, Multi-Variable Functional
  Interpolation and Adaptive Networks}}},\ \bibinfo {type} {Tech. Rep.}\
  (\bibinfo  {institution} {Royal Signals and Radar Establishment Malvern
  (United Kingdom)},\ \bibinfo {year} {1988})\BibitemShut {NoStop}%
\bibitem [{\citenamefont {Samarakoon}\ \emph {et~al.}(2021)\citenamefont
  {Samarakoon}, \citenamefont {Tennant}, \citenamefont {Ye}, \citenamefont
  {Zhang},\ and\ \citenamefont {Grigera}}]{pressure_paper}%
  \BibitemOpen
  \bibfield  {author} {\bibinfo {author} {\bibfnamefont {A.~M.}\ \bibnamefont
  {Samarakoon}}, \bibinfo {author} {\bibfnamefont {D.~A.}\ \bibnamefont
  {Tennant}}, \bibinfo {author} {\bibfnamefont {F.}~\bibnamefont {Ye}},
  \bibinfo {author} {\bibfnamefont {Q.}~\bibnamefont {Zhang}},\ and\ \bibinfo
  {author} {\bibfnamefont {S.~A.}\ \bibnamefont {Grigera}},\ }\Eprint
  {https://arxiv.org/abs/https://arxiv.org/abs/2110.15817}
  {https://arxiv.org/abs/2110.15817}  (\bibinfo {year} {2021}),\ \bibinfo
  {note} {arXiv:2110.15817}\BibitemShut {NoStop}%
\bibitem [{\citenamefont {Sugiura}\ and\ \citenamefont
  {Shimizu}(2012)}]{PhysRevLett.108.240401}%
  \BibitemOpen
  \bibfield  {author} {\bibinfo {author} {\bibfnamefont {S.}~\bibnamefont
  {Sugiura}}\ and\ \bibinfo {author} {\bibfnamefont {A.}~\bibnamefont
  {Shimizu}},\ }\href {https://doi.org/10.1103/PhysRevLett.108.240401}
  {\bibfield  {journal} {\bibinfo  {journal} {Phys. Rev. Lett.}\ }\textbf
  {\bibinfo {volume} {108}},\ \bibinfo {pages} {240401} (\bibinfo {year}
  {2012})}\BibitemShut {NoStop}%
\bibitem [{\citenamefont {Kawamura}\ \emph {et~al.}(2017)\citenamefont
  {Kawamura}, \citenamefont {Yoshimi}, \citenamefont {Misawa}, \citenamefont
  {Yamaji}, \citenamefont {Todo},\ and\ \citenamefont
  {Kawashima}}]{Kawamura2017}%
  \BibitemOpen
  \bibfield  {author} {\bibinfo {author} {\bibfnamefont {M.}~\bibnamefont
  {Kawamura}}, \bibinfo {author} {\bibfnamefont {K.}~\bibnamefont {Yoshimi}},
  \bibinfo {author} {\bibfnamefont {T.}~\bibnamefont {Misawa}}, \bibinfo
  {author} {\bibfnamefont {Y.}~\bibnamefont {Yamaji}}, \bibinfo {author}
  {\bibfnamefont {S.}~\bibnamefont {Todo}},\ and\ \bibinfo {author}
  {\bibfnamefont {N.}~\bibnamefont {Kawashima}},\ }\href
  {https://doi.org/10.1016/j.cpc.2017.04.006} {\bibfield  {journal} {\bibinfo
  {journal} {Comp. Phys. Comms.}\ }\textbf {\bibinfo {volume} {217}},\ \bibinfo
  {pages} {180} (\bibinfo {year} {2017})}\BibitemShut {NoStop}%
\end{thebibliography}%


\begin{thebibliography}{31}%
\makeatletter
\providecommand \@ifxundefined [1]{%
 \@ifx{#1\undefined}
}%
\providecommand \@ifnum [1]{%
 \ifnum #1\expandafter \@firstoftwo
 \else \expandafter \@secondoftwo
 \fi
}%
\providecommand \@ifx [1]{%
 \ifx #1\expandafter \@firstoftwo
 \else \expandafter \@secondoftwo
 \fi
}%
\providecommand \natexlab [1]{#1}%
\providecommand \enquote  [1]{``#1''}%
\providecommand \bibnamefont  [1]{#1}%
\providecommand \bibfnamefont [1]{#1}%
\providecommand \citenamefont [1]{#1}%
\providecommand \href@noop [0]{\@secondoftwo}%
\providecommand \href [0]{\begingroup \@sanitize@url \@href}%
\providecommand \@href[1]{\@@startlink{#1}\@@href}%
\providecommand \@@href[1]{\endgroup#1\@@endlink}%
\providecommand \@sanitize@url [0]{\catcode `\\12\catcode `\$12\catcode
  `\&12\catcode `\#12\catcode `\^12\catcode `\_12\catcode `\%12\relax}%
\providecommand \@@startlink[1]{}%
\providecommand \@@endlink[0]{}%
\providecommand \url  [0]{\begingroup\@sanitize@url \@url }%
\providecommand \@url [1]{\endgroup\@href {#1}{\urlprefix }}%
\providecommand \urlprefix  [0]{URL }%
\providecommand \Eprint [0]{\href }%
\providecommand \doibase [0]{https://doi.org/}%
\providecommand \selectlanguage [0]{\@gobble}%
\providecommand \bibinfo  [0]{\@secondoftwo}%
\providecommand \bibfield  [0]{\@secondoftwo}%
\providecommand \translation [1]{[#1]}%
\providecommand \BibitemOpen [0]{}%
\providecommand \bibitemStop [0]{}%
\providecommand \bibitemNoStop [0]{.\EOS\space}%
\providecommand \EOS [0]{\spacefactor3000\relax}%
\providecommand \BibitemShut  [1]{\csname bibitem#1\endcsname}%
\let\auto@bib@innerbib\@empty
%</preamble>
\bibitem [{\citenamefont {Glamazda}\ \emph {et~al.}(2017)\citenamefont
  {Glamazda}, \citenamefont {Lemmens}, \citenamefont {Do}, \citenamefont
  {Kwon},\ and\ \citenamefont {Choi}}]{PhysRevB.95.174429}%
  \BibitemOpen
  \bibfield  {author} {\bibinfo {author} {\bibfnamefont {A.}~\bibnamefont
  {Glamazda}}, \bibinfo {author} {\bibfnamefont {P.}~\bibnamefont {Lemmens}},
  \bibinfo {author} {\bibfnamefont {S.-H.}\ \bibnamefont {Do}}, \bibinfo
  {author} {\bibfnamefont {Y.~S.}\ \bibnamefont {Kwon}},\ and\ \bibinfo
  {author} {\bibfnamefont {K.-Y.}\ \bibnamefont {Choi}},\ }\href
  {https://doi.org/10.1103/PhysRevB.95.174429} {\bibfield  {journal} {\bibinfo
  {journal} {Phys. Rev. B}\ }\textbf {\bibinfo {volume} {95}},\ \bibinfo
  {pages} {174429} (\bibinfo {year} {2017})}\BibitemShut {NoStop}%
\bibitem [{\citenamefont {Mu}\ \emph {et~al.}(2022)\citenamefont {Mu},
  \citenamefont {Dixit}, \citenamefont {Wang}, \citenamefont {Abernathy},
  \citenamefont {Cao}, \citenamefont {Nagler}, \citenamefont {Yan},
  \citenamefont {Lampen-Kelley}, \citenamefont {Mandrus}, \citenamefont
  {Polanco}, \citenamefont {Liang}, \citenamefont {Hal\'asz}, \citenamefont
  {Cheng}, \citenamefont {Banerjee},\ and\ \citenamefont {Berlijn}}]{Mu2021}%
  \BibitemOpen
  \bibfield  {author} {\bibinfo {author} {\bibfnamefont {S.}~\bibnamefont
  {Mu}}, \bibinfo {author} {\bibfnamefont {K.~D.}\ \bibnamefont {Dixit}},
  \bibinfo {author} {\bibfnamefont {X.}~\bibnamefont {Wang}}, \bibinfo {author}
  {\bibfnamefont {D.~L.}\ \bibnamefont {Abernathy}}, \bibinfo {author}
  {\bibfnamefont {H.}~\bibnamefont {Cao}}, \bibinfo {author} {\bibfnamefont
  {S.~E.}\ \bibnamefont {Nagler}}, \bibinfo {author} {\bibfnamefont
  {J.}~\bibnamefont {Yan}}, \bibinfo {author} {\bibfnamefont {P.}~\bibnamefont
  {Lampen-Kelley}}, \bibinfo {author} {\bibfnamefont {D.}~\bibnamefont
  {Mandrus}}, \bibinfo {author} {\bibfnamefont {C.~A.}\ \bibnamefont
  {Polanco}}, \bibinfo {author} {\bibfnamefont {L.}~\bibnamefont {Liang}},
  \bibinfo {author} {\bibfnamefont {G.~B.}\ \bibnamefont {Hal\'asz}}, \bibinfo
  {author} {\bibfnamefont {Y.}~\bibnamefont {Cheng}}, \bibinfo {author}
  {\bibfnamefont {A.}~\bibnamefont {Banerjee}},\ and\ \bibinfo {author}
  {\bibfnamefont {T.}~\bibnamefont {Berlijn}},\ }\href
  {https://doi.org/10.1103/PhysRevResearch.4.013067} {\bibfield  {journal}
  {\bibinfo  {journal} {Phys. Rev. Research}\ }\textbf {\bibinfo {volume}
  {4}},\ \bibinfo {pages} {013067} (\bibinfo {year} {2022})}\BibitemShut
  {NoStop}%
\bibitem [{\citenamefont {Samarakoon}\ \emph {et~al.}(2020)\citenamefont
  {Samarakoon}, \citenamefont {Barros}, \citenamefont {Li}, \citenamefont
  {Eisenbach}, \citenamefont {Zhang}, \citenamefont {Ye}, \citenamefont
  {Sharma}, \citenamefont {Dun}, \citenamefont {Zhou}, \citenamefont {Grigera},
  \citenamefont {Batista},\ and\ \citenamefont {Tennant}}]{Samarakoon2020}%
  \BibitemOpen
  \bibfield  {author} {\bibinfo {author} {\bibfnamefont {A.~M.}\ \bibnamefont
  {Samarakoon}}, \bibinfo {author} {\bibfnamefont {K.}~\bibnamefont {Barros}},
  \bibinfo {author} {\bibfnamefont {Y.~W.}\ \bibnamefont {Li}}, \bibinfo
  {author} {\bibfnamefont {M.}~\bibnamefont {Eisenbach}}, \bibinfo {author}
  {\bibfnamefont {Q.}~\bibnamefont {Zhang}}, \bibinfo {author} {\bibfnamefont
  {F.}~\bibnamefont {Ye}}, \bibinfo {author} {\bibfnamefont {V.}~\bibnamefont
  {Sharma}}, \bibinfo {author} {\bibfnamefont {Z.~L.}\ \bibnamefont {Dun}},
  \bibinfo {author} {\bibfnamefont {H.}~\bibnamefont {Zhou}}, \bibinfo {author}
  {\bibfnamefont {S.~A.}\ \bibnamefont {Grigera}}, \bibinfo {author}
  {\bibfnamefont {C.~D.}\ \bibnamefont {Batista}},\ and\ \bibinfo {author}
  {\bibfnamefont {D.~A.}\ \bibnamefont {Tennant}},\ }\href
  {https://doi.org/10.1038/s41467-020-14660-y} {\bibfield  {journal} {\bibinfo
  {journal} {Nat. Commun.}\ }\textbf {\bibinfo {volume} {11}},\ \bibinfo
  {pages} {892} (\bibinfo {year} {2020})}\BibitemShut {NoStop}%
\bibitem [{\citenamefont {Metropolis}\ \emph {et~al.}(1953)\citenamefont
  {Metropolis}, \citenamefont {Rosenbluth}, \citenamefont {Rosenbluth},
  \citenamefont {Teller},\ and\ \citenamefont {Teller}}]{Metropolis1953}%
  \BibitemOpen
  \bibfield  {author} {\bibinfo {author} {\bibfnamefont {N.}~\bibnamefont
  {Metropolis}}, \bibinfo {author} {\bibfnamefont {A.~W.}\ \bibnamefont
  {Rosenbluth}}, \bibinfo {author} {\bibfnamefont {M.~N.}\ \bibnamefont
  {Rosenbluth}}, \bibinfo {author} {\bibfnamefont {A.~H.}\ \bibnamefont
  {Teller}},\ and\ \bibinfo {author} {\bibfnamefont {E.}~\bibnamefont
  {Teller}},\ }\href {https://doi.org/10.1063/1.1699114} {\bibfield  {journal}
  {\bibinfo  {journal} {J. Chem. Phys.}\ }\textbf {\bibinfo {volume} {21}},\
  \bibinfo {pages} {1087} (\bibinfo {year} {1953})}\BibitemShut {NoStop}%
\bibitem [{\citenamefont {Huberman}\ \emph {et~al.}(2008)\citenamefont
  {Huberman}, \citenamefont {Tennant}, \citenamefont {Cowley}, \citenamefont
  {Coldea},\ and\ \citenamefont {Frost}}]{Huberman_2008}%
  \BibitemOpen
  \bibfield  {author} {\bibinfo {author} {\bibfnamefont {T.}~\bibnamefont
  {Huberman}}, \bibinfo {author} {\bibfnamefont {D.~A.}\ \bibnamefont
  {Tennant}}, \bibinfo {author} {\bibfnamefont {R.~A.}\ \bibnamefont {Cowley}},
  \bibinfo {author} {\bibfnamefont {R.}~\bibnamefont {Coldea}},\ and\ \bibinfo
  {author} {\bibfnamefont {C.~D.}\ \bibnamefont {Frost}},\ }\href
  {https://doi.org/10.1088/1742-5468/2008/05/p05017} {\bibfield  {journal}
  {\bibinfo  {journal} {J. Stat. Mech.: Theory Exp.}\ }\textbf {\bibinfo
  {volume} {2008}}\bibinfo  {number} { (05)},\ \bibinfo {pages}
  {P05017}}\BibitemShut {NoStop}%
\bibitem [{\citenamefont {Chollet}\ \emph {et~al.}(2015)\citenamefont {Chollet}
  \emph {et~al.}}]{chollet2015keras}%
  \BibitemOpen
\bibfield  {number} {  }\bibfield  {author} {\bibinfo {author} {\bibfnamefont
  {F.}~\bibnamefont {Chollet}} \emph {et~al.},\ }\href
  {https://github.com/keras-team/keras} {\bibinfo {title} {Keras}} (\bibinfo
  {year} {2015}),\ \bibinfo {note} {github.com/keras-team/keras}\BibitemShut
  {NoStop}%
\bibitem [{\citenamefont {Tennant}\ and\ \citenamefont
  {Samarakoon}(2021)}]{Samarakoon2021}%
  \BibitemOpen
  \bibfield  {author} {\bibinfo {author} {\bibfnamefont {A.}~\bibnamefont
  {Tennant}}\ and\ \bibinfo {author} {\bibfnamefont {A.}~\bibnamefont
  {Samarakoon}},\ }\href
  {http://iopscience.iop.org/article/10.1088/1361-648X/abe818} {\bibfield
  {journal} {\bibinfo  {journal} {J. Phys. Condens. Matter}\ }\textbf {\bibinfo
  {volume} {34}},\ \bibinfo {pages} {044002} (\bibinfo {year}
  {2021})}\BibitemShut {NoStop}%
\bibitem [{\citenamefont {Kawamura}\ \emph {et~al.}(2017)\citenamefont
  {Kawamura}, \citenamefont {Yoshimi}, \citenamefont {Misawa}, \citenamefont
  {Yamaji}, \citenamefont {Todo},\ and\ \citenamefont
  {Kawashima}}]{Kawamura2017}%
  \BibitemOpen
  \bibfield  {author} {\bibinfo {author} {\bibfnamefont {M.}~\bibnamefont
  {Kawamura}}, \bibinfo {author} {\bibfnamefont {K.}~\bibnamefont {Yoshimi}},
  \bibinfo {author} {\bibfnamefont {T.}~\bibnamefont {Misawa}}, \bibinfo
  {author} {\bibfnamefont {Y.}~\bibnamefont {Yamaji}}, \bibinfo {author}
  {\bibfnamefont {S.}~\bibnamefont {Todo}},\ and\ \bibinfo {author}
  {\bibfnamefont {N.}~\bibnamefont {Kawashima}},\ }\href
  {https://doi.org/10.1016/j.cpc.2017.04.006} {\bibfield  {journal} {\bibinfo
  {journal} {Comp. Phys. Comms.}\ }\textbf {\bibinfo {volume} {217}},\ \bibinfo
  {pages} {180} (\bibinfo {year} {2017})}\BibitemShut {NoStop}%
\bibitem [{\citenamefont {Dagotto}(1994)}]{RevModPhys.66.763}%
  \BibitemOpen
  \bibfield  {author} {\bibinfo {author} {\bibfnamefont {E.}~\bibnamefont
  {Dagotto}},\ }\href {https://doi.org/10.1103/RevModPhys.66.763} {\bibfield
  {journal} {\bibinfo  {journal} {Rev. Mod. Phys.}\ }\textbf {\bibinfo {volume}
  {66}},\ \bibinfo {pages} {763} (\bibinfo {year} {1994})}\BibitemShut
  {NoStop}%
\bibitem [{\citenamefont {Laurell}\ and\ \citenamefont
  {Okamoto}(2020)}]{Laurell2020}%
  \BibitemOpen
  \bibfield  {author} {\bibinfo {author} {\bibfnamefont {P.}~\bibnamefont
  {Laurell}}\ and\ \bibinfo {author} {\bibfnamefont {S.}~\bibnamefont
  {Okamoto}},\ }\href {https://doi.org/10.1038/s41535-019-0203-y} {\bibfield
  {journal} {\bibinfo  {journal} {npj Quantum Mater.}\ }\textbf {\bibinfo
  {volume} {5}},\ \bibinfo {pages} {2} (\bibinfo {year} {2020})}\BibitemShut
  {NoStop}%
\bibitem [{\citenamefont {Moshtagh}\ \emph {et~al.}(2005)\citenamefont
  {Moshtagh} \emph {et~al.}}]{moshtagh2005minimum}%
  \BibitemOpen
  \bibfield  {author} {\bibinfo {author} {\bibfnamefont {N.}~\bibnamefont
  {Moshtagh}} \emph {et~al.},\ }\href@noop {} {\bibfield  {journal} {\bibinfo
  {journal} {Convex optimization}\ }\textbf {\bibinfo {volume} {111}},\
  \bibinfo {pages} {1} (\bibinfo {year} {2005})}\BibitemShut {NoStop}%
\bibitem [{\citenamefont {Winter}\ \emph {et~al.}(2017)\citenamefont {Winter},
  \citenamefont {Riedl}, \citenamefont {Maksimov}, \citenamefont {Chernyshev},
  \citenamefont {Honecker},\ and\ \citenamefont
  {Valent\'i}}]{10.1038/s41467-017-01177-0}%
  \BibitemOpen
  \bibfield  {author} {\bibinfo {author} {\bibfnamefont {S.~M.}\ \bibnamefont
  {Winter}}, \bibinfo {author} {\bibfnamefont {K.}~\bibnamefont {Riedl}},
  \bibinfo {author} {\bibfnamefont {P.~A.}\ \bibnamefont {Maksimov}}, \bibinfo
  {author} {\bibfnamefont {A.~L.}\ \bibnamefont {Chernyshev}}, \bibinfo
  {author} {\bibfnamefont {A.}~\bibnamefont {Honecker}},\ and\ \bibinfo
  {author} {\bibfnamefont {R.}~\bibnamefont {Valent\'i}},\ }\href
  {https://doi.org/10.1038/s41467-017-01177-0} {\bibfield  {journal} {\bibinfo
  {journal} {Nat. Commun.}\ }\textbf {\bibinfo {volume} {8}},\ \bibinfo {pages}
  {1152} (\bibinfo {year} {2017})}\BibitemShut {NoStop}%
\bibitem [{\citenamefont {Maksimov}\ and\ \citenamefont
  {Chernyshev}(2020)}]{PhysRevResearch.2.033011}%
  \BibitemOpen
  \bibfield  {author} {\bibinfo {author} {\bibfnamefont {P.~A.}\ \bibnamefont
  {Maksimov}}\ and\ \bibinfo {author} {\bibfnamefont {A.~L.}\ \bibnamefont
  {Chernyshev}},\ }\href {https://doi.org/10.1103/PhysRevResearch.2.033011}
  {\bibfield  {journal} {\bibinfo  {journal} {Phys. Rev. Research}\ }\textbf
  {\bibinfo {volume} {2}},\ \bibinfo {pages} {033011} (\bibinfo {year}
  {2020})}\BibitemShut {NoStop}%
\bibitem [{\citenamefont {Li}\ \emph {et~al.}(2021)\citenamefont {Li},
  \citenamefont {Zhang}, \citenamefont {Wang}, \citenamefont {Wu},
  \citenamefont {Gao}, \citenamefont {Qu}, \citenamefont {Liu}, \citenamefont
  {Gong},\ and\ \citenamefont {Li}}]{Li2021}%
  \BibitemOpen
  \bibfield  {author} {\bibinfo {author} {\bibfnamefont {H.}~\bibnamefont
  {Li}}, \bibinfo {author} {\bibfnamefont {H.-K.}\ \bibnamefont {Zhang}},
  \bibinfo {author} {\bibfnamefont {J.}~\bibnamefont {Wang}}, \bibinfo {author}
  {\bibfnamefont {H.-Q.}\ \bibnamefont {Wu}}, \bibinfo {author} {\bibfnamefont
  {Y.}~\bibnamefont {Gao}}, \bibinfo {author} {\bibfnamefont {D.-W.}\
  \bibnamefont {Qu}}, \bibinfo {author} {\bibfnamefont {Z.-X.}\ \bibnamefont
  {Liu}}, \bibinfo {author} {\bibfnamefont {S.-S.}\ \bibnamefont {Gong}},\ and\
  \bibinfo {author} {\bibfnamefont {W.}~\bibnamefont {Li}},\ }\href
  {https://doi.org/https://doi.org/10.1038/s41467-021-24257-8} {\bibfield
  {journal} {\bibinfo  {journal} {Nat. Commun.}\ }\textbf {\bibinfo {volume}
  {12}},\ \bibinfo {pages} {4007} (\bibinfo {year} {2021})}\BibitemShut
  {NoStop}%
\bibitem [{\citenamefont {Hou}\ \emph {et~al.}(2017)\citenamefont {Hou},
  \citenamefont {Xiang},\ and\ \citenamefont {Gong}}]{PhysRevB.96.054410}%
  \BibitemOpen
  \bibfield  {author} {\bibinfo {author} {\bibfnamefont {Y.~S.}\ \bibnamefont
  {Hou}}, \bibinfo {author} {\bibfnamefont {H.~J.}\ \bibnamefont {Xiang}},\
  and\ \bibinfo {author} {\bibfnamefont {X.~G.}\ \bibnamefont {Gong}},\ }\href
  {https://doi.org/10.1103/PhysRevB.96.054410} {\bibfield  {journal} {\bibinfo
  {journal} {Phys. Rev. B}\ }\textbf {\bibinfo {volume} {96}},\ \bibinfo
  {pages} {054410} (\bibinfo {year} {2017})}\BibitemShut {NoStop}%
\bibitem [{\citenamefont {Suzuki}\ \emph {et~al.}(2021)\citenamefont {Suzuki},
  \citenamefont {Liu}, \citenamefont {Bertinshaw}, \citenamefont {Ueda},
  \citenamefont {Kim}, \citenamefont {Laha}, \citenamefont {Weber},
  \citenamefont {Yang}, \citenamefont {Wang}, \citenamefont {H.~Takahash~and},
  \citenamefont {Minola}, \citenamefont {Lotsch}, \citenamefont {Kim},
  \citenamefont {Yavaş}, \citenamefont {Daghofer}, \citenamefont {Chaloupka},
  \citenamefont {Khaliullin}, \citenamefont {Gretarsson},\ and\ \citenamefont
  {Keimer}}]{Suzuki2021}%
  \BibitemOpen
  \bibfield  {author} {\bibinfo {author} {\bibfnamefont {H.}~\bibnamefont
  {Suzuki}}, \bibinfo {author} {\bibfnamefont {H.}~\bibnamefont {Liu}},
  \bibinfo {author} {\bibfnamefont {J.}~\bibnamefont {Bertinshaw}}, \bibinfo
  {author} {\bibfnamefont {K.}~\bibnamefont {Ueda}}, \bibinfo {author}
  {\bibfnamefont {H.}~\bibnamefont {Kim}}, \bibinfo {author} {\bibfnamefont
  {S.}~\bibnamefont {Laha}}, \bibinfo {author} {\bibfnamefont {D.}~\bibnamefont
  {Weber}}, \bibinfo {author} {\bibfnamefont {Z.}~\bibnamefont {Yang}},
  \bibinfo {author} {\bibfnamefont {L.}~\bibnamefont {Wang}}, \bibinfo {author}
  {\bibfnamefont {K.~F.}\ \bibnamefont {H.~Takahash~and}}, \bibinfo {author}
  {\bibfnamefont {M.}~\bibnamefont {Minola}}, \bibinfo {author} {\bibfnamefont
  {B.~V.}\ \bibnamefont {Lotsch}}, \bibinfo {author} {\bibfnamefont {B.~J.}\
  \bibnamefont {Kim}}, \bibinfo {author} {\bibfnamefont {H.}~\bibnamefont
  {Yavaş}}, \bibinfo {author} {\bibfnamefont {M.}~\bibnamefont {Daghofer}},
  \bibinfo {author} {\bibfnamefont {J.}~\bibnamefont {Chaloupka}}, \bibinfo
  {author} {\bibfnamefont {G.}~\bibnamefont {Khaliullin}}, \bibinfo {author}
  {\bibfnamefont {H.}~\bibnamefont {Gretarsson}},\ and\ \bibinfo {author}
  {\bibfnamefont {B.}~\bibnamefont {Keimer}},\ }\href
  {https://doi.org/https://doi.org/10.1038/s41467-021-24722-4} {\bibfield
  {journal} {\bibinfo  {journal} {Nat. Commun.}\ }\textbf {\bibinfo {volume}
  {12}},\ \bibinfo {pages} {4512} (\bibinfo {year} {2021})}\BibitemShut
  {NoStop}%
\bibitem [{\citenamefont {Winter}\ \emph {et~al.}(2016)\citenamefont {Winter},
  \citenamefont {Li}, \citenamefont {Jeschke},\ and\ \citenamefont
  {Valent\'{\i}}}]{PhysRevB.93.214431}%
  \BibitemOpen
  \bibfield  {author} {\bibinfo {author} {\bibfnamefont {S.~M.}\ \bibnamefont
  {Winter}}, \bibinfo {author} {\bibfnamefont {Y.}~\bibnamefont {Li}}, \bibinfo
  {author} {\bibfnamefont {H.~O.}\ \bibnamefont {Jeschke}},\ and\ \bibinfo
  {author} {\bibfnamefont {R.}~\bibnamefont {Valent\'{\i}}},\ }\href
  {https://doi.org/10.1103/PhysRevB.93.214431} {\bibfield  {journal} {\bibinfo
  {journal} {Phys. Rev. B}\ }\textbf {\bibinfo {volume} {93}},\ \bibinfo
  {pages} {214431} (\bibinfo {year} {2016})}\BibitemShut {NoStop}%
\bibitem [{\citenamefont {Wu}\ \emph {et~al.}(2018)\citenamefont {Wu},
  \citenamefont {Little}, \citenamefont {Aldape}, \citenamefont {Rees},
  \citenamefont {Thewalt}, \citenamefont {Lampen-Kelley}, \citenamefont
  {Banerjee}, \citenamefont {Bridges}, \citenamefont {Yan}, \citenamefont
  {Boone}, \citenamefont {Patankar}, \citenamefont {Goldhaber-Gordon},
  \citenamefont {Mandrus}, \citenamefont {Nagler}, \citenamefont {Altman},\
  and\ \citenamefont {Orenstein}}]{PhysRevB.98.094425}%
  \BibitemOpen
  \bibfield  {author} {\bibinfo {author} {\bibfnamefont {L.}~\bibnamefont
  {Wu}}, \bibinfo {author} {\bibfnamefont {A.}~\bibnamefont {Little}}, \bibinfo
  {author} {\bibfnamefont {E.~E.}\ \bibnamefont {Aldape}}, \bibinfo {author}
  {\bibfnamefont {D.}~\bibnamefont {Rees}}, \bibinfo {author} {\bibfnamefont
  {E.}~\bibnamefont {Thewalt}}, \bibinfo {author} {\bibfnamefont
  {P.}~\bibnamefont {Lampen-Kelley}}, \bibinfo {author} {\bibfnamefont
  {A.}~\bibnamefont {Banerjee}}, \bibinfo {author} {\bibfnamefont {C.~A.}\
  \bibnamefont {Bridges}}, \bibinfo {author} {\bibfnamefont {J.-Q.}\
  \bibnamefont {Yan}}, \bibinfo {author} {\bibfnamefont {D.}~\bibnamefont
  {Boone}}, \bibinfo {author} {\bibfnamefont {S.}~\bibnamefont {Patankar}},
  \bibinfo {author} {\bibfnamefont {D.}~\bibnamefont {Goldhaber-Gordon}},
  \bibinfo {author} {\bibfnamefont {D.}~\bibnamefont {Mandrus}}, \bibinfo
  {author} {\bibfnamefont {S.~E.}\ \bibnamefont {Nagler}}, \bibinfo {author}
  {\bibfnamefont {E.}~\bibnamefont {Altman}},\ and\ \bibinfo {author}
  {\bibfnamefont {J.}~\bibnamefont {Orenstein}},\ }\href
  {https://doi.org/10.1103/PhysRevB.98.094425} {\bibfield  {journal} {\bibinfo
  {journal} {Phys. Rev. B}\ }\textbf {\bibinfo {volume} {98}},\ \bibinfo
  {pages} {094425} (\bibinfo {year} {2018})}\BibitemShut {NoStop}%
\bibitem [{\citenamefont {Ozel}\ \emph {et~al.}(2019)\citenamefont {Ozel},
  \citenamefont {Belvin}, \citenamefont {Baldini}, \citenamefont {Kimchi},
  \citenamefont {Do}, \citenamefont {Choi},\ and\ \citenamefont
  {Gedik}}]{PhysRevB.100.085108}%
  \BibitemOpen
  \bibfield  {author} {\bibinfo {author} {\bibfnamefont {I.~O.}\ \bibnamefont
  {Ozel}}, \bibinfo {author} {\bibfnamefont {C.~A.}\ \bibnamefont {Belvin}},
  \bibinfo {author} {\bibfnamefont {E.}~\bibnamefont {Baldini}}, \bibinfo
  {author} {\bibfnamefont {I.}~\bibnamefont {Kimchi}}, \bibinfo {author}
  {\bibfnamefont {S.}~\bibnamefont {Do}}, \bibinfo {author} {\bibfnamefont
  {K.-Y.}\ \bibnamefont {Choi}},\ and\ \bibinfo {author} {\bibfnamefont
  {N.}~\bibnamefont {Gedik}},\ }\href
  {https://doi.org/10.1103/PhysRevB.100.085108} {\bibfield  {journal} {\bibinfo
   {journal} {Phys. Rev. B}\ }\textbf {\bibinfo {volume} {100}},\ \bibinfo
  {pages} {085108} (\bibinfo {year} {2019})}\BibitemShut {NoStop}%
\bibitem [{\citenamefont {Kim}\ and\ \citenamefont
  {Kee}(2016)}]{PhysRevB.93.155143}%
  \BibitemOpen
  \bibfield  {author} {\bibinfo {author} {\bibfnamefont {H.-S.}\ \bibnamefont
  {Kim}}\ and\ \bibinfo {author} {\bibfnamefont {H.-Y.}\ \bibnamefont {Kee}},\
  }\href {https://doi.org/10.1103/PhysRevB.93.155143} {\bibfield  {journal}
  {\bibinfo  {journal} {Phys. Rev. B}\ }\textbf {\bibinfo {volume} {93}},\
  \bibinfo {pages} {155143} (\bibinfo {year} {2016})}\BibitemShut {NoStop}%
\bibitem [{\citenamefont {Kim}\ \emph {et~al.}(2015)\citenamefont {Kim},
  \citenamefont {V.}, \citenamefont {Catuneanu},\ and\ \citenamefont
  {Kee}}]{PhysRevB.91.241110}%
  \BibitemOpen
  \bibfield  {author} {\bibinfo {author} {\bibfnamefont {H.-S.}\ \bibnamefont
  {Kim}}, \bibinfo {author} {\bibfnamefont {V.~S.}\ \bibnamefont {V.}},
  \bibinfo {author} {\bibfnamefont {A.}~\bibnamefont {Catuneanu}},\ and\
  \bibinfo {author} {\bibfnamefont {H.-Y.}\ \bibnamefont {Kee}},\ }\href
  {https://doi.org/10.1103/PhysRevB.91.241110} {\bibfield  {journal} {\bibinfo
  {journal} {Phys. Rev. B}\ }\textbf {\bibinfo {volume} {91}},\ \bibinfo
  {pages} {241110} (\bibinfo {year} {2015})}\BibitemShut {NoStop}%
\bibitem [{\citenamefont {Janssen}\ \emph {et~al.}(2017)\citenamefont
  {Janssen}, \citenamefont {Andrade},\ and\ \citenamefont
  {Vojta}}]{PhysRevB.96.064430}%
  \BibitemOpen
  \bibfield  {author} {\bibinfo {author} {\bibfnamefont {L.}~\bibnamefont
  {Janssen}}, \bibinfo {author} {\bibfnamefont {E.~C.}\ \bibnamefont
  {Andrade}},\ and\ \bibinfo {author} {\bibfnamefont {M.}~\bibnamefont
  {Vojta}},\ }\href {https://doi.org/10.1103/PhysRevB.96.064430} {\bibfield
  {journal} {\bibinfo  {journal} {Phys. Rev. B}\ }\textbf {\bibinfo {volume}
  {96}},\ \bibinfo {pages} {064430} (\bibinfo {year} {2017})}\BibitemShut
  {NoStop}%
\bibitem [{\citenamefont {Banerjee}\ \emph {et~al.}(2016)\citenamefont
  {Banerjee}, \citenamefont {Bridges}, \citenamefont {Yan}, \citenamefont
  {Aczel}, \citenamefont {Li}, \citenamefont {Stone}, \citenamefont {Granroth},
  \citenamefont {Lumsden}, \citenamefont {Yiu}, \citenamefont {Knolle},
  \citenamefont {Bhattacharjee}, \citenamefont {Kovrizhin}, \citenamefont
  {Moessner}, \citenamefont {Tennant}, \citenamefont {Mandrus},\ and\
  \citenamefont {Nagler}}]{Banerjee2016}%
  \BibitemOpen
  \bibfield  {author} {\bibinfo {author} {\bibfnamefont {A.}~\bibnamefont
  {Banerjee}}, \bibinfo {author} {\bibfnamefont {C.~A.}\ \bibnamefont
  {Bridges}}, \bibinfo {author} {\bibfnamefont {J.-Q.}\ \bibnamefont {Yan}},
  \bibinfo {author} {\bibfnamefont {A.~A.}\ \bibnamefont {Aczel}}, \bibinfo
  {author} {\bibfnamefont {L.}~\bibnamefont {Li}}, \bibinfo {author}
  {\bibfnamefont {M.~B.}\ \bibnamefont {Stone}}, \bibinfo {author}
  {\bibfnamefont {G.~E.}\ \bibnamefont {Granroth}}, \bibinfo {author}
  {\bibfnamefont {M.~D.}\ \bibnamefont {Lumsden}}, \bibinfo {author}
  {\bibfnamefont {Y.}~\bibnamefont {Yiu}}, \bibinfo {author} {\bibfnamefont
  {J.}~\bibnamefont {Knolle}}, \bibinfo {author} {\bibfnamefont
  {S.}~\bibnamefont {Bhattacharjee}}, \bibinfo {author} {\bibfnamefont {D.~L.}\
  \bibnamefont {Kovrizhin}}, \bibinfo {author} {\bibfnamefont {R.}~\bibnamefont
  {Moessner}}, \bibinfo {author} {\bibfnamefont {D.~A.}\ \bibnamefont
  {Tennant}}, \bibinfo {author} {\bibfnamefont {D.~G.}\ \bibnamefont
  {Mandrus}},\ and\ \bibinfo {author} {\bibfnamefont {S.~E.}\ \bibnamefont
  {Nagler}},\ }\href {https://doi.org/10.1038/nmat4604} {\bibfield  {journal}
  {\bibinfo  {journal} {Nat. Mat.}\ }\textbf {\bibinfo {volume} {15}},\
  \bibinfo {pages} {733} (\bibinfo {year} {2016})}\BibitemShut {NoStop}%
\bibitem [{\citenamefont {Cookmeyer}\ and\ \citenamefont
  {Moore}(2018)}]{PhysRevB.98.060412}%
  \BibitemOpen
  \bibfield  {author} {\bibinfo {author} {\bibfnamefont {J.}~\bibnamefont
  {Cookmeyer}}\ and\ \bibinfo {author} {\bibfnamefont {J.~E.}\ \bibnamefont
  {Moore}},\ }\href {https://doi.org/10.1103/PhysRevB.98.060412} {\bibfield
  {journal} {\bibinfo  {journal} {Phys. Rev. B}\ }\textbf {\bibinfo {volume}
  {98}},\ \bibinfo {pages} {060412} (\bibinfo {year} {2018})}\BibitemShut
  {NoStop}%
\bibitem [{\citenamefont {Ran}\ \emph {et~al.}(2017)\citenamefont {Ran},
  \citenamefont {Wang}, \citenamefont {Wang}, \citenamefont {Dong},
  \citenamefont {Ren}, \citenamefont {Bao}, \citenamefont {Li}, \citenamefont
  {Ma}, \citenamefont {Gan}, \citenamefont {Zhang}, \citenamefont {Park},
  \citenamefont {Deng}, \citenamefont {Danilkin}, \citenamefont {Yu},
  \citenamefont {Li},\ and\ \citenamefont {Wen}}]{PhysRevLett.118.107203}%
  \BibitemOpen
  \bibfield  {author} {\bibinfo {author} {\bibfnamefont {K.}~\bibnamefont
  {Ran}}, \bibinfo {author} {\bibfnamefont {J.}~\bibnamefont {Wang}}, \bibinfo
  {author} {\bibfnamefont {W.}~\bibnamefont {Wang}}, \bibinfo {author}
  {\bibfnamefont {Z.-Y.}\ \bibnamefont {Dong}}, \bibinfo {author}
  {\bibfnamefont {X.}~\bibnamefont {Ren}}, \bibinfo {author} {\bibfnamefont
  {S.}~\bibnamefont {Bao}}, \bibinfo {author} {\bibfnamefont {S.}~\bibnamefont
  {Li}}, \bibinfo {author} {\bibfnamefont {Z.}~\bibnamefont {Ma}}, \bibinfo
  {author} {\bibfnamefont {Y.}~\bibnamefont {Gan}}, \bibinfo {author}
  {\bibfnamefont {Y.}~\bibnamefont {Zhang}}, \bibinfo {author} {\bibfnamefont
  {J.~T.}\ \bibnamefont {Park}}, \bibinfo {author} {\bibfnamefont
  {G.}~\bibnamefont {Deng}}, \bibinfo {author} {\bibfnamefont {S.}~\bibnamefont
  {Danilkin}}, \bibinfo {author} {\bibfnamefont {S.-L.}\ \bibnamefont {Yu}},
  \bibinfo {author} {\bibfnamefont {J.-X.}\ \bibnamefont {Li}},\ and\ \bibinfo
  {author} {\bibfnamefont {J.}~\bibnamefont {Wen}},\ }\href
  {https://doi.org/10.1103/PhysRevLett.118.107203} {\bibfield  {journal}
  {\bibinfo  {journal} {Phys. Rev. Lett.}\ }\textbf {\bibinfo {volume} {118}},\
  \bibinfo {pages} {107203} (\bibinfo {year} {2017})}\BibitemShut {NoStop}%
\bibitem [{\citenamefont {Suzuki}\ and\ \citenamefont
  {Suga}(2018)}]{PhysRevB.97.134424}%
  \BibitemOpen
  \bibfield  {author} {\bibinfo {author} {\bibfnamefont {T.}~\bibnamefont
  {Suzuki}}\ and\ \bibinfo {author} {\bibfnamefont {S.-i.}\ \bibnamefont
  {Suga}},\ }\href {https://doi.org/10.1103/PhysRevB.97.134424} {\bibfield
  {journal} {\bibinfo  {journal} {Phys. Rev. B}\ }\textbf {\bibinfo {volume}
  {97}},\ \bibinfo {pages} {134424} (\bibinfo {year} {2018})}\BibitemShut
  {NoStop}%
\bibitem [{\citenamefont {Suzuki}\ and\ \citenamefont
  {Suga}(2019)}]{Suzuki2019}%
  \BibitemOpen
  \bibfield  {author} {\bibinfo {author} {\bibfnamefont {T.}~\bibnamefont
  {Suzuki}}\ and\ \bibinfo {author} {\bibfnamefont {S.-i.}\ \bibnamefont
  {Suga}},\ }\href {https://doi.org/10.1103/PhysRevB.99.249902} {\bibfield
  {journal} {\bibinfo  {journal} {Phys. Rev. B}\ }\textbf {\bibinfo {volume}
  {99}},\ \bibinfo {pages} {249902} (\bibinfo {year} {2019})}\BibitemShut
  {NoStop}%
\bibitem [{\citenamefont {Ran}\ \emph {et~al.}(2022)\citenamefont {Ran},
  \citenamefont {Wang}, \citenamefont {Bao}, \citenamefont {Cai}, \citenamefont
  {Shangguan}, \citenamefont {Ma}, \citenamefont {Wang}, \citenamefont {Dong},
  \citenamefont {Čermák}, \citenamefont {Schneidewind}, \citenamefont {Meng},
  \citenamefont {Lu}, \citenamefont {Yu}, \citenamefont {Li},\ and\
  \citenamefont {Wen}}]{KejingRan:27501}%
  \BibitemOpen
  \bibfield  {author} {\bibinfo {author} {\bibfnamefont {K.}~\bibnamefont
  {Ran}}, \bibinfo {author} {\bibfnamefont {J.}~\bibnamefont {Wang}}, \bibinfo
  {author} {\bibfnamefont {S.}~\bibnamefont {Bao}}, \bibinfo {author}
  {\bibfnamefont {Z.}~\bibnamefont {Cai}}, \bibinfo {author} {\bibfnamefont
  {Y.}~\bibnamefont {Shangguan}}, \bibinfo {author} {\bibfnamefont
  {Z.}~\bibnamefont {Ma}}, \bibinfo {author} {\bibfnamefont {W.}~\bibnamefont
  {Wang}}, \bibinfo {author} {\bibfnamefont {Z.-Y.}\ \bibnamefont {Dong}},
  \bibinfo {author} {\bibfnamefont {P.}~\bibnamefont {Čermák}}, \bibinfo
  {author} {\bibfnamefont {A.}~\bibnamefont {Schneidewind}}, \bibinfo {author}
  {\bibfnamefont {S.}~\bibnamefont {Meng}}, \bibinfo {author} {\bibfnamefont
  {Z.}~\bibnamefont {Lu}}, \bibinfo {author} {\bibfnamefont {S.-L.}\
  \bibnamefont {Yu}}, \bibinfo {author} {\bibfnamefont {J.-X.}\ \bibnamefont
  {Li}},\ and\ \bibinfo {author} {\bibfnamefont {J.}~\bibnamefont {Wen}},\
  }\href {https://doi.org/10.1088/0256-307X/39/2/027501} {\bibfield  {journal}
  {\bibinfo  {journal} {Chin. Phys. Lett.}\ }\textbf {\bibinfo {volume} {39}},\
  \bibinfo {eid} {027501} (\bibinfo {year} {2022})}\BibitemShut {NoStop}%
\bibitem [{\citenamefont {Wang}\ \emph {et~al.}(2017)\citenamefont {Wang},
  \citenamefont {Dong}, \citenamefont {Yu},\ and\ \citenamefont
  {Li}}]{PhysRevB.96.115103}%
  \BibitemOpen
  \bibfield  {author} {\bibinfo {author} {\bibfnamefont {W.}~\bibnamefont
  {Wang}}, \bibinfo {author} {\bibfnamefont {Z.-Y.}\ \bibnamefont {Dong}},
  \bibinfo {author} {\bibfnamefont {S.-L.}\ \bibnamefont {Yu}},\ and\ \bibinfo
  {author} {\bibfnamefont {J.-X.}\ \bibnamefont {Li}},\ }\href
  {https://doi.org/10.1103/PhysRevB.96.115103} {\bibfield  {journal} {\bibinfo
  {journal} {Phys. Rev. B}\ }\textbf {\bibinfo {volume} {96}},\ \bibinfo
  {pages} {115103} (\bibinfo {year} {2017})}\BibitemShut {NoStop}%
\bibitem [{\citenamefont {Sugiura}\ and\ \citenamefont
  {Shimizu}(2012)}]{PhysRevLett.108.240401}%
  \BibitemOpen
  \bibfield  {author} {\bibinfo {author} {\bibfnamefont {S.}~\bibnamefont
  {Sugiura}}\ and\ \bibinfo {author} {\bibfnamefont {A.}~\bibnamefont
  {Shimizu}},\ }\href {https://doi.org/10.1103/PhysRevLett.108.240401}
  {\bibfield  {journal} {\bibinfo  {journal} {Phys. Rev. Lett.}\ }\textbf
  {\bibinfo {volume} {108}},\ \bibinfo {pages} {240401} (\bibinfo {year}
  {2012})}\BibitemShut {NoStop}%
\bibitem [{\citenamefont {Widmann}\ \emph {et~al.}(2019)\citenamefont
  {Widmann}, \citenamefont {Tsurkan}, \citenamefont {Prishchenko},
  \citenamefont {Mazurenko}, \citenamefont {Tsirlin},\ and\ \citenamefont
  {Loidl}}]{PhysRevB.99.094415}%
  \BibitemOpen
  \bibfield  {author} {\bibinfo {author} {\bibfnamefont {S.}~\bibnamefont
  {Widmann}}, \bibinfo {author} {\bibfnamefont {V.}~\bibnamefont {Tsurkan}},
  \bibinfo {author} {\bibfnamefont {D.~A.}\ \bibnamefont {Prishchenko}},
  \bibinfo {author} {\bibfnamefont {V.~G.}\ \bibnamefont {Mazurenko}}, \bibinfo
  {author} {\bibfnamefont {A.~A.}\ \bibnamefont {Tsirlin}},\ and\ \bibinfo
  {author} {\bibfnamefont {A.}~\bibnamefont {Loidl}},\ }\href
  {https://doi.org/10.1103/PhysRevB.99.094415} {\bibfield  {journal} {\bibinfo
  {journal} {Phys. Rev. B}\ }\textbf {\bibinfo {volume} {99}},\ \bibinfo
  {pages} {094415} (\bibinfo {year} {2019})}\BibitemShut {NoStop}%
\end{thebibliography}%

\end{document}

%\title{High-dimensional modeling of the proximate Kitaev spin liquid material RuCl$_3$}
% \title{Extraction of the interaction parameters for \texorpdfstring{$\alpha$-RuCl$_3$}{RuCl3} from neutron data using machine learning}
\title{Supplemental material for: Extraction of the interaction parameters for \texorpdfstring{$\alpha$-RuCl$_3$}{RuCl3} from neutron data using machine learning}
\author{Anjana M. Samarakoon}
\affiliation{Shull-Wollan Center, Oak Ridge National Laboratory, Oak Ridge, Tennessee 37831, USA}
\affiliation{Neutron Scattering Division, Oak Ridge National Laboratory, Oak Ridge, Tennessee 37831, USA}
\affiliation{Materials Science Division, Argonne National Laboratory, Lemont, Illinois 60439, USA}
% \thanks{Current address: Materials Science Division, Argonne National Laboratory, Argonne, IL, USA.}
\author{Pontus Laurell}
\affiliation{Center for Nanophase Materials Sciences, Oak Ridge National Laboratory, Oak Ridge, Tennessee 37831, USA}
\affiliation{Computational Sciences and Engineering Division, Oak Ridge National Laboratory, Oak Ridge, Tennessee 37831, USA}
\affiliation{Department of Physics and Astronomy, University of Tennessee, Knoxville, Tennessee 37996, USA.}
\author{Christian Balz}
\affiliation{Neutron Scattering Division, Oak Ridge National Laboratory, Oak Ridge, Tennessee 37831, USA}
\affiliation{ISIS Neutron and Muon Source, Rutherford Appleton Laboratory, Didcot, OX11 0QX, United Kingdom}
\author{Arnab Banerjee}
\affiliation{Neutron Scattering Division, Oak Ridge National Laboratory, Oak Ridge, Tennessee 37831, USA}
\affiliation{Department of Physics and Astronomy, Purdue University, West Lafayette, Indiana 47906, USA}
\author{Paula Lampen-Kelley}
\affiliation{Department of Materials Science and Engineering, University of Tennessee, Knoxville, Tennessee 37996, USA}
\affiliation{Materials Science and Technology Division, Oak Ridge National Laboratory, Oak Ridge, Tennessee 37831, USA}
\author{David Mandrus}
\affiliation{Department of Materials Science and Engineering, University of Tennessee, Knoxville, Tennessee 37996, USA}
\affiliation{Materials Science and Technology Division, Oak Ridge National Laboratory, Oak Ridge, Tennessee 37831, USA}
\author{Stephen E. Nagler}
\affiliation{Neutron Scattering Division, Oak Ridge National Laboratory, Oak Ridge, Tennessee 37831, USA}
\affiliation{Quantum Science Center, Oak Ridge National Laboratory, Oak Ridge, Tennessee 37831, USA}
\author{Satoshi Okamoto}
\affiliation{Materials Science and Technology Division, Oak Ridge National Laboratory, Oak Ridge, Tennessee 37831, USA}
\affiliation{Quantum Science Center, Oak Ridge National Laboratory, Oak Ridge, Tennessee 37831, USA}
\author{D. Alan Tennant}
\affiliation{Materials Science and Technology Division, Oak Ridge National Laboratory, Oak Ridge, Tennessee 37831, USA}
\affiliation{Quantum Science Center, Oak Ridge National Laboratory, Oak Ridge, Tennessee 37831, USA}
\affiliation{Shull-Wollan Center, Oak Ridge National Laboratory, Oak Ridge, Tennessee 37831, USA}
\affiliation{Department of Physics and Astronomy, University of Tennessee, Knoxville, Tennessee 37996, USA.}
\date{\today}

\begin{abstract}
\hfill \break
\hfill \break
\hfill \break
\hfill \break
\hfill \break
\hfill \break
\hfill \break
\hfill \break
\hfill \break
\hfill \break
\hfill \break
   
\noindent This PDF file includes:
\begin{itemize}
\item Details on numerical methods and training of networks
\item More information on Neutron data analysis and treatments
\item Real-space spin configurations 
\item Phase map generation and SSF predictions using machine-learning
\item Additional predictions of dynamical structure factor
\item Details of ED calculations and comparisons with MC results
\item Formulation of the optimal region 
\item Comparison of machine-learned parameter set with some of the previously reported values
\item Magnetic specific heat
\end{itemize}
Supplementary Figures 1-13\\
Supplementary Tables I-III
\end{abstract}
\maketitle

\newpage

%%%%%%%%%%%%%%%%%%%%%%%%%%%%%%%%%%%%%%%%%%%%%%%
% Contents
%%%%%%%%%%%%%%%%%%%%%%%%%%%%%%%%%%%%%%%%%%%%%%%

\section{Details on numerical methods and training of networks}

\subsection{Monte Carlo and Landau-Lifshitz Solver for scattering}

Neutron scattering from the magnetic system directly probes the magnetic two-point correlations of the material and the scattering cross section can be formulated as:
\begin{equation}
\frac{d^2\sigma}{d\hbar\omega\Omega}=\frac{k_f}{k_i}r_m^2 \sum_{\alpha,\beta} \frac{g_{\alpha}g_{\beta}}{4} \left(\delta_{\alpha\beta}-
\frac{q_{\alpha}q_{\beta}}{q^2}\right)
|F(\textbf{Q})|^2
\mathcal{S}^{\alpha\beta}\left( \textbf{Q},\omega \right)
\label{eq:two_point}
\end{equation}
where $\textbf{Q}$ and $\omega$ are the wavevector and energy transfer in the scattering process, $k_i$ and $k_f$ are the initial and final wavevectors of the neutrons, $r_m$ is a scattering factor, $\alpha,\beta=x,y,z$ are Cartesian coordinates indicating initial and final spin polarization of the neutron, $F(\textbf{Q})$ is the magnetic form factor and $\mathcal{S}^{\alpha\beta}\left( \textbf{Q},\omega \right)$ is the spin correlation function. The pre-factors are set by measurement conditions or atomic properties of the material and so are straightforwardly evaluated allowing quantitative experimental determination of $\mathcal{S}\left( \textbf{Q}, \omega \right)$.  

The dynamical correlation function $\mathcal{S}^{\alpha\beta}(\textbf{Q},\omega)$ is equal to the Fourier transform of the spin-spin correlation functions in space and time:% and is equal to:
\begin{equation}
\mathcal{S}^{\alpha\beta}(\textbf{Q},\omega)=\frac{1}{2\pi N}
\sum_{i,j} e^{i\textbf{Q}.\left( \textbf{R}_j-\textbf{R}_i \right)}
\int_{-\infty}^{\infty} e^{-i\omega t} \langle S_i^{\alpha}(t_0)S_j^{\beta}(t_0+t) \rangle dt.
\label{eq:correlations}
\end{equation}
\begin{figure}[!htb]
\centering\includegraphics[width=0.5\textwidth]{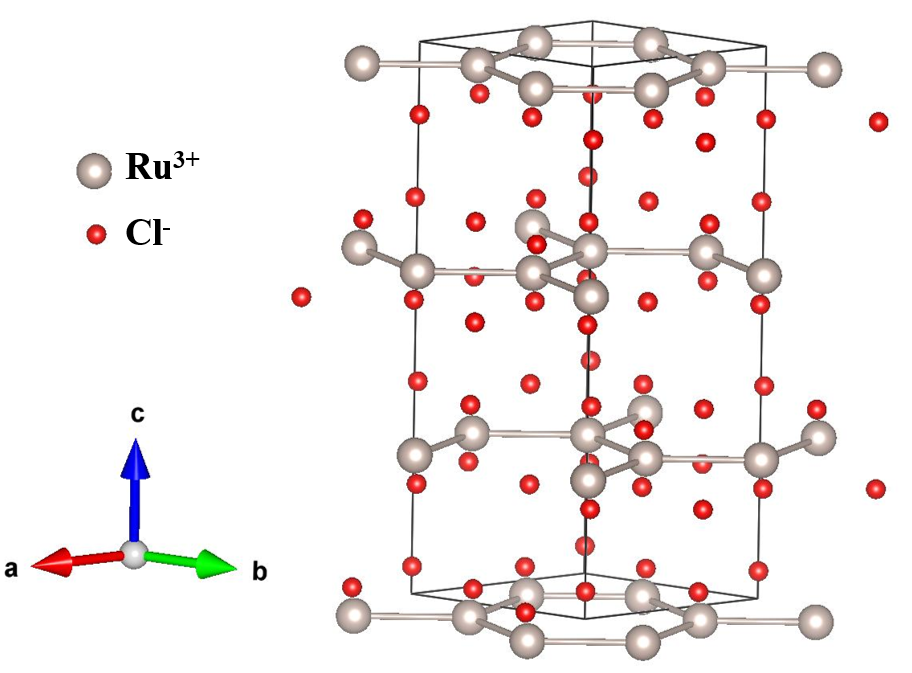}
% \centering\includegraphics[width=0.95\textwidth]{}
	\caption{
	The nuclear structure of RuCl$_3$ as reported in Refs. \cite{PhysRevB.95.174429, Mu2021}. The space group is {\it R$\bar{3}$} (148). Note that there are 6 Ru-ions per unit-cell corresponding to three stacked honeycomb layers. We used this structure in our MC/LLD simulations even though the Hamiltonian of this study does not contain out-of-plane exchanges. Thus, our simulation setup can be readily generalized to 3D models with inter-layer exchange interactions.   
	}
\label{figure_structure}
\end{figure}

The Monte Carlo solver used in Ref. \cite{Samarakoon2020} is deployed for a network of Heisenberg spins on a honeycomb lattice. Although we neglect out-of-plane exchanges in this study, the unit cell with three honeycomb layers shown in Fig.~\ref{figure_structure} was used throughout the simulations. The spins are treated as vectors $\mathbf{S}_i=[ S_i^x,S_i^y,S_i^z ]$ of fixed length of $\sqrt{S(S+1)}(=\sqrt{3}/2)$ at positions $\mathbf{R}_i$. The energy due to interactions, $\lbrace p \rbrace$ is given by the spin Hamiltonian, $\mathcal{H}=\mathcal{H}(\lbrace p \rbrace,\mathbf{S},\mathbf{R})$ given in Eq.~(1) of the main text. Spin configurations which are representative of the system  in thermal equilibrium at chosen temperature $T$ are found using simulated annealing based on the Metropolis algorithm, a Markov Chain Monte Carlo method \cite{Metropolis1953}. From these configurations a full range of physical properties can be calculated including $S^{\rm sim}(\textbf{Q})$. 
The time development of a thermally equilibrated spin configuration, $ \mathbf{S}(t)$ can be evaluated using the Landau-Lifshitz (LL) equations of motion which are readily derived through Poisson brackets involving $\mathbf{S}_i$ and  $\mathcal{H}$. 

% \begin{equation}
% {\dot{S}}_i\left({\{S\}}_{ij}\right)=\ {{dS_i}\over {dt}}=\ \gamma \left[S_i\times F_i\right]-\ \gamma \alpha \left[S_i\times \left(S_i\times F_i\right)\right]\ \ \ ;\ \ \ \ \ \ \ \ \ \ \ \ \ \ F_i=\ H_i\left({\{S\}}_{ij}\right)+\ {\xi }_i(t)
% \label{eq:LL}
% \end{equation}

\begin{equation}
\frac{d\mathbf{S}_i}{dt}= \gamma \left[S_i\times  \frac{\partial \mathcal{H}}{\partial \mathbf{S}_i}\right]
\label{eq:LL}
\end{equation}
The LL equation is solved numerically using a fourth-order Runge-Kutta algorithm with adaptive step size \cite{Huberman_2008}. The  $ \mathbf{S}(t)$
is calculated  for an appropriately spaced set of discrete times from $t_0$ to $t_N$ with spacing $\Delta t$ and used to calculate $\mathcal{S}^{\alpha\beta}(\textbf{Q},\omega)$. The $t_N$ is determined by the target energy resolution of 0.5 meV. For a given $\{p\}$, the $S(\textbf{Q}, \omega)$ was calculated from 256 independent simulations and averaged to yield good statistics. The actual measured cross section $\mathcal{S}^{\rm exp}\left( \nu \right)$ depends on experimental conditions including resolution and the MC/LLD solver then undertakes the transformation $\lbrace \mathcal{H}(p) \rbrace \rightarrow \mathcal{S}^{\rm sim}(\textbf{Q},\lbrace p \rbrace)$ to replicate expected scattering.

\subsection{Parameterization of the Hamiltonian and optimization}\label{app:parameterization}
The original Hamiltonian space is re-parameterized as follows to cover the entire parameter-space (both positive and negative sides of each parameter):
% \begin{align}
\begin{equation} \label{eq1}
\begin{split}
    % \begin{bmatrix}
    J_1 &= \left(\frac{U}{3}\right){\rm sin}(\phi){\rm sin}(\theta){\rm sin}(\beta){\rm sin}(\gamma),\\
    K &= \left(U\right){\rm cos}(\phi){\rm sin}(\theta){\rm sin}(\beta){\rm sin}(\gamma),\\
    \Gamma &= \left(\frac{U}{2}\right){\rm cos}(\theta){\rm sin}(\beta){\rm sin}(\gamma),\\
    J_2 &= \left(\frac{U}{6}\right){\rm cos}(\beta){\rm sin}(\gamma),\\
    J_3 &= \left(\frac{U}{3}\right){\rm cos}(\gamma),
    % \end{bmatrix}
\end{split}
\end{equation}
% \end{align}
where $U$ is the overall energy scale. 
The hyper-parameter space of $\{p\}=\{\phi, \theta, \beta, \gamma\}$ was explored with fixed $U =$ 10 meV, using the iterative mapping algorithm (IMA), which is a variant of the Efficient Global Optimization algorithm as described in Ref. \cite{Samarakoon2020}. The IMA is employed to minimize the multi-experiment error measure, ${\hat \chi}^2_{\rm com}={\hat \chi}^2_{S(\textbf{Q})}\times{\hat \chi}^2_{S(\textbf{Q},\omega)}$.  The ${\hat \chi}^2_{S(\textbf{Q})}$ and ${\hat \chi}^2_{S(\textbf{Q},\omega)}$ are the low-cost estimators of the mean squared error, $\chi^2_{S(\nu)}=\sum_{\nu}[S^{\rm exp}(\nu)-S^{\rm sim}(\nu)]^2$ with respect to static and dynamical structure factors. In contrast to the Gaussian process regression in Ref. \cite{Samarakoon2020}, we here calculate a low cost estimators using the surrogate networks to determine the squared distance between prediction and experimental data, (Eq. \ref{eq:low-cost}). 
\begin{equation} \label{eq:low-cost}
\begin{split}
{\hat \chi}^2_{S(\textbf{Q})}&=\sum_{\textbf{Q}}[S^{\rm exp}(\textbf{Q})-S^{\rm sur}(\textbf{Q})]^2\\
{\hat \chi}^2_{S(\textbf{Q}, \omega)}&=\sum_{\{\textbf{Q}, \omega\}}[S^{\rm exp}(\textbf{Q}, \omega)-S^{\rm sur}(\textbf{Q}, \epsilon\omega)]^2
\end{split}
\end{equation}
Here, $\epsilon$ renormalizes the energy scale $U$ and is found by minimizing ${\hat \chi}^2_{S(\textbf{Q}, \omega)}$ within the interval of $\epsilon\in[1/3,3]$ for a given Hamiltonian parameter set $\{p\}$. The $\epsilon$ is necessary to account for the overall energy scale. 

The IMA iteratively samples parameter space subject to the condition that ${\hat \chi}^2_{\rm com}$ is below the error tolerance threshold, $C_{\rm com}$, and the surrogates are retrained. As more data is collected, the prediction accuracy of the surrogates towards minimum of the ${\hat \chi}^2_{\rm com}$ becomes reliable.

\subsection{Training of Autoencoders}\label{app:autoencoder}
In this work, we have trained two Autoencoders; for static and dynamic structure factors. The training includes $S^{\rm Sim}(\textbf{Q})$ and $S^{\rm Sim}(\textbf{Q}, \omega)$ generated from the MC/LLD solver for random sets of Hamiltonian parameter produced by the IMA. By the last iteration of IMA, we have employed 10000 models for Hamiltonian  Eq.~(1). For each model, both $S^{\rm Sim}(\textbf{Q})$ and $S^{\rm Sim}(\textbf{Q}, \omega)$ are calculated. The training data includes simulated data of 90$\%$ of randomly selected existing samples and the remaining data was used as ML training validation. The autoencoder tries to minimize the deviation between its input $S^{\rm sim}(\textbf{Q})$ and filtered output, summed over all random models in the dataset.

The Autoencoder corresponding to $S(\textbf{Q})$, is constructed with three hidden layers of logistic neurons and latent space of 3-dimensions. The number of neurons for the three layers are empirically found to be 300-3-300. A single hidden layer autoencoder architecture of 200 linear neurons is found to be performing reasonably well with $S(\textbf{Q},\omega)$ data. 

Training the autoencoder corresponds to determining the weight matrix $(W)$ and bias $(b)$ for each layer including the output layer. A loss function, $\mathcal{L}$

\begin{eqnarray}
\mathcal{L}&=& \sum_{\{p\}}\left[\frac{1}{N_\nu}\sum_{\nu}(S^{\rm sim}(\nu) - S^{\rm sim}_{\rm AE}(\nu))^2+\mathcal{L}_R\right]
\end{eqnarray}
 is minimized using the Adam optimization algorithm as readily accessible in Keras \cite{chollet2015keras}, a deep learning API written in Python. The term $\mathcal{L}_R$ is relatively weak, and includes two types of regularization: An $L_2$ regularization on the weight matrices, and an activity regularizer for uncorrelated feature constraint. This regularization seems to improve the physical interpretability of the latent space representation. Despite having millions of trainable parameters in the neural network, the autoencoder does not seem prone to overfitting; the low-dimensionality of the latent space itself acts as a strong regularizer.

\subsection{Training of the generative model}\label{app:generative}
 
 The RBN is constructed with two layers: A layer of radial basis (RB) neurons followed by an output layer of logistic neurons. The latent space predictions for a given parameter set $\lbrace p \rbrace$ are defined as: \begin{eqnarray}
\mathcal{L}_i&=& f_2(w^{(2)}_{ij}.h_j(\lbrace p \rbrace) + b_2)
\end{eqnarray}\begin{eqnarray}h_j(\lbrace p \rbrace) = exp\left\lbrack -\frac{\sum_{k} (p_k - c_{jk})^2}{\sigma^2} \right\rbrack\end{eqnarray}where  $f_2$ is the logistic activation  function $f(x) = 1/(1+e^{-x})$  similar to the output layer of the NLAE {\it Encoder}. The weight matrix $w^{(2)}_{ij}$, and bias vector $b_2$ of the output layer and the clustering centers $c_{jk}$  of the RB layer are to be determined from the training process. The spread of RB function, $\sigma$  is preset to 0.05. The network is trained with the $S^{\rm sim}(L)$  as the target and the corresponding  $\lbrace p \rbrace$ as the input. Thus the input and the output dimensionality is set by the dimensionality of the $\lbrace H(p)\rbrace$ and the $\mathcal{L}_Q$. The number of neutrons in the RB layer is determined during the training process. The training starts with no neurons in the hidden layer and iteratively adding neurons to minimize the error between output and the target.

%\subsection{Experiment Details}\label{app:experiment}
%PL: These are in the main text
%
%\subsubsection{Neutron Diffraction Experiment}
% Christian's text
%Elastic neutron studies were performed at Spallation Neutron Source (SNS) using the CORELLI beamline. A 125 mg $\alpha$-RuCl$_3$ crystal was mounted on an aluminium plate and aligned with the (H,0,L) plane horizontal. The crystal was rotated through $170$ degrees in $2^\degree$ steps about the vertical axis. The temperature of the measurement was 2 K and the perpendicular wave vector transfer was integrated in the range L = [0.92, 1.08] r.l.u..
%
%
%\subsubsection{Neutron Spectroscopy Experiment}
% Christian's text
%Inelastic neutron scattering was performed on a single crystal with a mass of 0.7 g. This was sealed in a thin-walled aluminium can with 1 atmosphere of Helium gas for thermal contact. Measurements were carried out using the SEQUOIA spectrometer at the Spallation Neutron Source (SNS). The incident energy was set to $E_i = 22.5$~meV and the temperature to 4 K. The crystal was mounted with the (H, 0, 0) and (0, 0, L) axes in the horizontal plane, and the orthogonal (0.5K, -K, 0) axis pointing vertically upwards. Data were collected by rotating the crystal about the vertical axis over $290^\degree$ in $1^\degree$ steps. The data are integrated over ranges (0.5K, -K, 0)=[-0.05, 0.05] and (0, 0, L)=[-2, 2].

\section{More information on Neutron data analysis and treatments}
{
Fig. \ref{figure_experiment}(a) shows a 2D slice of 4D raw data from the inelastic neutron experiment executed at the Fine-Resolution Fermi Chopper Spectrometer (SEQUOIA), SNS, ORNL, under temperature of $\sim$ 4 K and zero field conditions. The level of experimental noise is comparable to the magnetic signal even though we have integrated over a wide range of $l=0\pm 3.5$ . %Therefore, we a 3D Gaussian filtering to the $l$-integrated 3D data volume as a pre-processing step to the ML integrated workflow. 
We therefore perform a 3D Gaussian filtering on the $l$-integrated 3D data volume as a pre-processing step to the ML integrated workflow. 
As shown in Fig. \ref{figure_experiment}(b), the Gaussian filter helps to get rid of statistical noise up to some extent and makes the spectrum more clearer.  A Gaussian kernel with standard deviation ($\sigma$) of 1 was used here. A lower value of $\sigma$ will not do better on statistical noise while a higher value would make the features broader than instrument resolution.

\begin{figure}[!htb]
\centering\includegraphics[width=0.8\columnwidth]{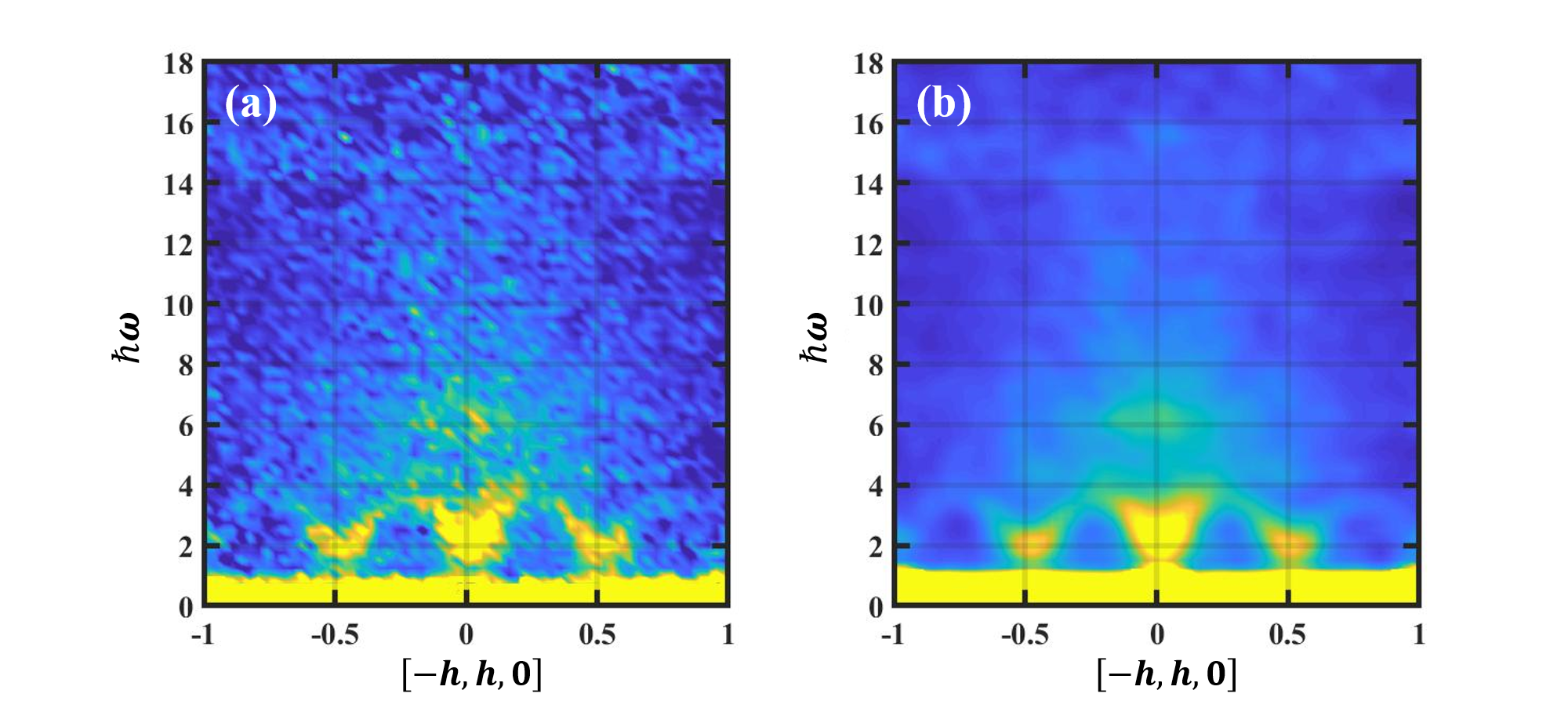}
% \centering\includegraphics[width=0.95\textwidth]{}
	\caption{
	Pre-processing treatment on SEQUOIA data. (a) the raw data sliced in $[-h,h,0]$ and $\hbar\omega$ plane with the integration of$[k,k,0] :k=0\pm0.02$,   $[0,0,l]:l = 0\pm 3.5$ and (b) the corresponding slice of smoothed data are shown here. The data was collected at ~4 K with a single crystal of mass ~0.7 g. % as reported in the Ref. \cite{?}. 
	The data was smoothed by filtering $l$-integrated 3D image with a 3D Gaussian smoothing kernel with standard deviation of 1.0. The Gaussian filter helps to get rid of statistical noise and sharpen up the features in the raw data.  
	}
\label{figure_experiment}
\end{figure}
}
\newpage
%%%%%%%%%%%%%%%%%%%%%%%%%%%%%%%%%%%%%%%%%%%%%%%%%%%%%%%%%%%%%%%%%%%%%%%%%%%%%%%%%%%%%%%%%%%%%%%%%%
\section{Real-space spin configurations}
\begin{figure}[!htb]
\centering\includegraphics[width=0.6\columnwidth]{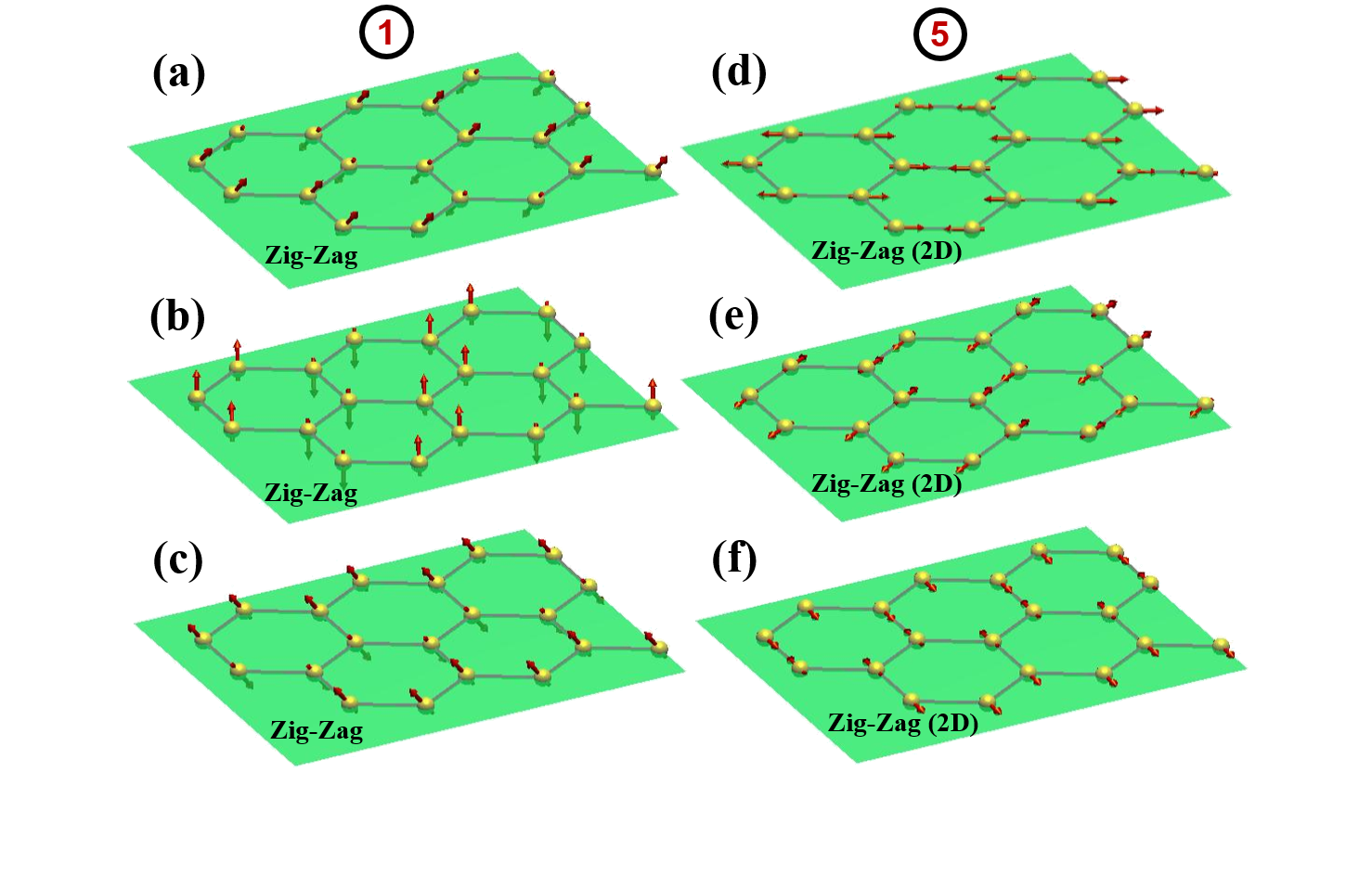}
% \centering\includegraphics[width=0.95\textwidth]{}
	\caption{
	Magnetic structures of the two zig-zag (Z.Z.) phases mentioned in the main text. Panel (a), (b) and (c) show the three of the six magnetic domains for the 3D Z.Z. phase. Panel (d), (e) and (f) are the three of the six magnetic domains for the 2D Z.Z.  phase. The yellow spheres and red arrows represents the Ru-ions and its magnetic moments respectively. The green plane cuts through the vertical center of the lattice for better perspective on the out of plane spin components. Both phases can be found in parameter space plane shown in Fig. 2(a). $\Gamma>0$ favors 3D Z.Z., while $\Gamma<0$ favors the 2D Z.Z. phase. 
	%The anti-ferromagnetic $\Gamma (>0)$ favors 3D Z.Z. while ferromagnetic $\Gamma (<0)$ favors the 2D Z.Z. phase. 
	The magnetic configuration of both phases consist of zig-zag chains, but the spin orientations different. The directions of zig-zag chains in 3D Z.Z. phase aligns with the principal axis of the anisotropic models while it is in-plane and perpendicular to the propagation vector of the chains. The corresponding labels to map the $S(\textbf{Q})$ shown in Fig. 2 are also displayed. Due to the difference in spin orientations, neutron structure factor is also distinguishable.   
	}
\label{figure_MagStrZZ}
\end{figure}

\begin{figure}[!htb]
\centering\includegraphics[width=0.6\columnwidth]{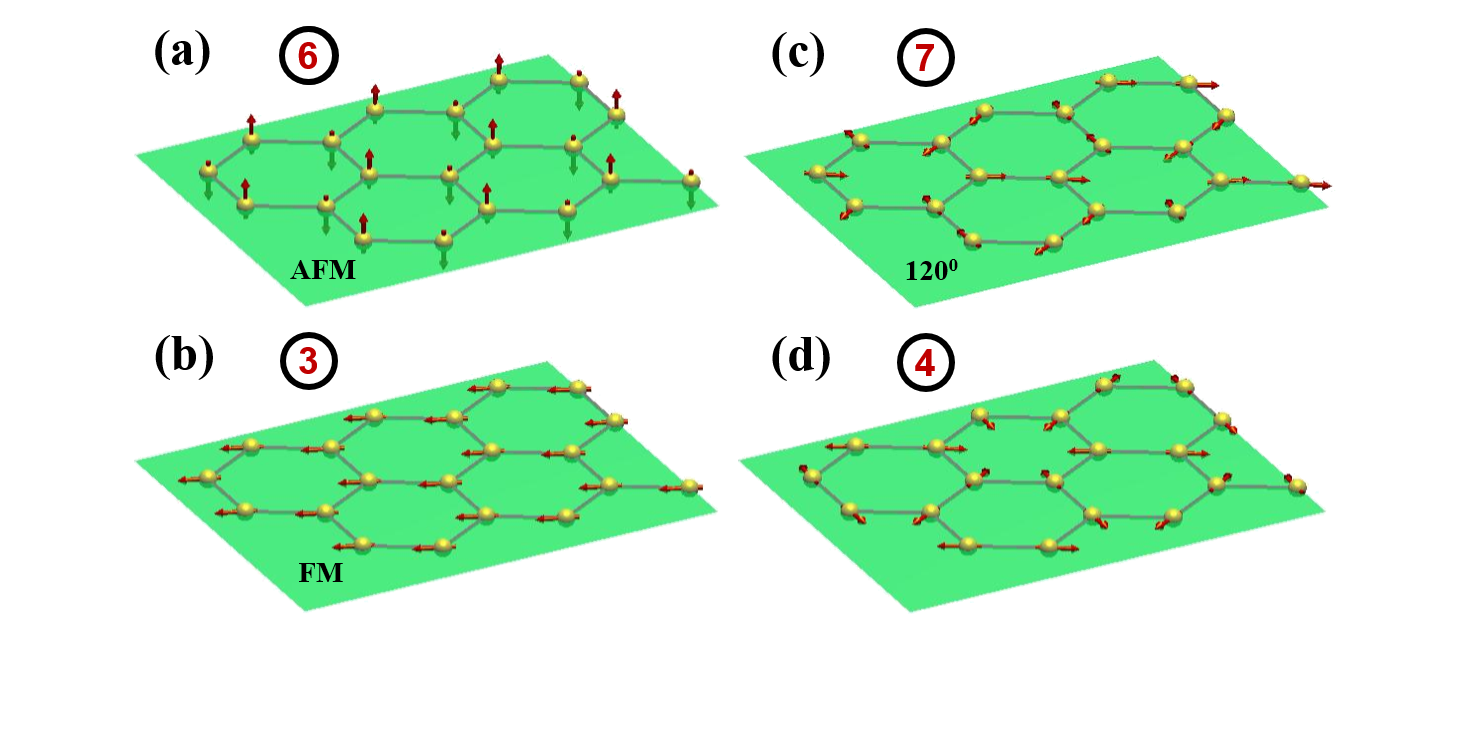}
% \centering\includegraphics[width=0.95\textwidth]{}
	\caption{
	Magnetic structure of some of the phases shown in Fig. 2. The spin configuration for the $S(\textbf{Q})$ labeled as (a) 6 , (b) 3, (c) 7 and (d) 4 in Fig. 2 are shown here. The yellow spheres and red arrows represents the Ru-ions and its magnetic moments respectively. The green plane cuts through the vertical center of the lattice for better perspective on the out of plane spin components. The corresponding phases for label 6, 3 and 7 can be easily identify as anti-ferromagnetic, ferromagnetic and 120$^\degree$ orders respectively. The spin-structure corresponding to the label 4 [panel (d)] is planar with ``6-in'' type plaquettes forming a triangular network. 
	}
\label{figure_MagStrOther}
\end{figure}

Fig.~\ref{figure_MagStrZZ} compares the real-space spin configurations of the 3D and planar zigzag phases. Fig.~\ref{figure_MagStrOther} shows spin configurations for some other phases identified in Fig.~2.

\newpage
%%%%%%%%%%%%%%%%%%%%%%%%%%%%%%%%%%%%%%%%%%%%%%%%%%%%%%%%%%%%%%%%%%%%%%%%%%%%%%%%%%%%%%%%%%%%%%%%%%%%%%%%
\section{Phase map generation and SSF predictions using machine-learning}

\begin{figure}[!htb]
% \centering\includegraphics[width=\columnwidth]{FigureS03_ML.jpg}
%\centering\includegraphics[width=\columnwidth]{FigureS03_ML.png}
\centering\includegraphics[width=0.9\columnwidth]{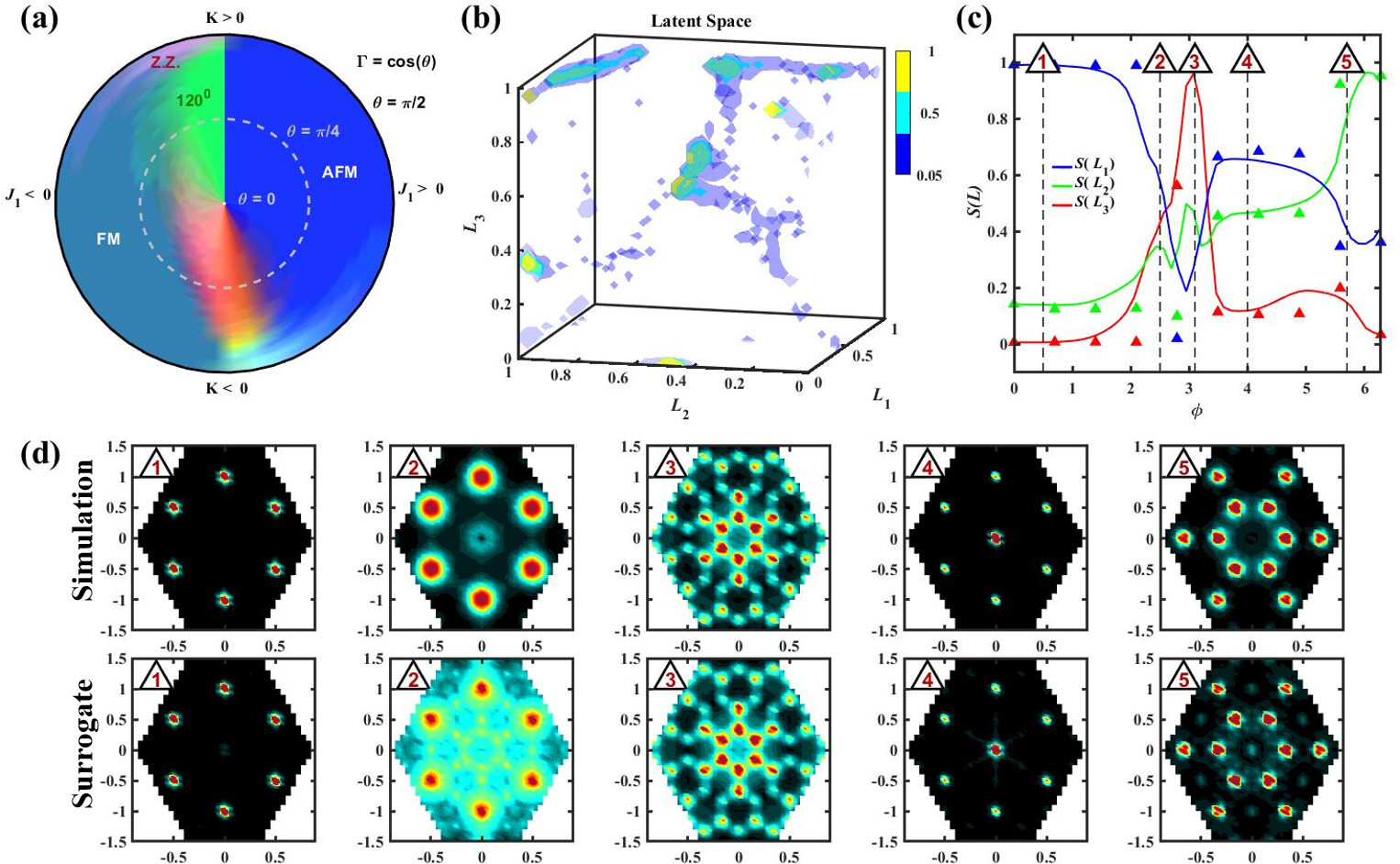}
% \centering\includegraphics[width=0.95\textwidth]{}
	\caption{
	Latent space representations and generative model predictability. (a) A predicted map of magnetic orderings in the $J_1-K-\Gamma$ space on the hyper-surface of $ J_1^2 + K^2 + \Gamma^2 = 1$ at a fixed temperature of $T/|K|=0.02$. The phase map is generated for $\rm \Gamma>0$ and depicted according to $\rm \Gamma=cos(\theta)$. (b) 3D histogram (density map) of the $\{p\}$ manifold in latent space. The surfaces are
the 3D contours of the histogram and their color represents the bin-height (density) used to generate the contour. In this latent space representation, a broad phase in parameter space appears as a dense point and a higher order phase transition / crossover appears as lines of points connecting dense clusters. First-order phase transitions appear as isolated clusters and discontinuities in the latent space projection \cite{Samarakoon2021}. (c) $S(L)$ for (filled triangles) the simulated structure factors calculated along the white dashed line shown in panel (a) at $\theta=\pi/4$ along with (solid lines) the prediction from the GN. (d) The high-symmetry-plane slices of simulated and surrogate-predicted $S(\textbf{Q})$ data at multiple places in parameter space as indexed on panel (c).
	}
\label{figure_ML}
\end{figure}

As explained in the main text, an Autoencoder with latent space of $N_L = 3$ is trained to compress $S(\textbf{Q})$ data and used to predict phase maps of magnetic ordering (see Fig. \ref{figure_ML}(a)).  The $S(\textbf{Q})$ (and consequently $S(L)$) encodes correlations of the systems and provides natural classification with an adequate size of reciprocal space. %On the  
Parameter sets corresponding to the same structure would cluster together in either $\textbf{Q}$ or $L$ space. (see Fig. \ref{figure_ML}(b)). Thus, it is technically possible to perform a clustering analysis to identify and label different clusters. However, such clustering analysis would fail when the system undergoes continuous transitions or crossovers rather than abrupt first-order like changes. In this case it is still possible to easily construct a graphical phase diagram. If the $\textbf{Q}$ space can be reduced to an $L$ space with $N_L = 3$, then the latent vectors can be treated as the RGB color components of a phase map.

In order to construct high-dimensional phase maps, we used a {\it Generator} network (see Fig. 3(b)) to predict $S_{\rm GN}(L)$ for the parameters set located in between existing $\{p\}$samples. Figure \ref{figure_ML}(a) is generated in such fashion in the hyper plane of $\sqrt{J_1^2+K^2+\Gamma^2}=1$ and $J_2=J_3=0$ and Figure \ref{figure_ML}(c) shows the $S_{\rm GN}(L)$ in comparison to $S(L)$ not included in GN training, along the dashed circle in panel (a). 

As shown in Fig. 3 (b) and (c), a faster surrogate for the MC/LLD solver can be constructed by connecting the GN with the {\it Decoder} network of the Autoencoder. Figure \ref{figure_ML} (d) shows the surrogate-predicted structure factors, $S^{\rm sur}(\textbf{Q})$ in comparison to the simulated for the parameter sets, not used as a part of training data. The surrogate-predictions are worse in some regions where the the $S^{\rm sim}(\textbf{Q})$ change rapidly and the $\{p\}$ samples are taken sparsely. The predictability of both {\it Decoder} and the {\it Generator} contribute the total prediction error of the surrogate. As more $\{p\}$ are evaluated on demand in the regions of interest, the surrogate predictions will be more reliable. Even though the surrogate predictions are not accurate in some regions, it can be used to do guide parameter space searches and even draw quick conclusions for experiment planning. However, any prediction leading to critical decisions or conclusion, should always be validated with direct calculations.

\newpage
%%%%%%%%%%%%%%%%%%%%%%%%%%%%%%%%%%%%%%%%%%%%%%%%%%%%%%%%%%%%%%%%%%%%%%%%%%%%%%%%%%%%%%%%%%%%%%%%%%%%%%%%%%%%%%%%%%
\section{Additional predictions of dynamical structure factor}
\begin{figure}[!htb]
\centering\includegraphics[width=\columnwidth]{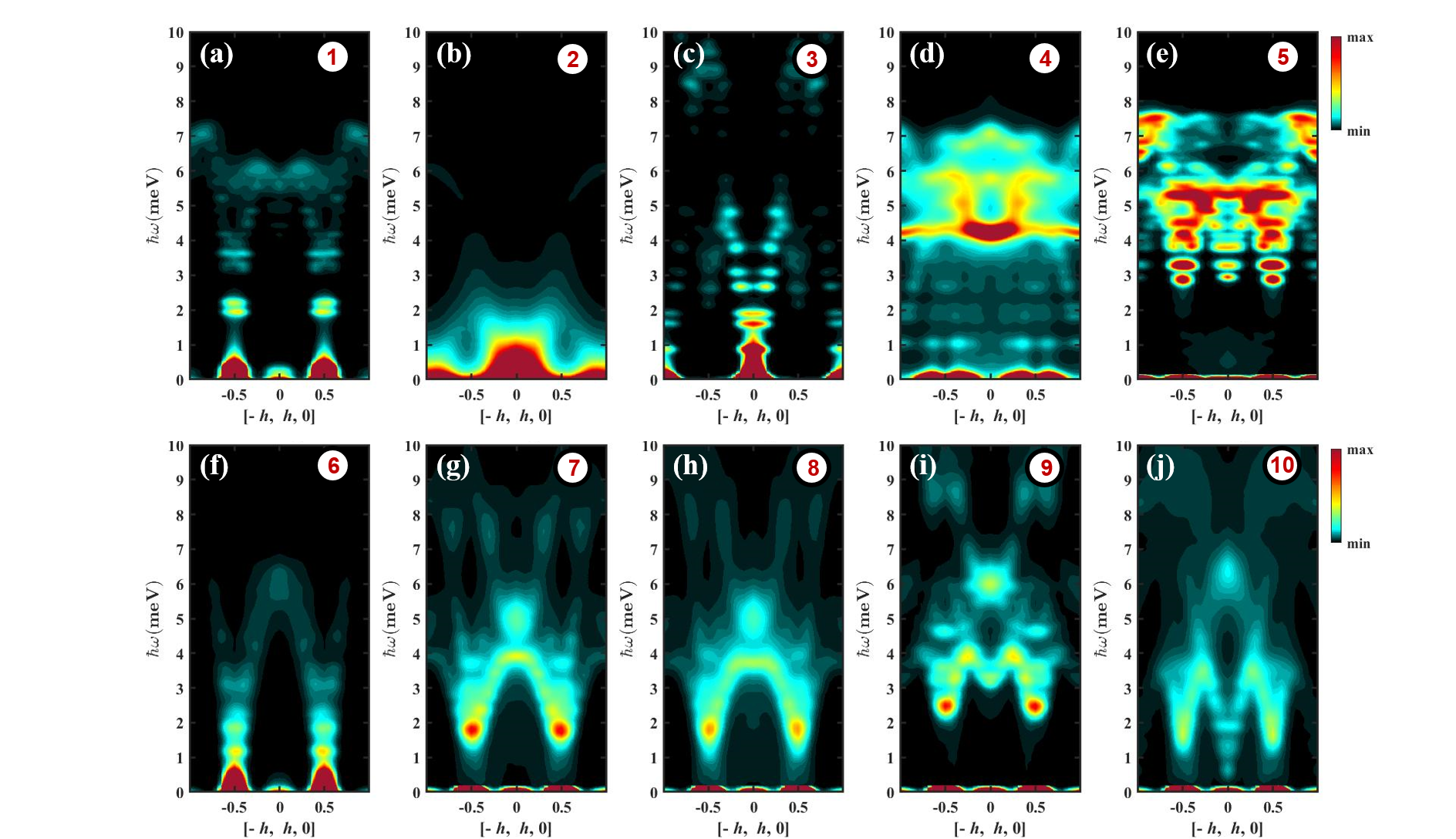}
% \centering\includegraphics[width=0.95\textwidth]{}
	\caption{
	The dynamical structure factors for multiple parameter sets. All the panels shows $[-h,h,0]-\hbar\omega$ slices of $S^{\rm Sur}(\textbf{Q}, \omega) $ at $[k,k,0] :k=0$,   $[0,0,l]:l = 0$  for the parameter sets listed in the Table \ref{tbl_sqwsets:1}. The ratios between parameter for the $S^{\rm Sur}(\textbf{Q}, \omega) $ shown in panel (a)-(e) also match for the sets labeled as 1-5 in the Fig. 2(a). 
	}
\label{figure_SQW}
\end{figure}

Fig.~\ref{figure_SQW} shows predicted dynamical structure factors for the parameter sets specified in Table~\ref{tbl_sqwsets:1}.

\begin{table}[!htb]
% \begin{tabular}{|c c c c c c|} 
\begin{tabular}{| m{2em} | m{3em} | m{3em} | m{3em} | m{3em} | m{3em}|} 
 \hline
 Set $\#$  &$J_1$ (meV)&$K$ (meV)& $\Gamma$ (meV) & $J_2$ (meV) & $J_3$ (meV)\\ [0.5ex] 
 \hline
   
   1 & -0.28  &   -1.39   &   0.87  &   0   &   1.75 \\
 \hline
  2 &  -0.47   &  -2.33   &   1.89   &  -0.07   &   0.04 \\
 \hline
  3 &  -0.21   &  -2.50   &   0.99   &   0.08   &  -1.58 \\
 \hline
  4 &   0.12   &  -3.40   &  -2.23   &   0.01   &   0.43 \\
 \hline
   5 & -0.28   &  -1.74   &  -0.83   &  -0.04   &   1.71 \\
 \hline
    6 &  -0.04   &  -1.66   &   0.99   &   0.04   &   0.11 \\
 \hline
  7 &  -0.34   &  -7.72   &   2.34   &  -0.1   &   0.74 \\
 \hline
   8 & -0.52   &  -5.02   &   1.46   &  -0.12   &   0.74 \\
 \hline
  9 &  -0.84   &  -4.45   &   0.92   &  -0.08   &   1.05 \\
 \hline
  10 &  -0.18   &  -5.89    &  0.15 &  -0.23  &  1.45 \\
 \hline

\end{tabular}
\caption{The corresponding parameter sets for the dynamical structure factors shown in Fig. \ref{figure_SQW}.}
\label{tbl_sqwsets:1}
\end{table}

\newpage
%%%%%%%%%%%%%%%%%%%%%%%%%%%%%%%%%%%%%%%%%%%%%%%%%%%%%%%%%%%%%%%%%%%%%%%%%%%%%%%%%%%%%%%%%%%%%%%%%%%%%%%%%%%
\section{Details of ED calculations and comparisons with MC results}
\begin{figure}[!htb]
\centering\includegraphics[width=0.3\textwidth]{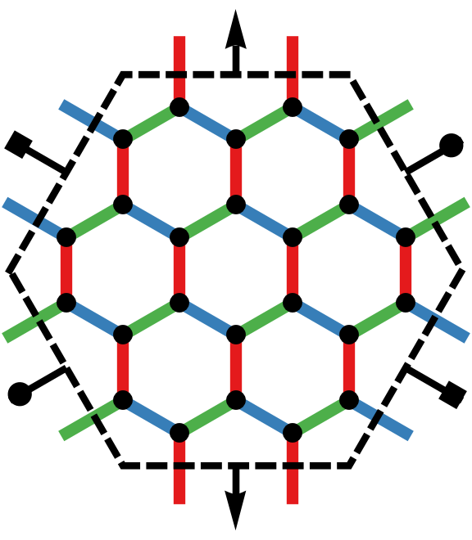}
% \centering\includegraphics[width=0.95\textwidth]{}
	\caption{
	Finite size cluster with periodic boundary conditions used for Lanczos ED calculations. Sides with matching symbols are identified. This figure is reproduced without changes from Ref.~\cite{Laurell2020}; licensed under a Creative Commons Attribution (CC BY) license \footnote{\url{http://creativecommons.org/licenses/by/4.0/}}.
	}
\label{figure_EDnet}
\end{figure}

\begin{figure}[!htb]
\centering\includegraphics[width=0.8\textwidth]{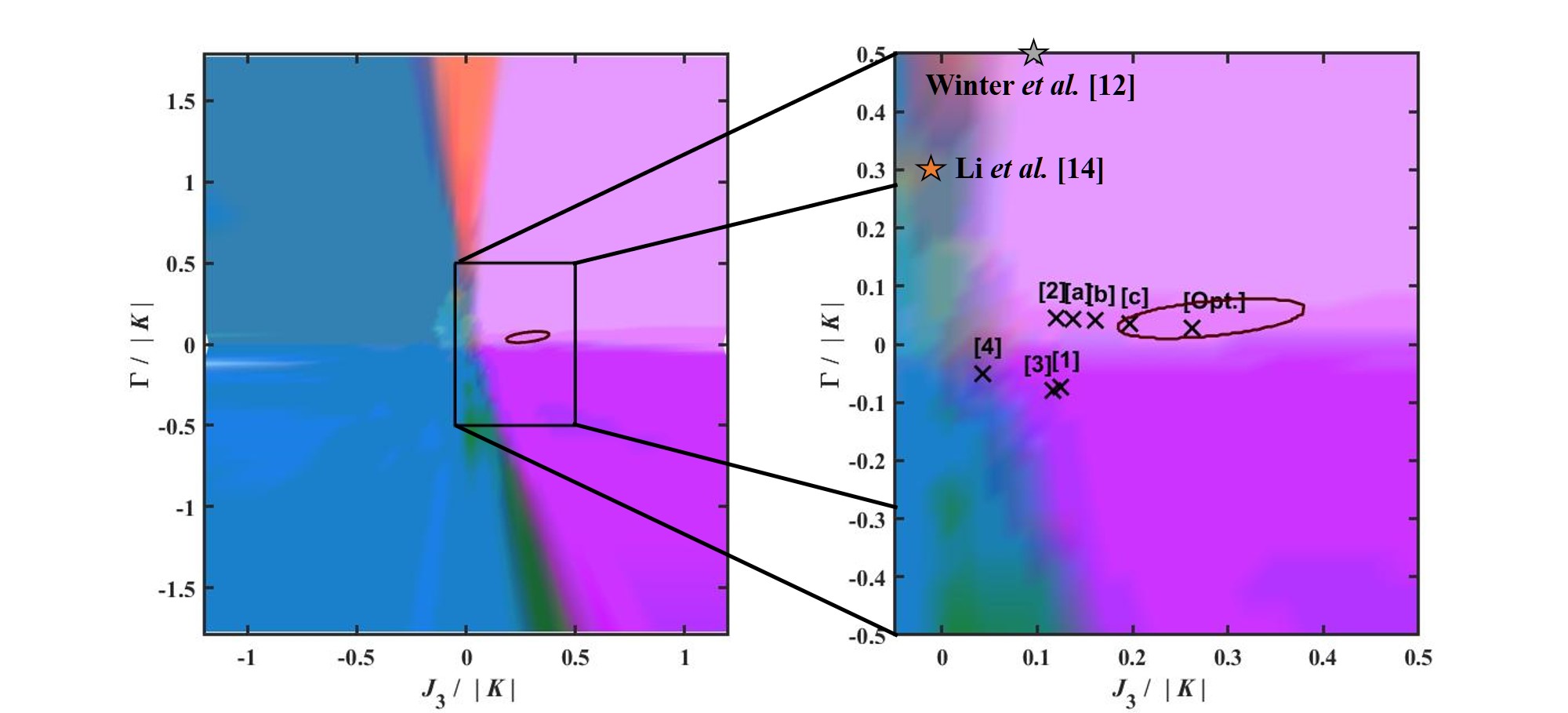}
% \centering\includegraphics[width=0.95\textwidth]{}
	\caption{
	The parameters sets listed in Table \ref{tbl_valsets:1} are shown in the machine-learned phase map varying $\Gamma/|K|$ and $J_3/|K|$  through the optimal solution for $\alpha$-RuCl$_3$ at fixed $J_1/|K|=-0.1$ and $J_2/|K|=0$, with $K<0$. Parameter sets 1-4 were chosen to be located close to phase transitions, and parameter sets a-c were chosen to lie on a line connecting the optimized parameter set with parameter set 2.
	}
\label{figure_GJ3PhaseMap}
\end{figure}

Lanczos exact diagonalization calculations were performed at zero temperature using the $\mathcal{H}\Phi$ library \cite{Kawamura2017} on the $24$-site $C_3$-symmetric finite size cluster shown in Fig.~\ref{figure_EDnet}. It has the same rotational symmetry as the full honeycomb lattice. Static structure factors were evaluated directly as ground state expectation values, whereas dynamical structure factors were calculated using the continued fraction expansion method (CFE) \cite{RevModPhys.66.763}. As in Ref.~\cite{Laurell2020} we used 500 Lanczos steps in the CFE. We used a Lorentzian broadening $\eta$ of $0.05$ meV (half width at half maximum) for the optimized parameters, or $0.05/sqrt(3)$ meV for parameter sets 1-4 defined in Table~\ref{tbl_valsets:1} and represented visually in Fig.~\ref{figure_GJ3PhaseMap}. The broadening was intentionally made small to allow experimental resolution effects to be implemented as in the LLD calculations. The frequency step size was $\Delta \omega = \eta/5$.

\begin{figure}[!htb]
\centering\includegraphics[width=\columnwidth]{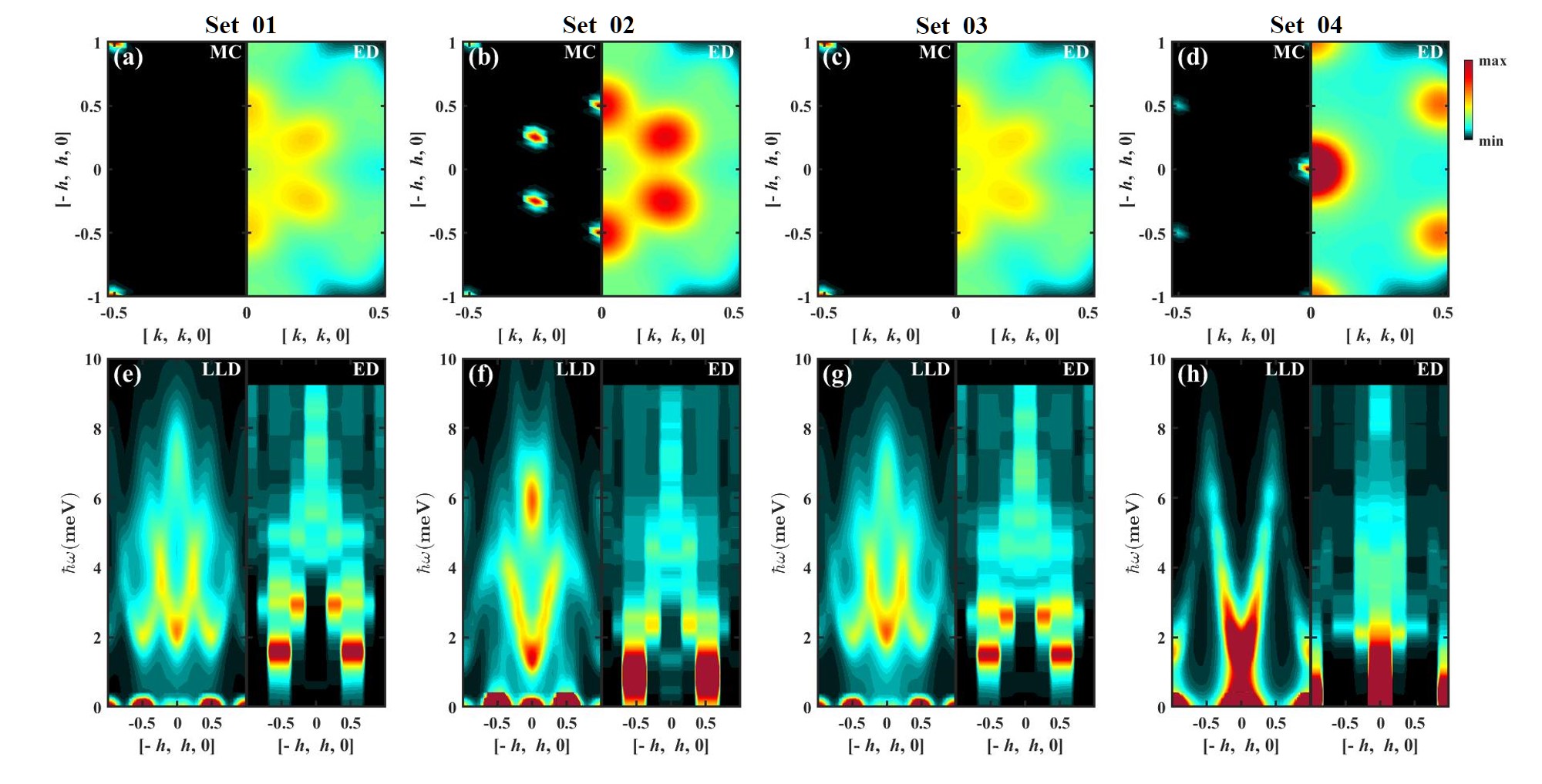}
% \centering\includegraphics[width=0.95\textwidth]{}
	\caption{
	The validation of MC/LLD with Lanczos ED for four different parameter sets listed in Table \ref{tbl_valsets:1}.  The comparisons of (a - d) static and (e - h) dynamic structure factors from (left side) Monte-Carlo simulations/LLD and (right side) ED are shown. The ED calculations use a 24-site cluster as shown in Fig. \ref{figure_EDnet} and structure factors are calculated up to the 2$^{\rm nd}$ Brillouin Zone of the honeycomb lattice.  The MC/LLD structure factors are calculated on a super cell of $20\times20$ RuCl$_3$ nuclear unit cells (equivalent to 2400 sites).  Even though the the Q-resolutions of two calculations are not comparable, the magnetic peak structure in $S(\textbf{Q})$ agrees well for sets (b)2 and (d)4 but not for set (a)1 or (c)3. According to the MC/LLD calculations, the corresponding magnetic structure for set 1 and 3 is 2D Z.Z.. For this phase, M-point peaks in the 1$\rm ^{st}$ B.Z. are not allowed by the polarization factor, but M-point peaks appear in higher Brillouin zones, which are not covered in the ED calculation. Thus, the diffuse patterns in the $S_{\rm ED}(\textbf{Q})$ for set 1 and 3 are due to the finite-size effect.
	}
\label{figure_validation}
\end{figure}

\begin{figure}
\centering\includegraphics[width=\columnwidth]{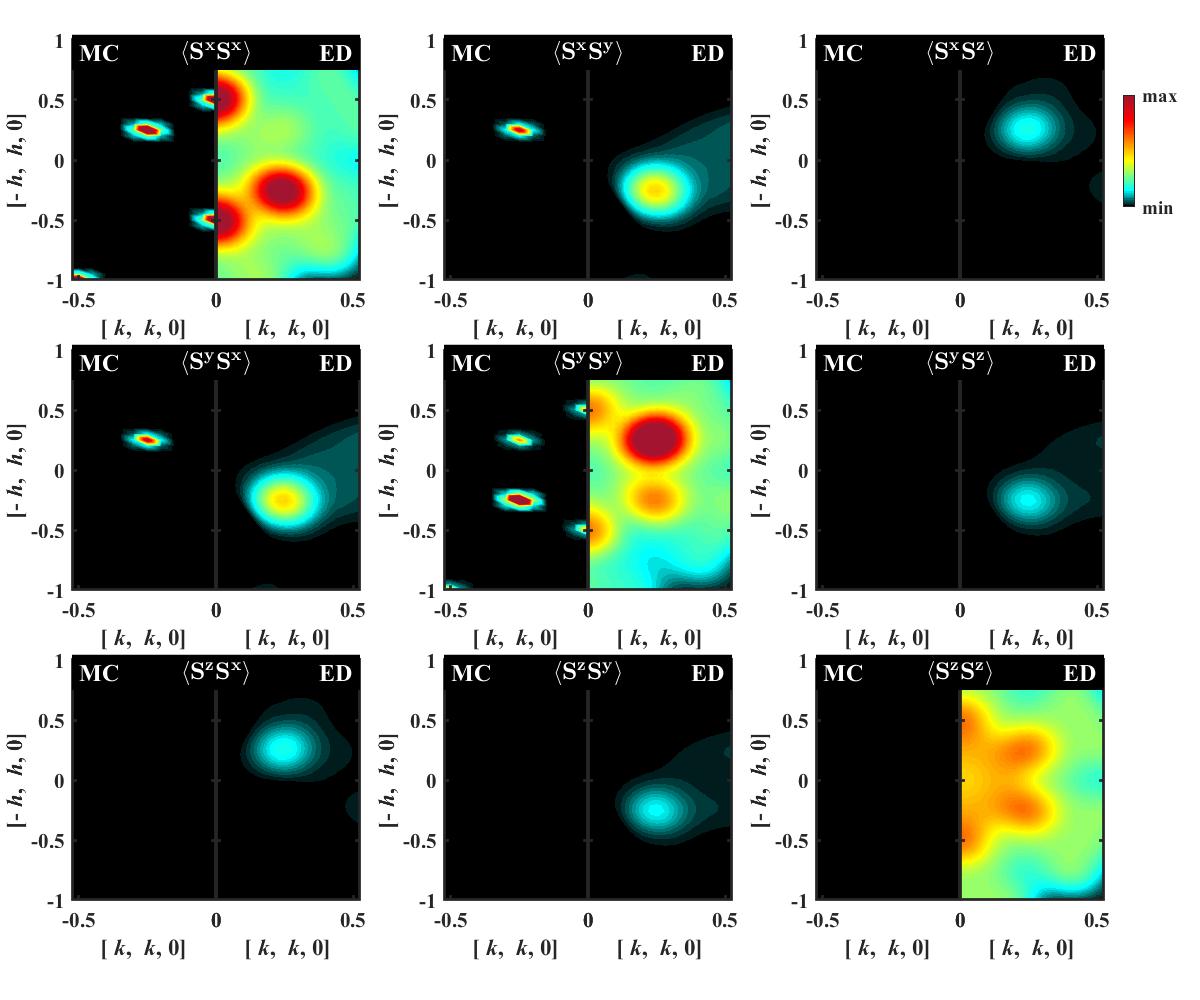}
% \centering\includegraphics[width=0.95\textwidth]{}
	\caption{
	The comparison of the individual SSF matrix elements, $\langle S^{\alpha}S^{\beta}\rangle$ for the parameter set 1 listed in Table  \ref{tbl_valsets:1}.
	}
\label{figure_Com_set01}
\end{figure}

\begin{figure}
\centering\includegraphics[width=\columnwidth]{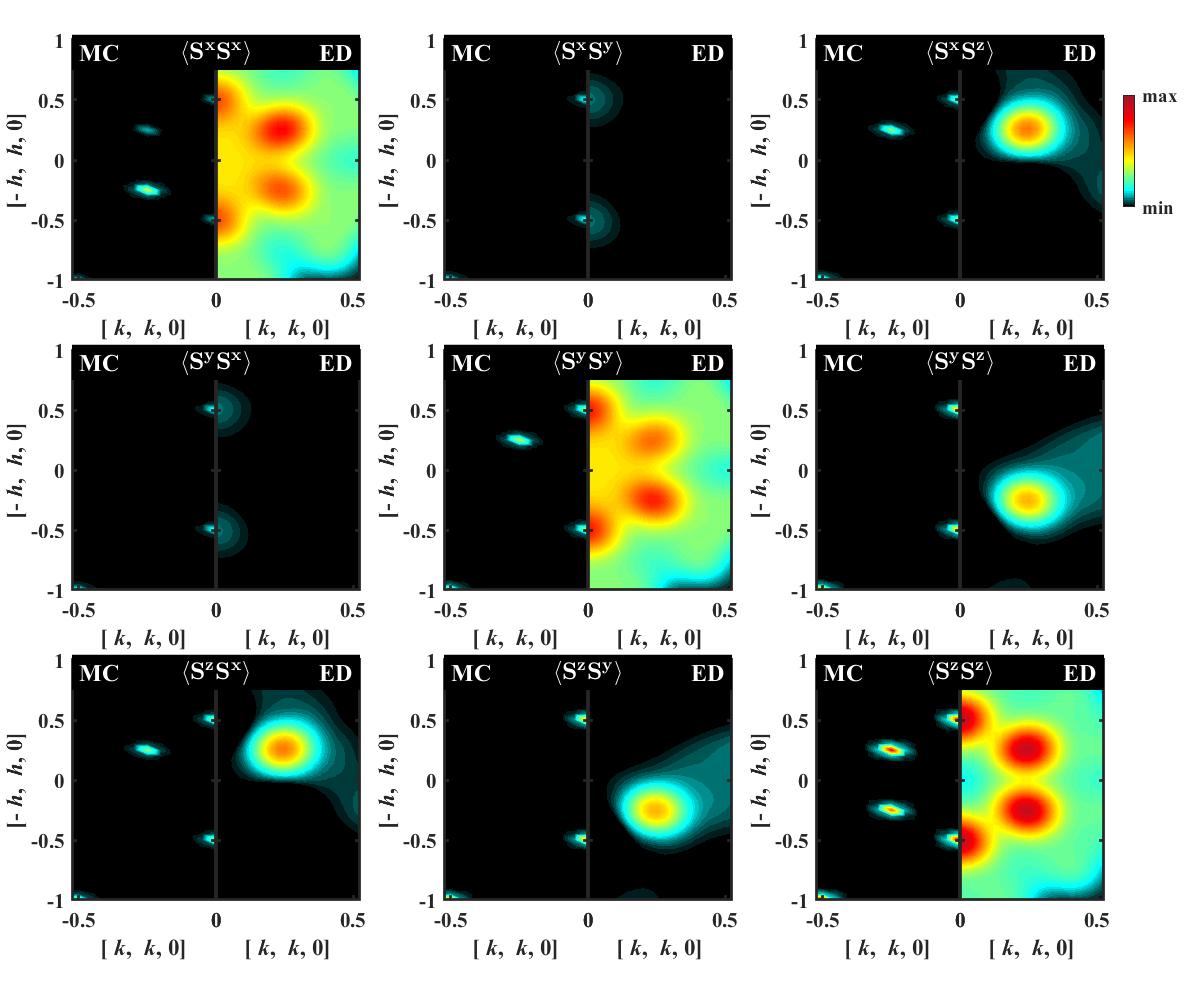}
% \centering\includegraphics[width=0.95\textwidth]{}
	\caption{
	The comparison of the individual SSF matrix elements, $\langle S^{\alpha}S^{\beta}\rangle$ for the parameter set 2 listed in Table  \ref{tbl_valsets:1}.
	}
\label{figure_Com_set02}
\end{figure}

\begin{figure}
\centering\includegraphics[width=\columnwidth]{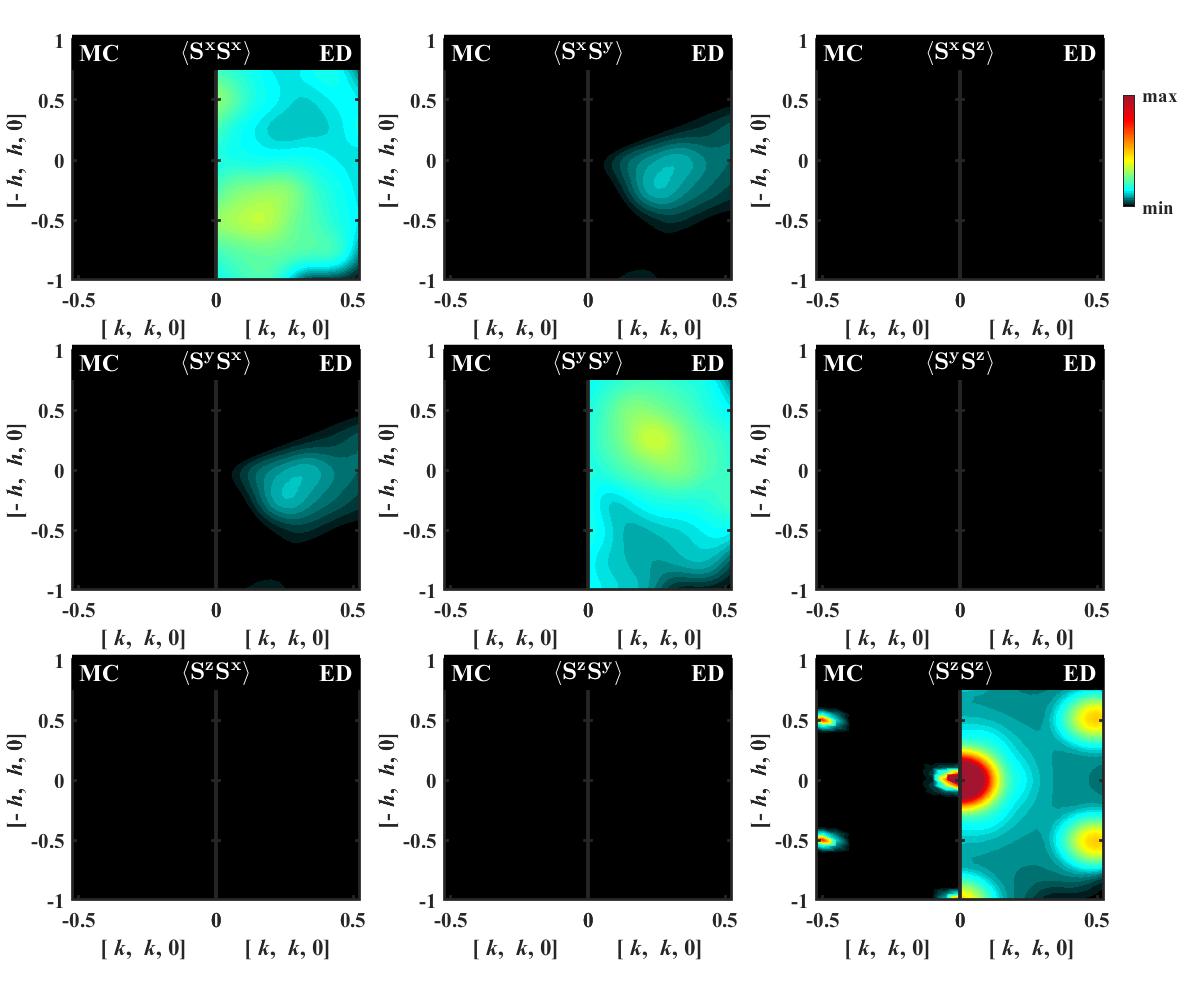}
% \centering\includegraphics[width=0.95\textwidth]{}
	\caption{
	The comparison of the individual SSF matrix elements, $\langle S^{\alpha}S^{\beta}\rangle$ for the parameter set 4 listed in Table  \ref{tbl_valsets:1}.
	}
\label{figure_Com_set04}
\end{figure}

Comparisons of $S(\mathbf{Q})$ and $S(\mathbf{Q},\omega)$ are shown in Fig.~\ref{figure_validation}. A comparison of the individual matrix elements, $\langle S^{\alpha}S^{\beta}\rangle$ is also shown in Fig. \ref{figure_Com_set01}, \ref{figure_Com_set02} and \ref{figure_Com_set04} for parameter sets 1,2,4 listed in Table~\ref{tbl_valsets:1}.

\begin{table}[!htb]
% \begin{tabular}{|c c c c c c|} 
\begin{tabular}{| m{2em} | m{3em} | m{3em} | m{3em} | m{3em} | m{3em}|} 
 \hline
 Set $\#$  &$J_1$ (meV)&$K$ (meV)& $\Gamma$ (meV) & $J_2$ (meV) & $J_3$ (meV)\\ [0.5ex] 
 \hline
   1 & -2.1 &  -13.6   & -1.0   & -0.2   &  1.7 \\
 \hline
   2 & -2.5 &  -13.3    & 0.6   & -0.2   &  1.6 \\
 \hline
   3 & -2.0  & -13.8   & -1.1  &  -0.2   &  1.6 \\
 \hline
   4 & -2.5  & -13.9   & -0.7  &  -0.1   &  0.6 \\
   \hline
   a & -1.98  & -11.29   & 0.49  &  -0.19  &  1.55 \\
   \hline
   b & -1.45  & -9.29   & 0.38  &  -0.195   &  1.49 \\
   \hline
   c & -0.93  & -7.28   & 0.26  &  -0.193   &  1.435 \\
   \hline
   Opt. & -0.4  & -5.27   & 0.15  &  -0.19   &  1.38 \\[1ex] 
 \hline
\end{tabular}
\caption{Parameter sets used to compare ED calculation with LLD. }
\label{tbl_valsets:1}
\end{table}

\pagebreak

\pagebreak
\hfill \break
\hfill \break
\hfill \break
\hfill \break
\hfill \break
\hfill \break
\hfill \break
\hfill \break
\hfill \break
\hfill \break
\hfill \break
\pagebreak

\pagebreak
\hfill \break
\hfill \break
\hfill \break
\hfill \break
\hfill \break
\hfill \break
\hfill \break
\hfill \break
\hfill \break
\hfill \break
\hfill \break
\pagebreak

\pagebreak
\hfill \break
\hfill \break
\hfill \break
\hfill \break
\hfill \break
\hfill \break
\hfill \break
\hfill \break
\hfill \break
\hfill \break
\hfill \break
\pagebreak

%%%%%%%%%%%%%%%%%%%%%%%%%%%%%%%%%%%%%%%%%%%%%%%%%%%%%%%%%%%%%%%%%%%%%%%%%%%%%%%%%%%%%%%%%%%%%%%%%%%%%%%%%%%%%%%%%%%%%%%%%%%%%%%%%%%%
\section{Formulation of the optimal region}
The five-dimensional optimal region for which ${\hat \chi}_{\rm Com}^2<C_{\rm Com}^2$ as illustrated in Fig. 4 (blue contours) can be formulated by fitting the region to a minimum volume ellipsoid \cite{moshtagh2005minimum} as,

\begin{align}
% \begin{eqnarray}
V\times M\times V^{\dagger}\leq1,\label{eq:ellipsoid}
% \end{eqnarray}
\end{align}where, 
\begin{center}
$M = \begin{bmatrix}
	0.0108   &  0.0011  &  -0.0202   &  0.0062   &  0.0069\\
    0.0011   &  0.0004    & 0.0003   & -0.0011   &  0.0007\\
   -0.0202  &   0.0003    & 0.1101   & -0.0269   & -0.0131\\
    0.0062   & -0.0011   & -0.0269   &  0.0508   &  0.0239\\
    0.0069    & 0.0007   & -0.0131   &  0.0239   &  0.0213\\
\end{bmatrix}, $    

$V = \begin{bmatrix}
   J_1 + 0.4481 &  K + 5.5136  &  \Gamma  - 0.1627 &  J_2 + 0.2228 &  J_3 - 1.4543
\end{bmatrix},     
$
\end{center}

%%%%%%%%%%%%%%%%%%%%%%%%%%%%%%%%%%%%%%%%%%%%%%%%%%%%%%%%%%%%%%%%%%%%%%%%%%%%%%%%%%%%%%%%%%%%%%%%%%%%%%%%%%%%%%%%%%%%%%%%%%%%%%%%%%%%
\section{Comparison of machine-learned parameter set with some of the previously reported values}

In Table~\ref{tbl_litsets:3} we show $\chi^2_\mathrm{Com}$ evaluated for some models in the literature. 
By this metric the optimized parameter set obtained by our machine learning method performs best with a value of $\chi^2_\mathrm{Com}\approx 5$ --- but it is expected to do so by design. However, no other model produces $\chi^2_\mathrm{Com}<10$, while three models are found to produce $\chi^2_\mathrm{Com}$ values in the $[10,20]$ range. It is remarkable that our optimized parameter set and the model of Ref.~\cite{10.1038/s41467-017-01177-0} have significantly different $\chi^2_\mathrm{Com}$ values despite appearing close in terms of interaction parameters---perhaps $J_2$ plays an important role? We stress that our treatment here neglects other small terms --- both in some of the models in Table~\ref{tbl_litsets:3} and in determination of the fitness landscape. This may be too drastic an approximation, as suggested particularly by models with larger $\Gamma'$ values such as in Ref.~\cite{PhysRevResearch.2.033011}. For this reason we suggest the $\chi^2_\mathrm{Com}$ values be interpreted with care. 
In addition, our SCS predicts a ferromagnetic ground state for the model of Ref.~\cite{Li2021} rather than zigzag, which likely contributes substantially to the high $\chi^2_\mathrm{Com}$ value.

\begin{table}[!htb]
\begin{tabular}{| m{16em} | m{4em} | m{4em} | m{4em} | m{4em}| m{4em} |m{4em} |} 
% \begin{tabular}{|c c c c c c c c c c c|} 
 \hline
{\bf Reference} & 	{\bf $J_1$ }& 	{\bf $K$ }& 	{\bf $\Gamma$ }& 	{\bf $J_2$ }& 	{\bf $J_3$  }& {\bf $\hat{\chi}^2_{\rm Com}$ }\\
 \hline
    Optimal ML parameters (this work)        &               -0.3961  &   -5.2731  &   0.15  &    -0.1935   &  1.3761   &  4.95 \\
 \hline
    Hou et al. \cite{PhysRevB.96.054410}           &         -1.87    &   -10.7   &   3.8       &     0     &  1.27    &  13.78 \\
 \hline
    Suzuki et al. \cite{Suzuki2021}    & -3    &   -5  & 2.5   & 0 & 0.75  & 18.6\\
 \hline
    Winter et al. PRB \cite{PhysRevB.93.214431}       &       -5.5    &     7.6    &  8.4     &       0    &    2.3    & 19.58 \\
 \hline
    Winter et al. NC \cite{10.1038/s41467-017-01177-0}       &         -0.5      &    -5     & 2.5      &      0      &  0.5   &  26.96 \\
 \hline
    Wu et al. \cite{PhysRevB.98.094425}            &         -0.35     &   -2.8   &   2.4  &          0    &   0.34   &  28.70 \\
 \hline
 Ozel et al. \cite{PhysRevB.100.085108}         &      -0.95     &   1.15   &  3.8     &       0       &   0    & 35.75 \\
 \hline
    Kim and Kee \cite{PhysRevB.93.155143}           &         -3.5    &     4.6   &  6.42    &        0    &      0   &   36.27 \\
 \hline
    Kim et al. \cite{PhysRevB.91.241110, PhysRevB.96.064430}               &       -12      &    17    &   12     &       0    &      0    & 38.23 \\
 \hline
    Li et al. \cite{Li2021}    &   -2.5    & -25   &   7.5 &   0   &   0   &   38.424\\
 \hline
    Banerjee et al. \cite{Banerjee2016}      &         -4.6     &      7    &    0   &         0   &       0  &   38.54 \\
 \hline
    Laurell and Okamoto \cite{Laurell2020}    &    -1.3     &  -15.1   &    10     &       0     &   0.9  &   38.65 \\
 \hline
    Cookmeyer and Moore \cite{PhysRevB.98.060412}     &       -0.5       &   -5   &   2.5  &          0    & 0.1125  &   38.70 \\
 \hline
    Ran et al. (2017) \cite{PhysRevLett.118.107203}               &        0    &    -6.8  &    9.5  &          0   &       0    & 39.04 \\
 \hline
    Suzuki and Suga \cite{PhysRevB.97.134424, Suzuki2019}      &          -1.53     &  -24.4     &5.25      &      0   &       0  &   40.03 \\
 \hline
    Ozel et al. \cite{PhysRevB.100.085108}           &     0.46     &   -3.5   &  2.35       &     0     &     0  &    40.12 \\
 \hline
    Ran et al. (2022) \cite{KejingRan:27501}   & 0 & -7.2  & 5.6   & 0 &   0   & 40.339\\
 \hline
    Wang et al. \cite{PhysRevB.96.115103}          &          -0.3   &    -10.9  &    6.1     &       0   &    0.03  &   40.65 \\
 \hline
    \end{tabular}
\caption{$\chi^2_\mathrm{Com}$ for different models approximately described by Eq.~(1) in the main text. For some of the models we have  neglected additional small terms not captured by our model, such as $\Gamma'$. Those models are those of Refs.~\cite{Laurell2020, PhysRevB.93.214431, PhysRevB.97.134424, PhysRevB.96.054410, Suzuki2019, PhysRevB.93.155143}. It is possible that inclusion of those terms and/or usage of parameters that have not been bond-averaged would improve those models' $\chi^2_\mathrm{Com}$ performance.}
\label{tbl_litsets:3}
\end{table}

\newpage
\newpage

\section{Magnetic specific heat}

\begin{figure}[!htb]
    \centering
	\includegraphics[width=0.7\textwidth]{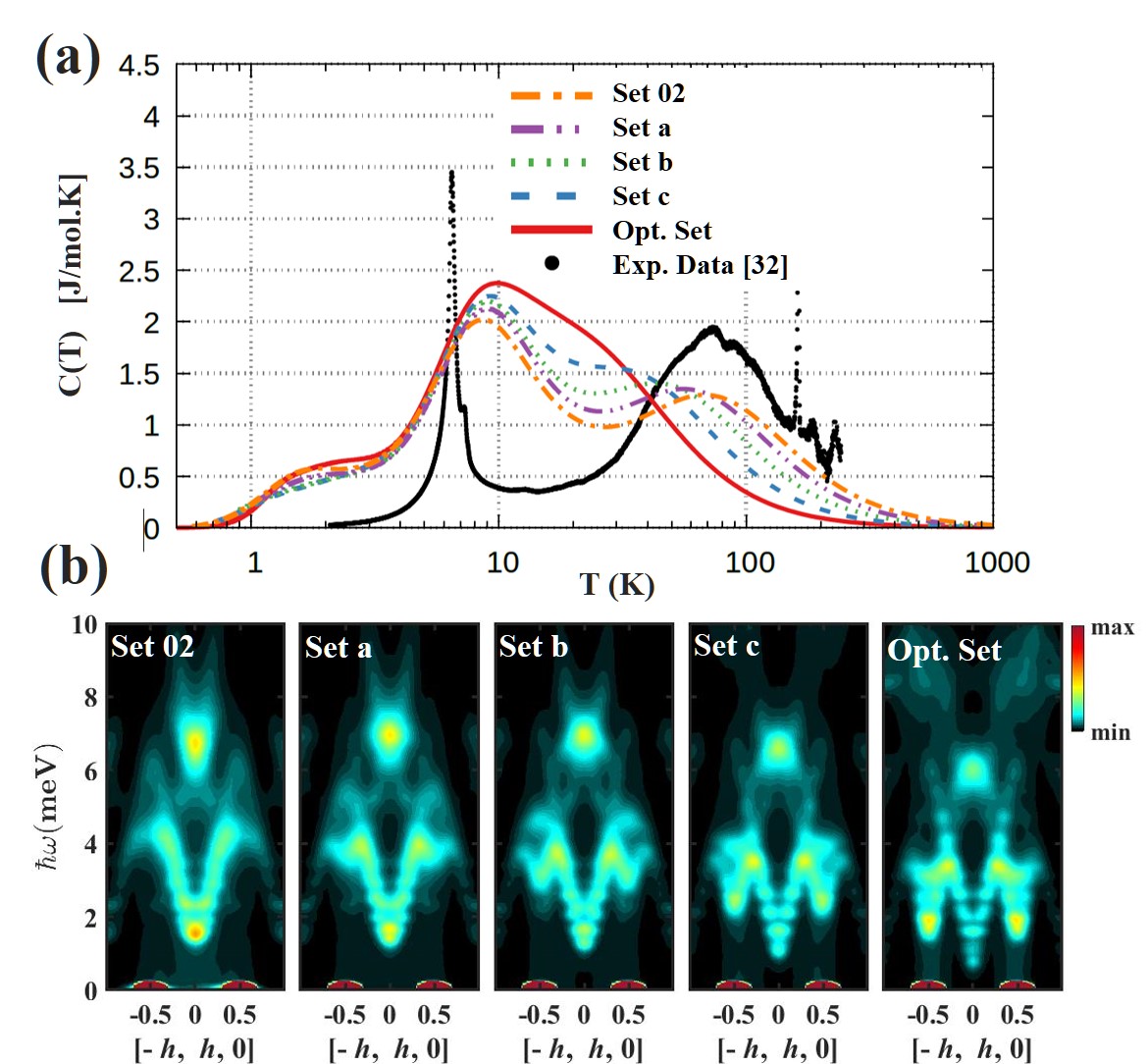}
    \caption{(a) Magnetic specific heat for some of the parameter sets listed in Table \ref{tbl_valsets:1} and (b) the corresponding $S(\mathbf{Q},\omega)$ are shown. In panel (a), the black line shows experimentally determined $C_\mathrm{mag}(T)$ from Ref.~\cite{PhysRevB.99.094415}, which features a peak near $6.5$ K due to onset of magnetic order and a high-temperature peak around $70$ K. The sharp peak at $\approx 170$ K is not a feature of the specific heat, but an artifact of the background subtraction due to a structural transition not present in the nonmagnetic analog. Shown are also mTPQ calculations for five parameters sets near the machine-learning-predicted optimal region. 
%     The solid lines indicate the average, and shaded areas indicate the standard deviation.
The other lines indicates the statistical average over initial vectors. We see that our optimized parameters (Opt. Set) do not capture the high-temperature peak well, whereas Set 2 [see Table~\ref{tbl_valsets:1}] performs better in this regard. This matches the general trend noticed in Ref.~\cite{Laurell2020} that the $C(T)$ behavior is largely controlled by the overall energy scale of the spin Hamiltonian, whereas $S(\mathbf{Q},\omega)$ depends crucially on the balance of different interactions.  Note that Sets a,b and c are linearly connected points between Set 2 and Opt. Set in parameter space as shown in Fig. \ref{figure_GJ3PhaseMap}.  Even though, Set 02 is doing well on $C(T)$, the corresponding $S(\mathbf{Q},\omega)$ does not capture the dispersion at $M-$points as shown in panel (b). Evidently, it is unlikely to find a single region to capture both experimental observations in the current parameter space. However the two regions are not far away in parameter space as shown in Fig. \ref{figure_GJ3PhaseMap} and it might be possible to find a single solution in a further extended parameter space with appropriate extra exchanges. 
    }
    \label{fig:TPQ}
\end{figure}

Magnetic specific heat for some models was calculated using the microcanonical thermal pure quantum state (mTPQ) method \cite{PhysRevLett.108.240401} implemented in the $\mathcal{H}\Phi$ library \cite{Kawamura2017}. 
Similarly to the ED calculations, the $24$-site cluster in Fig.~\ref{figure_EDnet} was used. To reduce statistical error, results were averaged over 15 initial random vectors. The results are shown in Fig.~\ref{fig:TPQ}. These findings are not compared with SCS results since the specific heat of classical and quantum models are fundamentally different. In future work one may co-optimize parameters to fit both neutron data and other observables, e.g. $C(T)$.

\newpage

\bibliography{kitaev,MachineLearning,otherRef}